\documentclass[hyper]{JHEP3} 

\usepackage{epsfig}
\usepackage{amsmath}
\usepackage{cite}
\usepackage{axodraw}
\usepackage{lscape}
\usepackage{multirow}

\newcommand{\beq}{\begin{equation}}
\newcommand{\eeq}{\end{equation}}

\title{Using Subsystem $M_{T2}$ for Complete Mass Determinations in Decay Chains %
with Missing Energy at Hadron Colliders}

\author{Michael Burns\\
        Physics Department, University of Florida,
        Gainesville, FL 32611, USA
        }

\author{Kyoungchul Kong\\
        Theoretical Physics Department, Fermilab,
        Batavia, IL 60510, USA
        }

\author{Konstantin T.~Matchev, Myeonghun Park\\
        Physics Department, University of Florida,
        Gainesville, FL 32611, USA
        }

\received{\today}               
\accepted{\today}               

\preprint{FERMILAB-PUB-08-497-T\\
          UFIFT-HEP-08-16 \\
          February 9, 2009
          } 

\abstract{We propose to use the $M_{T2}$ concept to measure
the masses of {\em all} particles in SUSY-like events with two 
unobservable, identical particles. To this end we generalize the 
usual notion of $M_{T2}$ and define a new $M_{T2}^{(n,p,c)}$
variable, which can be applied 
to various subsystem topologies, as well as the full event topology. 
We derive analytic formulas for its endpoint 
$M_{T2,max}^{(n,p,c)}$ as a function of the unknown test mass $\tilde M_c$
of the final particle in the subchain and the transverse momentum $p_T$
due to radiation from the initial state.
We show that the endpoint functions $M_{T2,max}^{(n,p,c)}(\tilde M_c,p_T)$
may exhibit three different types of kinks and 
discuss the origin of each type. We prove that 
the subsystem $M_{T2}^{(n,p,c)}$ variables by themselves already
yield a sufficient number of measurements for a {\em complete}
determination of the mass spectrum (including the overall mass scale).
As an illustration, we consider the simple
case of a decay chain with up to three heavy 
particles, $X_2\to X_1\to X_0$, which is rather problematic
for all other mass measurement methods. We propose three 
different $M_{T2}$-based methods, each of which allows 
a complete determination of the masses of particles $X_0$, $X_1$ and $X_2$. 
The first method only uses $M_{T2}^{(n,p,c)}$ endpoint measurements
at a single fixed value of the test mass $\tilde M_c$. 
In the second method the unknown mass spectrum is fitted to one or more
endpoint functions $M_{T2,max}^{(n,p,c)}(\tilde M_c,p_T)$  
exhibiting a kink. The third method is hybrid, combining 
$M_{T2}$ endpoints with measurements of kinematic edges in
invariant mass distributions.
As a practical application of our methods, 
we show that the dilepton $W^+W^-$ and $t\bar{t}$ samples at the Tevatron 
can be used for an independent determination of 
the masses of the top quark, the $W$ boson and the neutrino,
without any prior assumptions. 
}


\begin{document} 

\section{Introduction}
\label{sec:intro}

The ongoing Run II of the Fermilab Tevatron and the
now commencing run of the Large Hadron Collider (LHC) 
at CERN are on the hunt for new physics beyond the 
Standard Model (BSM) at the TeV scale. Arguably the 
most compelling {\em phenomenological} evidence for 
BSM particles and interactions at the TeV scale is 
provided by the dark matter problem \cite{Bertone:2004pz},
whose solution requires new particles and interactions BSM.
A typical particle dark matter candidate does not interact
in the detector and can only manifest itself as missing energy. 
The dark matter problem therefore greatly motivates the
study of missing energy signatures at the Tevatron and 
the LHC \cite{Hubisz:2008gg}. 

The long lifetime of the dark matter particle is typically ensured by 
some new exact symmetry\footnote{Some well known examples are:
$R$-parity in supersymmetry \cite{Jungman:1995df}, 
KK parity in models with extra dimensions 
\cite{Appelquist:2000nn,Cheng:2002ab,Agashe:2007jb,Arrenberg:2008wy},
$T$-parity in Little Higgs models \cite{Cheng:2003ju,Birkedal:2006fz}, 
$U$-parity \cite{Hur:2007ur,Lee:2008pc} etc.}, under which the SM
particles are neutral, while the BSM particles are charged.
This setup implies that the new particles will be pair-produced, 
and each of the two cascades will terminate in the dark matter
candidate, giving rise to missing energy in the detector.
(A generic example of this topology is shown in Fig.~\ref{fig:metevent}.)
Since the energies and momenta of the
final two invisible particles $X_0$ are not measured, one cannot
directly apply resonance mass reconstruction 
techniques\footnote{See, however, Section~\ref{sec:exact}.}.
This represents a significant challenge for determining the masses $M_i$
of the new particles $X_i$. 
In recognition of this problem, there has been a recent 
resurgence of interest in the development of different methods 
for mass measurements in
cascade decays with missing energy \cite{Hinchliffe:1996iu,Lester:1999tx,
Bachacou:1999zb,Hinchliffe:1999zc,Allanach:2000kt,Barr:2003rg,Nojiri:2003tu,
Kawagoe:2004rz,Gjelsten:2004ki,Gjelsten:2005aw,Birkedal:2005cm,Miller:2005zp,
Meade:2006dw,Gjelsten:2006tg,Matsumoto:2006ws,Cheng:2007xv,Lester:2007fq,
Cho:2007qv,Gripaios:2007is,Barr:2007hy,Cho:2007dh,Ross:2007rm,Nojiri:2007pq,Huang:2008ae,
Nojiri:2008hy,Tovey:2008ui,Nojiri:2008ir,Cheng:2008mg,Cho:2008cu,Serna:2008zk,Bisset:2008hm,
Barr:2008ba,Kersting:2008qn,Nojiri:2008vq,Burns:2008cp,Cho:2008tj,Cheng:2008hk}. 
Most of these techniques fall into one of the following three categories:

\FIGURE[ht]{
{
\unitlength=1.5 pt
\SetScale{1.5}
\SetWidth{1.0}      
\normalsize    
{} \qquad\allowbreak
\begin{picture}(250,100)(0,0)
\SetColor{Gray}
\Line( 30,65)( 50,95)
\Line( 80,65)(100,95)
\Line(120,65)(140,95)
\Line(160,65)(180,95)
\Line(200,65)(220,95)
\Line( 30,35)( 50, 5)
\Line( 80,35)(100, 5)
\Line(120,35)(140, 5)
\Line(160,35)(180, 5)
\Line(200,35)(220, 5)
\Line(0,65)(50,65)
\Line(0,35)(50,35)
\SetColor{Red}
\Text( 70, 70)[c]{\Red{$X_n$}}
\Text(140, 70)[c]{\Red{$X_2$}}
\Text(180, 70)[c]{\Red{$X_1$}}
\Text(220, 70)[c]{\Red{$X_0$}}
\Text( 70, 40)[c]{\Red{$X_n$}}
\Text(140, 40)[c]{\Red{$X_2$}}
\Text(180, 40)[c]{\Red{$X_1$}}
\Text(220, 40)[c]{\Red{$X_0$}}
\SetWidth{1.2}      
\Line(50,65)(80,65)
\Line(50,35)(80,35)
\DashLine(80,65)(120,65){2}
\DashLine(80,35)(120,35){2}
\Line(120,65)(240,65)
\Line(120,35)(240,35)
\Text( 45,95)[r]{\Black{ISR}}
\Text( 45, 5)[r]{\Black{ISR}}
\Text( 97,95)[r]{\Black{$x_n$}}
\Text(137,95)[r]{\Black{$x_3$}}
\Text(177,95)[r]{\Black{$x_2$}}
\Text(217,95)[r]{\Black{$x_1$}}
\Text( 97, 5)[r]{\Black{$x_n$}}
\Text(137, 5)[r]{\Black{$x_3$}}
\Text(177, 5)[r]{\Black{$x_2$}}
\Text(217, 5)[r]{\Black{$x_1$}}
\Text(15,70)[c]{\Black{$p(\bar{p})$}}
\Text(15,40)[c]{\Black{$p(\bar{p})$}}
\COval(50,50)(30,10)(0){Blue}{Green}
\end{picture}
}
\caption{The generic event topology under consideration in this paper.
The particles $X_i, 1\le i\le n$, are new BSM particles which appear as 
promptly decaying, on-shell intermediate resonances. 
The particles $x_i$ are the corresponding SM decay products, 
which are all visible in the detector, i.e.~we assume that 
there are no neutrinos among them. ISR stands for generic 
initial state radiation with total transverse momentum $\vec{p}_T$.
$X_0$ is a BSM particle which is invisible in the detector. 
The integer $n$ counts the total number of {\em intermediate} BSM particles 
in each chain, so that the {\em total} number of BSM particles 
in each chain is $n+1$. 
For simplicity, in this paper we shall only consider 
symmetric events, in which the two decay chains are identical.
The generalization of our methods to asymmetric decay chains 
is straightforward.}
\label{fig:metevent} 
}

\begin{itemize}
\item {\bf I. Endpoint methods.} They rely on the kinematic endpoints 
\cite{Hinchliffe:1996iu,Bachacou:1999zb,Hinchliffe:1999zc,%
Allanach:2000kt,Gjelsten:2004ki,Gjelsten:2005aw,Gjelsten:2006tg}
or shapes \cite{Birkedal:2005cm,Miller:2005zp,Burns:2008cp}
of various invariant mass distributions constructed out of the 
visible (SM) decay products $x_i$ in the cascade chain.
\item {\bf II. Polynomial methods.} Here one attempts exact event reconstruction 
using the measured momenta of the SM particles and the measured
missing transverse momentum
\cite{Nojiri:2003tu,Kawagoe:2004rz,Cheng:2007xv,Nojiri:2008ir,Cheng:2008mg}.
\item {\bf III. $M_{T2}$ methods.} These methods explore the 
transverse invariant mass variable $M_{T2}$
originally proposed in \cite{Lester:1999tx} and later used and developed in
\cite{Barr:2003rg,Meade:2006dw,Matsumoto:2006ws,Lester:2007fq,Tovey:2008ui,Cho:2008cu,Serna:2008zk,Nojiri:2008vq}.
Recently it was shown that under certain circumstances,
the endpoint of the $M_{T2}$ distribution, when considered 
as a function of the unknown test mass $\tilde M_0$
of the lightest new particle $X_0$,
exhibits a kink and the true mass $M_0$ of $X_0$, i.e. at $\tilde M_0=M_0$
\cite{Cho:2007qv,Gripaios:2007is,Barr:2007hy,Cho:2007dh,Nojiri:2008hy}.
\end{itemize}
One could also combine two or more of these techniques
into a hybrid method, e.g.~a mixed polynomial and endpoint method \cite{Nojiri:2007pq},
a mixed $M_{T2}$ and endpoint method \cite{Ross:2007rm,Barr:2008ba},
or a mixed $M_{T2}$ and polynomial method \cite{Cho:2008tj,Cheng:2008hk}.
In Section~\ref{sec:methods} we shall describe in detail
each of these three basic approaches {\bf I - III}. 
We shall then contrast them to each other and discuss
their pros and cons. In particular, we shall
concentrate on their applicability as a function of the length of the decay chain,
i.e. the number $n$ of intermediate resonances in Fig.~\ref{fig:metevent}.
We shall find that for sufficiently long decay chains, namely $n\ge 3$,
each method {\bf I} - {\bf III} {\em by itself} is able 
to completely determine the unknown particle spectrum, 
at least as a matter of principle. Therefore, if Nature 
is so kind to us as to present us with such a long decay chain, 
it does not really matter which of the three methods above 
we decide to use -- sooner or later, success will be guaranteed 
with each one. 

However, if the decay chain happens to be relatively short,
i.e. $n\le 2$, neither method {\bf I}, nor method {\bf II},
nor a hybrid combination of {\bf I} and {\bf II} will 
be able to completely determine the unknown particle mass spectrum.
In contrast, method {\bf III} {\em by itself} can still
provide a sufficient number of measurements for a {\em complete}
determination of the mass spectrum of the new particles.
We argue that in order to achieve this, the conventional 
$M_{T2}$ variable needs to be promoted to a more general 
quantity $M_{T2}^{(n,p,c)}$, 
which can be applied not only to the whole event, but also to 
a particular sub-chain starting at $X_p$ and ending in 
$X_c$ \cite{Serna:2008zk,Nojiri:2008vq}.
We present the basic steps for this generalization in 
Section~\ref{sec:mt2def}, 
where we also introduce our conventions and notation.
Then in Section~\ref{sec:example} we concentrate on the 
problematic case of $n\le 2$ and discuss what type of 
$M_{T2}^{(n,p,c)}$ measurements are available in that case.
We then show that the newly defined $M_{T2}^{(n,p,c)}$ 
may also exhibit a kink in the graph of its endpoint $M_{T2,max}^{(n,p,c)}$  
as a function of the test mass $\tilde M_c$. 
In order to be able to properly interpret this kink, we derive 
analytic expressions for the function $M_{T2,max}^{(n,p,c)}(\tilde M_c,p_T)$,
including the effect of initial state radiation (ISR)
with some arbitrary transverse momentum $p_T$ (see Fig.~\ref{fig:metevent}).
In all cases, a kink would always appear at $\tilde M_c = M_c$:
\begin{equation}
\left(\frac{\partial M_{T2,max}^{(n,p,c)}(\tilde M_c,p_T)}{\partial \tilde M_c}\right)_{\tilde M_c=M_c-\epsilon}
\ne
\left(\frac{\partial M_{T2,max}^{(n,p,c)}(\tilde M_c,p_T)}{\partial \tilde M_c}\right)_{\tilde M_c=M_c+\epsilon} ,
\label{kink_npc}
\end{equation}
and the value of $M_{T2,max}^{(n,p,c)}$ at that point
reveals the true mass $M_p$ of the mother particle $X_p$:
\begin{equation}
M_{T2,max}^{(n,p,c)}(M_c,p_T) = M_p\ .
\label{truenpc}
\end{equation}
However, there may be up to three different reasons for the
origin of the kink (\ref{kink_npc}).
For example, in the case of $M_{T2,max}^{(1,1,0)}(\tilde M_0,p_T)$
and $M_{T2,max}^{(2,2,1)}(\tilde M_1,p_T)$
with non-zero $p_T$, the kink arises due to recoils against the 
ISR jets, as explained in \cite{Gripaios:2007is,Barr:2007hy}
(see Sections \ref{sec:mt2_110} and \ref{sec:mt2_221} below).
On the other hand,
in the case of $M_{T2,max}^{(2,2,0)}(\tilde M_0,p_T=0)$, 
the kink is due to the variable mass of the 
composite system of SM particles $\{x_1x_2\}$, as already observed in 
Refs.~\cite{Cho:2007qv,Cho:2007dh}
(see Section \ref{sec:mt2_220} below). 
Finally, for $M_{T2,max}^{(2,1,0)}(\tilde M_0,p_T=0)$,
we encounter a new type of kink, which arises due to the decay
of a heavier particle (in this case $X_2$) upstream
(see Section \ref{sec:mt2_210}).

In Section~\ref{sec:our} we propose three different methods 
for measuring the masses of all the particles in 
the problematic case of an $n=2$ decay chain. With the first method, 
presented in Section~\ref{sec:ourmt2}, we always consider a 
fixed value of the test mass (for convenience we choose it to be zero),
and perform a sufficient number of $M_{T2}^{(n,p,c)}$ endpoint 
measurements for various $n$, $p$ and $c$. In our second method, 
described in Section~\ref{sec:ourkink}, we choose a suitable
$M_{T2}^{(n,p,c)}$ variable whose endpoint $M_{T2,max}^{(n,p,c)}$
exhibits a kink, and then fit for the function $M_{T2,max}^{(n,p,c)}(\tilde M_c,p_T)$.
Our last method, presented in Section~\ref{sec:ourim}, 
is hybrid, in the sense that we use combined information
from the measured endpoints of some $M_{T2}$ distributions,
as well as the measured endpoints of certain invariant mass 
distributions. Neither of our three methods relies on
reconstructing the actual momentum of each missing particle. 

All of our discussion throughout the paper
will be completely model-independent
and can be applied to any BSM scenario, including supersymmetry, 
extra dimensions, little Higgs theory etc. 
In Section \ref{sec:our}, however, we shall use a specific
example in order to illustrate 
each of our three proposed methods.
Instead of considering a decay chain of some BSM model,
we chose to select an example which is already present
in the Tevatron data, and will soon be tested at the LHC as well:
the dilepton event samples from top quark pair production
and from $W$-pair production.
Those two dilepton samples satisfy all of our assumptions, 
and would be a perfect
testing ground for any new ideas about mass measurements in
missing energy events from new physics.
In Section \ref{sec:our} we will show that using any one 
of our three $M_{T2}$-based mass measurement methods, one can 
in principle determine the mass of each of the three particles:
top quark, $W$-boson, and neutrino, independently 
and in a completely model-independent fashion.
Section \ref{sec:conclusions} contains a summary and a discussion 
of our main results. 
In Appendix \ref{app:mt2max} we collect all relevant formulas 
for the endpoint functions $M_{T2,max}^{(n,p,c)}(\tilde M_c,p_T)$. 

\section{Mass measurement methods in missing energy events}
\label{sec:methods}

Let us now discuss in some detail each of the three basic 
methods {\bf I} - {\bf III} for mass measurements in missing energy events.
The basic topology is shown in Fig.~\ref{fig:metevent}, where
the particles $X_i$, $0\le i\le n$, (denoted in red) are new BSM particles, and
the particles $x_i$, $1\le i\le n$, (denoted in black) are the corresponding SM decay products.
ISR stands for generic 
initial state radiation with total transverse momentum $\vec{p}_T$.
$X_0$ is a dark matter candidate which is invisible in the detector. 
For simplicity, in this paper we shall make two assumptions,
each of which can be easily relaxed without significantly 
changing our conclusions. First, we shall assume
that the intermediate particles $X_i$, $1\le i\le n$, are all on-shell,
i.e. their masses $M_i\equiv M_{X_i}$ obey the hierarchy
\begin{equation}
M_{n} > M_{n-1} > \ldots > M_1 > M_0\ .
\label{masshierarchy}
\end{equation}
Consequently, all decays along our decay chain are two-body, i.e.~each 
SM decay product $x_i$ is a single particle of mass $m_i\equiv m_{x_i}$. 
In this paper we will also only concentrate on the commonly encountered
case where $x_i$ is either a lepton, photon or jet, 
i.e.~massless:
\begin{equation}
m_i = 0, \qquad i=1,\ldots, n\ .
\end{equation}
Second, we shall also assume that our events are symmetric, 
i.e. the two decay chains are identical. Again, this assumption 
can be easily dropped and one could consider asymmetric events 
as well (see, for example, Ref.~\cite{Nojiri:2008vq}).
These assumptions are made only for simplicity. 
All of our subsequent discussion
can be easily generalized to include off-shell
decays, simply by promoting some of the visible
SM particles $x_i$ to composite particles with variable mass. 
For example, Appendix \ref{app:mt2max} already contains some 
results for an off-shell case, while the study of asymmetric
events is postponed for future work~\cite{BKMP}.

We shall use the integer $n$ to count the total number of 
{\em intermediate} on-shell BSM particles in each chain. 
Then, the total number of new BSM particles 
in the decay chain is $n+1$. 
With those preliminaries, we are ready to discuss each of the three 
different methods for mass measurements in missing energy events.

\subsection{Endpoint method}
\label{sec:imd}

With this method, one forms the invariant mass distributions
$M_{x_{i_1}x_{i_2}...x_{i_k}}$ of various groups of $k$
SM decay products $x_i$, where the number $k$ in principle 
can range from 2 to $n$. Each such distribution exhibits an 
{\em upper} kinematic endpoint, which can be related to the
underlying unknown masses $M_{i}$. If one makes a sufficient 
number of independent upper endpoint measurements, the system of equations
giving the kinematic endpoints $E_j$ in terms of the masses $M_i$
\begin{equation}
E_j = E_j(M_0,\ldots,M_n), \qquad j=0,\ldots,n
\end{equation}
can be solved for the masses $M_i$, although on some occasions the 
solution may not be unique --  see, e.g.~\cite{Gjelsten:2006as,MP1}. 

Clearly, the method will be fully successful only if the number of 
measurements $N_{m}$ is no less than the number of unknown parameters $N_p$.
For the decay chain of Fig.~\ref{fig:metevent}, the number of unknown
mass parameters $N_p$ is simply the total number of BSM particles:
\begin{equation}
N_p = n+1 \ .
\label{Np_im}
\end{equation}
How many measurements $N_m$ are available with this method?
The answer to this question depends on the length
of the decay chain. It is easy to see that,
if $n=1$, there are no endpoint measurements at all; 
if $n=2$, there is a single measurement of the endpoint of the $M_{x_1x_2}$
distribution, etc. In general, for an arbitrary fixed $n$,
the number of different invariant mass distributions 
$M_{x_{i_1}x_{i_2}...x_{i_k}}$ that one can form and study, is equal to the number 
of ways in which we can select a group of at least two objects 
from a set of $n$ objects, and is given by
\begin{equation}
N_m = 2^n - (n + 1)\ .
\label{Nm_im}
\end{equation}
Strictly speaking, eq.~(\ref{Nm_im}) only gives an upper bound
on the number of independent {\em upper} endpoints in the 
invariant mass distributions. Indeed, there are certain cases where
not all of the upper kinematic endpoints are independent. 
For example, consider the 
familiar case of a squark decay chain in supersymmetry:
$X_3=\tilde q$, $X_2=\tilde\chi^0_2$, $X_1=\tilde\ell$ and $X_0=\tilde\chi^0_1$.
The SM decay products consist of a quark jet $q$ and two leptons:
$\ell^+$ and $\ell^-$. It is well known
that in some regions of parameter space the upper endpoint
of the $M_{q\ell\ell}$ distribution does not provide an independent 
measurement, since it can be related to the upper endpoints of the 
$M_{\ell^+\ell^-}$ and $M_{q\ell(high)}\equiv \max\{M_{q\ell^+},M_{q\ell^-}\}$
distributions \cite{Gjelsten:2004ki}. Fortunately, 
one can use additional measurements from the {\em lower} endpoints
of suitably restricted invariant mass distributions, 
e.g.~$M_{qll(\theta>\frac{\pi}{2})}$ \cite{Bachacou:1999zb}.
We see that the precise count of the number of measurements $N_m$ available 
in the endpoint method is somewhat model-dependent, but
nevertheless, the estimate (\ref{Nm_im}) is sufficient
to make our main point below.

From eqs.~(\ref{Np_im}) and (\ref{Nm_im}) it readily follows 
that the number of undetermined parameters with this method is
\begin{equation}
N_p - N_m = 2(n+1) - 2^n \ .
\label{Ndif_im}
\end{equation}
The dependence of this quantity on the length of the 
decay chain is plotted in Fig.~\ref{fig:param} with red open circles,
connected with red line segments. The yellow-shaded region in the figure 
is where $N_m \ge N_p$, so that we have a sufficient number of measurements 
for a {\em complete} determination of the heavy particle spectrum. 
Conversely, whenever a symbol appears inside the white region, 
where $N_m < N_p$, there is only partial information about the mass spectrum
and the spectrum cannot be fully determined.

\FIGURE[t]{
\epsfig{file=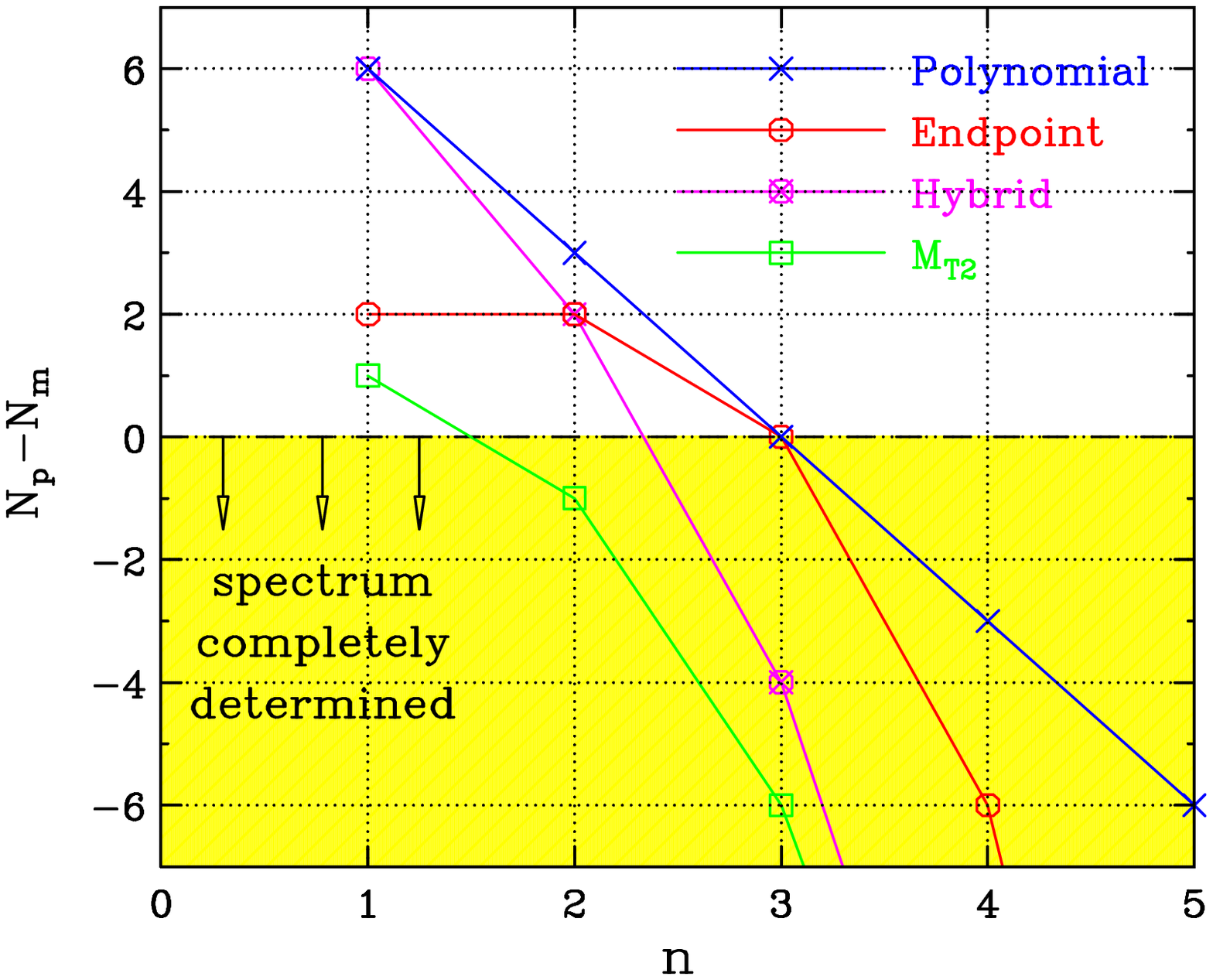,width=9cm}
\caption{\sl The dependence of the number of undetermined parameters 
$N_p-N_m$ as a function of the number $n$ of intermediate heavy resonances in
the decay chains of Fig.~\ref{fig:metevent}, for 
various mass determination methods: $M_{T2}$ method (green, open squares),
endpoint method (red, open circles), polynomial method for $N_{ev}=2$
(blue, $\times$ symbols), or a hybrid method which is a combination
of the latter two methods (magenta, $\otimes$ symbols). Within the 
yellow-shaded region the number of unknowns $N_p$ does not exceed the 
number of measurements $N_m$ for the corresponding method, and the 
mass spectrum can be completely determined.}
\label{fig:param}}

Fig.~\ref{fig:param} reveals that the endpoint method cannot
succeed unless $n\ge 3$. This conclusion has already 
been confirmed by numerous studies of various low-energy 
SUSY models, where one considers a decay chain of sufficient length:
$n=3$ as in the squark example mentioned above, or
$n=4$ as for a gluino chain \cite{Gjelsten:2005aw}.
On the other hand, if $n=1$ or $n=2$, 
with this method we are unable to pin down {\em all}
of the new particle masses, even as a matter of principle.
These are exactly the cases where the additional information
from mass measurements at future lepton colliders
has been seen as extremely useful \cite{Weiglein:2004hn}.

\subsection{Polynomial method}
\label{sec:exact}

The basic idea behind the method is to use all of the available experimental 
information in each event, and enforce a sufficient number of 
constraints, which would allow to actually solve for the 
unknown momenta of the missing particles $X_0$. Before
we analyze this method in more detail, let us introduce
some of our notations and conventions. We shall use lowercase 
letters to denote various quantities relating to the SM particles 
$x_i$, $1\le i\le n$. At the same time, we shall use a superscript $(k)$
to denote whether a particular quantity belongs to the upper ($k=1$) 
or lower ($k=2$) decay chain in Fig.~\ref{fig:metevent}.
For example, the 4-momentum of the SM particle $x_i$ in the $k$-th chain 
will be denoted as $p_i^{(k)}$, the corresponding transverse momentum 
will be $\vec{p}_{iT}^{\,(k)}$, while the mass of $x_i$ will simply be $m_i$.
On the other hand, uppercase letters will denote quantities relating 
to the BSM particles $X_i$, $0\le i\le n$. Thus the 4-momentum 
of the BSM particle $X_i$ appearing in the $k$-th chain is $P_i^{(k)}$, 
the corresponding transverse momentum is $\vec{P}_{iT}^{(k)}$, 
and the mass is $M_i$. One should keep in mind that for SM particles 
the index $i$ runs from $1$ to $n$, while for BSM particles 
$i$ runs from $0$ to $n$.

In order to apply the polynomial method, one uses the experimentally 
measured 4-momenta $p^{(k)}_i$ as well as the missing 
transverse momentum $\vec{p}_{T,miss}$ in the event.
Then, one imposes the mass shell constraints for 
the intermediate BSM particles $X_i$ and tries to 
solve the resulting system of equations for the 8 unknown
components of the 4-momenta $P_0^{(k)}$ of the missing particles $X_0$.
Including the $n+1$ unknown masses $M_i$, this amounts to a total of
\begin{equation}
N_p = 8 + (n + 1)= n+9 
\label{Np_pol1}
\end{equation}
unknown parameters.
How many measurements (constraints) are present in this case?
First, there is a total of $2(n+1)$ mass-shell conditions:
one for each BSM particle $X_i$ in each of the two decay chains 
in Fig.~\ref{fig:metevent}
\begin{equation}
M_i^2 = \left( P^{(1)}_i\right)^2 = \left( P^{(2)}_i\right)^2 \ ,
\qquad i=0,1,\ldots, n\ .
\label{massshell}
\end{equation}
Using energy and momentum conservation
\begin{equation}
P_i^{(k)}= P_0^{(k)}+\sum_{j=1}^i p^{(k)}_j\ , \qquad k=1,2\ ,
\label{EPcons}
\end{equation}
these constraints can be rewritten in terms of the unknown variables 
$P_0^{(k)}$:
\begin{equation}
M_i^2 
= \left( P_0^{(1)} + \sum_{j=1}^i p^{(1)}_j\right)^2 
= \left( P_0^{(2)} + \sum_{j=1}^i p^{(2)}_j\right)^2 \ ,
\qquad i=0,1,\ldots, n\ .
\label{massshell2}
\end{equation}
Furthermore, the measurement of the missing transverse momentum 
$\vec{p}_{T,miss}$ provides two additional constraints
\begin{equation}
\vec{P}^{(1)}_{0T} + \vec{P}^{(2)}_{0T} = \vec{p}_{T,miss} \label{misspt} 
\end{equation}
on the unknown transverse momentum components $\vec{P}^{(k)}_{0T}$. Therefore, 
the total number of measurements is
\begin{equation}
N_m = 2(n+1) + 2 = 2n + 4
\label{Nm_pol1}
\end{equation}
and the number of undetermined parameters for any given event is 
readily obtained from (\ref{Np_pol1}) and (\ref{Nm_pol1})
\begin{equation}
N_p - N_m = 5 - n \ .
\label{Ndif_pol1}
\end{equation}
However, one might do better than this, by combining the information from
two or more events \cite{Nojiri:2003tu,Kawagoe:2004rz}. 
For example, consider another event of the same type.
Since the $n+1$ unknown masses were already counted in eq.~(\ref{Np_pol1}),
the second event introduces only 8 new parameters 
(the 4-momenta of the two $X_0$ particles in the second event),
bringing up the total number of unknowns in the two events to
\begin{equation}
N_p = 8 + 8 + (n + 1)= n+17 \ .
\label{Np_pol2}
\end{equation}
At the same time, the constraints (\ref{massshell2})
and (\ref{misspt}) are still valid for the second event,
which results in $2n+4$ additional constraints. 
This brings the total number of constraints to
\begin{equation}
N_m = (2n+4) + (2n+4) = 4n + 8\ .
\label{Nm_pol2}
\end{equation}
Subtracting (\ref{Np_pol2}) and (\ref{Nm_pol2}), we get
\begin{equation}
N_p - N_m = 9 - 3n \ .
\label{Ndif_pol2}
\end{equation}
Comparing our previous result (\ref{Ndif_pol1}) 
with (\ref{Ndif_pol2}), we see that the latter
decreases much faster with $n$, therefore, when using the polynomial method,
combining information from two different events is beneficial for large $n$
(in this example, for $n \ge 3$).

Following the same logic, one can generalize this parameter counting 
to the case where the polynomial method is applied for sets of $N_{ev}$ 
different events at a time. The number of unknown parameters is
\begin{equation}
N_p = n + 1 + 8 N_{ev} \ ,
\label{Np_polNev}
\end{equation}
the number of constraints is
\begin{equation}
N_m = (2n+4)N_{ev}\ ,
\label{Nm_polNev}
\end{equation}
and therefore, the number of undetermined parameters is given by
\begin{equation}
N_p - N_m = n+1 -2(n-2)N_{ev} \ .
\label{Ndif_polNev}
\end{equation}
For $N_{ev}=1$ and $N_{ev}=2$ this equation reduces to (\ref{Ndif_pol1}) and 
(\ref{Ndif_pol2}), respectively. What is the optimal number of events 
$N_{ev}$ for the polynomial method? The answer can be readily obtained 
from eq.~(\ref{Ndif_polNev}), where $N_{ev}$ enters the last term on
the right-hand side. If this term is negative, increasing $N_{ev}$
would decrease the number of undetermined parameters, therefore 
it would be beneficial to combine information from more and more 
different events. From eq.~(\ref{Ndif_polNev}) we see that 
this would be case if the decay chain is sufficiently long, 
i.e.~$n \ge 3$. On the other hand, when $n = 1$, considering
more than one event at a time is actually detrimental - we are
adding more unknowns than constraints. In the case of $n=2$,
the number of undetermined parameters $N_p-N_m$ is 
actually independent of $N_{ev}$ and one might as well consider 
the simplest case of $N_{ev}=1$. 

Let us now analyze how successful the polynomial method will be for 
different decay chain lengths. The number of undetermined parameters 
(\ref{Ndif_pol2}) for $N_{ev}=2$ is plotted in Fig.~\ref{fig:param} 
with blue $\times$ symbols, connected with blue line segments. 
We see that the polynomial method will be successful in determining 
all the masses of the BSM particles only if $n\ge 3$.
For $n=1$ or $n=2$, there will not be enough measurements
for a complete mass determination\footnote{As can be seen from the more
general expression (\ref{Ndif_polNev}), this conclusion will not
change even if we consider arbitrarily large number of events $N_{ev}$.}
and the best one can do in that case is to obtain a {\em range of}
possible values for the masses $M_0$, $M_1$ and $M_2$ \cite{Cheng:2007xv}.
Recall that in the previous 
subsection we reached a similar conclusion regarding the endpoint method.
Therefore, we see that both the endpoint and the 
polynomial methods, when used in isolation, would
fail whenever the decay chain is rather short: $n=1$ or $n=2$.
This represents a definite problem, since there is no guarantee 
that the new physics would exhibit a long ($n\ge3$) decay chain.
Therefore it is worth investigating whether there is an alternative 
method which would be successful in those two cases, i.e.~$n\le 2$.

One immediate idea which comes to mind is to use a hybrid method, i.e.
combining the techniques of the polynomial and endpoint methods \cite{Nojiri:2007pq}.
The parameter count in that case is very easy to do.
The number of unknown parameters is the same 
as in the polynomial method:
\begin{equation}
N_p = n + 1 + 8 N_{ev} \ .
\label{Np_hybNev}
\end{equation}
Now, however, we need to account for the extra measurements (\ref{Nm_im})
which are available from the endpoint method. Therefore, 
the total number of measurements for a hybrid method of this type
is the sum of (\ref{Nm_im}) and (\ref{Nm_polNev}):
\begin{equation}
N_m = 2^n - (n+1) + (2n+4)N_{ev}\ .
\label{Nm_hybNev}
\end{equation}
Subtracting (\ref{Np_hybNev}) and (\ref{Nm_hybNev}), we get
\begin{equation}
N_p - N_m = 2(n+1) - 2^n - 2(n-2)N_{ev} \ .
\label{Ndif_hyb}
\end{equation}
In Fig.~\ref{fig:param}, this quantity is plotted for fixed $N_{ev}=2$
with magenta $\otimes$ symbols. We see that, even though
the hybrid method performs better than the individual
endpoint and polynomial methods, it still cannot 
solve the problem of masses for $n=2$!
Therefore, a different approach is needed.
We shall now argue that the $M_{T2}$ method might just 
provide the solution to this problem in the $n=2$ case.
What is more, we shall show that the $M_{T2}$
method can do that all {\em by itself}, without 
using any information derived from the endpoint 
or polynomial methods.

\subsection{$M_{T2}$ method}
\label{sec:mt2}

Here the number of unknown parameters is still
\begin{equation}
N_p = n+1 \ .
\label{Np_mt2}
\end{equation}
In the next section we shall prove that, once we consider
the notion of subsystem $M_{T2}^{(n,p,c)}$, the total number of 
$M_{T2}$-type endpoint measurements is
\begin{equation}
N_m = \frac{1}{6}n(n+1)(n+2)\ .
\label{Nm_mt2}
\end{equation}
Then the number of undetermined parameters with this method is
\begin{equation}
N_p - N_m = \frac{1}{6}(n+1)(6-2n-n^2) \ ,
\label{Ndif_mt2}
\end{equation}
which is plotted with green square symbols in 
Fig.~\ref{fig:param}. We see that for $n\le 3$, the
$M_{T2}$ method is by far the most powerful, and more importantly, 
it is the only method which is able to handle the problematic 
case of $n=2$!

\section{Defining a subsystem $M_{T2}$ variable}
\label{sec:mt2def}

The idea for a subsystem $M_{T2}$ was first discussed
in \cite{Serna:2008zk} and applied in \cite{Nojiri:2008vq}
for a specific supersymmetry example
(associated squark-gluino production and decay). 
Here we shall generalize that concept
for a completely general decay chain. 
For this purpose, let us redraw Fig.~\ref{fig:metevent}
as shown in Fig.~\ref{fig:metevent2}. The subsystem
$M_{T2}$ variable will be defined for the subchain
inside the blue (yellow-shaded) box in Fig.~\ref{fig:metevent2}.
Before we give a formal definition of the subsystem $M_{T2}$
variables, let us first introduce 
some terminology for the BSM particles appearing 
in the decay chain. We shall find it convenient to
distinguish the following types of BSM particles:
\FIGURE[ht]{
{
\unitlength=1.5 pt
\SetScale{1.5}
\SetWidth{1.0}      
\normalsize    
{} \qquad\allowbreak
\begin{picture}(350,100)(0,0)
\CBoxc(172,50)(102,100){Blue}{Yellow}
\SetColor{Gray}
\Line( 30,65)( 50,95)
\Line( 30,35)( 50, 5)
\Line( 80,65)(100,95)
\Line( 80,35)(100, 5)
\Line(140,65)(160,95)
\Line(140,35)(160, 5)
\Line(200,65)(220,95)
\Line(200,35)(220, 5)
\Line(260,65)(280,95)
\Line(260,35)(280, 5)
\Line(0,65)(50,65)
\Line(0,35)(50,35)
\SetColor{Red}
\Text( 70, 70)[c]{\Red{$X_n$}}
\Text( 70, 40)[c]{\Red{$X_n$}}
\Text( 95, 70)[c]{\Red{$X_{n-1}$}}
\Text( 95, 40)[c]{\Red{$X_{n-1}$}}
\Text(130, 70)[c]{\Red{$X_p$}}
\Text(130, 40)[c]{\Red{$X_p$}}
\Text(155, 70)[c]{\Red{$X_{p-1}$}}
\Text(155, 40)[c]{\Red{$X_{p-1}$}}
\Text(190, 70)[c]{\Red{$X_{c+1}$}}
\Text(190, 40)[c]{\Red{$X_{c+1}$}}
\Text(215, 70)[c]{\Red{$X_c$}}
\Text(215, 40)[c]{\Red{$X_c$}}
\Text(253, 70)[c]{\Red{$X_1$}}
\Text(253, 40)[c]{\Red{$X_1$}}
\Text(275, 70)[c]{\Red{$X_0$}}
\Text(275, 40)[c]{\Red{$X_0$}}
\SetWidth{1.2}      
\Line(50,65)(80,65)
\Line(50,35)(80,35)
\Line(80,65)(100,65)
\Line(80,35)(100,35)
\DashLine(100,65)(120,65){2}
\DashLine(100,35)(120,35){2}
\Line(120,65)(140,65)
\Line(120,35)(140,35)
\Line(140,65)(160,65)
\Line(140,35)(160,35)
\DashLine(160,65)(180,65){2}
\DashLine(160,35)(180,35){2}
\Line(180,65)(200,65)
\Line(180,35)(200,35)
\Line(200,65)(225,65)
\Line(200,35)(225,35)
\DashLine(225,65)(245,65){2}
\DashLine(225,35)(245,35){2}
\Line(245,65)(260,65)
\Line(245,35)(260,35)
\Line(260,65)(290,65)
\Line(260,35)(290,35)
\Text( 45,95)[r]{\Black{ISR}}
\Text( 45, 5)[r]{\Black{ISR}}
\Text( 97,95)[r]{\Black{$x_n$}}
\Text( 97, 5)[r]{\Black{$x_n$}}
\Text(157,95)[r]{\Black{$x_p$}}
\Text(157, 5)[r]{\Black{$x_p$}}
\Text(217,95)[r]{\Black{$x_{c+1}$}}
\Text(217, 5)[r]{\Black{$x_{c+1}$}}
\Text(277,95)[r]{\Black{$x_1$}}
\Text(277, 5)[r]{\Black{$x_1$}}
\Text(15,70)[c]{\Black{$p(\bar{p})$}}
\Text(15,40)[c]{\Black{$p(\bar{p})$}}
\COval(50,50)(30,10)(0){Blue}{Green}
\end{picture}
}
\caption{An alternative representation of Fig.~\ref{fig:metevent},
which illustrates the meaning of the subsystem 
$M_{T2}^{(n,p,c)}$ 
variable defined in eq.~(\ref{mt2inc}). }
\label{fig:metevent2} 
}
\begin{itemize}
\item {\em ``Grandparents''}. Those are the two BSM particles 
$X_n$ at the very top of the decay chains in Fig.~\ref{fig:metevent2}.
Since we have assumed symmetric events, the two grandparents
in each event are identical, and carry the same index $n$. 
Of course, one may relax this assumption, and consider 
asymmetric events, as was done in \cite{Nojiri:2008hy,Nojiri:2008vq}. 
Then, the two ``grandparents'' will be different, and 
one would simply need to keep track of two separate 
grandparent indices $n^{(1)}$ and $n^{(2)}$.
\item {\em ``Parents''}. Those are the two BSM particles 
$X_p$ at the top of the subchain used to define the subsystem 
$M_{T2}$ variable. In Fig.~\ref{fig:metevent2} this subchain is
identified by the blue (yellow-shaded) rectangular box.
The idea behind the subsystem $M_{T2}$ is simply to apply the
usual $M_{T2}$ definition for the subchain inside this box.
Notice that the $M_{T2}$ concept usually requires the parents 
to be identical, therefore here we will characterize them by
a single ``parent'' index $p$.
\item {\em ``Children''}. Those are the two BSM particles 
$X_c$ at the very end of the subchain used to define the subsystem 
$M_{T2}$ variable, as indicated by the blue (yellow-shaded) 
rectangular box in Fig.~\ref{fig:metevent2}. 
The children are also characterized by a single
index $c$. In general, the true mass $M_c$ of the two children 
is unknown. As usual, when calculating the value of the $M_{T2}$ variable, one 
needs to choose a child ``test'' mass, which we shall denote with a tilde,
$\tilde M_c$, in order to distinguish it from the true mass $M_c$
of $X_c$. 
\item {\em Dark matter candidates}. Those are the two stable neutral particles
$X_0$ appearing at the very end of the cascade chain. We see that while 
those are the particles responsible for the measured missing 
momentum in the event (see eq.~(\ref{misspt})),
they are relevant for $M_{T2}$ only in the special case of $c=0$. 
\end{itemize}

With those definitions, we are now ready to generalize
the conventional $M_{T2}$ definition \cite{Lester:1999tx,Barr:2003rg}.
From Fig.~\ref{fig:metevent2} we see that any subchain is
specified by the parent index $p$ and the child index $c$,
while the total length of the whole chain (and thus the type of event)
is given by the grandparent index $n$. 
Therefore, the subsystem $M_{T2}$ variable will
have to carry those three indices as well, and we shall use the
notation $M_{T2}^{(n,p,c)}$. In the following we shall refer to 
this generalized quantity as either ``subsystem'' or ``subchain'' 
$M_{T2}$. It is clear that
the set of three indices $(n,p,c)$ must be ordered as follows:
\begin{equation}
n\ge p > c \ge 0\ .
\end{equation}
We shall now give a formal definition 
of the quantity $M_{T2}^{(n,p,c)}$, generalizing the original
idea of $M_{T2}$ \cite{Lester:1999tx,Barr:2003rg}.
The parent and child indices $p$ and $c$ uniquely define a
subchain, within which one can form the transverse masses 
$M_T^{(1)}$ and $M_T^{(2)}$ of the two parents:
\begin{equation}
M_{T}^{(k)}(p^{(k)}_{p},p^{(k)}_{p-1},\ldots,p^{(k)}_{c+1},\vec{P}^{(k)}_{cT};\tilde M_c), \qquad k=1,2\ .
\label{MT_parents}
\end{equation}
Here $p^{(k)}_{i}$, $c+1\le i\le p$, are the measured 4-momenta
of the SM particles within the subchain, $\vec{P}^{(k)}_{cT}$
are the unknown transverse momenta of the children, while
$\tilde M_c$ is their unknown (test) mass. Then,
the subsystem $M_{T2}^{(n,p,c)}$ is defined by minimizing
the larger of the two transverse masses (\ref{MT_parents})
over the allowed values of the children's transverse momenta $\vec{P}^{(k)}_{cT}$:
\begin{equation}
M_{T2}^{(n,p,c)}(\tilde M_c)= 
\min_{\sum_{k=1}^2 \vec{P}^{(k)}_{cT}= - \sum_{k=1}^2 \sum_{j=c+1}^n \vec{p}^{\,(k)}_{jT} - {\vec{p}}_{T} }
\left\{\max\left\{M_{T}^{(1)},M_{T}^{(2)}\right\} \right\}\ ,
\label{mt2inc}
\end{equation}
where $\vec{p}_{T}$ indicates any additional transverse momentum 
due to initial state radiation (ISR) (see 
Figs.~\ref{fig:metevent} and \ref{fig:metevent2}).
Notice that in this definition, 
the dependence on the grandparent index $n$ enters only through 
the restriction on the children's transverse momenta $\vec{P}^{(k)}_{cT}$.
Using momentum conservation in the transverse plane
\begin{equation}
\sum_{k=1}^2 \vec{P}^{(k)}_{0T} + \sum_{k=1}^2 \sum_{j=1}^n \vec{p}^{\,(k)}_{jT}  
+ {\vec{p}}_{T}= 0\ ,
\label{balmom}
\end{equation}
we can rewrite the restriction on the children's transverse momenta $\vec{P}^{(k)}_{cT}$
as
\begin{equation}
\sum_{k=1}^2 \vec{P}^{(k)}_{cT} 
= \sum_{k=1}^2 \vec{P}^{(k)}_{0T} + \sum_{k=1}^2 \sum_{j=1}^c \vec{p}^{\,(k)}_{jT} 
= \vec{p}_{T,miss} + \sum_{k=1}^2 \sum_{j=1}^c \vec{p}^{\,(k)}_{jT}\ , \label{Pcsum}
\end{equation}
where in the last step we used eq.~(\ref{misspt}).
Eq.~(\ref{Pcsum}) allows us to rewrite the subsystem $M_{T2}^{(n,p,c)}$
definition (\ref{mt2inc}) in a form which does not manifestly 
depend on the grandparent index $n$:
\begin{equation}
M_{T2}^{(n,p,c)}(\tilde M_c)= 
\min_{\sum_{k=1}^2 \vec{P}^{(k)}_{cT}= \vec{p}_{T,miss} + \sum_{k=1}^2 \sum_{j=1}^c \vec{p}^{\,(k)}_{jT}  }
\left\{\max\left\{M_{T}^{(1)},M_{T}^{(2)}\right\} \right\}\ .
\label{mt2inc2}
\end{equation}
However, the grandparent index $n$ is still implicitly present through the 
global quantity $\vec{p}_{T,miss}$, which knows about the whole event.
We shall see below that the interpretation of the experimentally observable 
endpoints, kinks, etc., for the so defined subsystem 
$M_{T2}^{(n,p,c)}$ quantity, does depend on the grandparent index $n$,
which justifies our notation.

We are now in a position to compare our subsystem $M_{T2}^{(n,p,c)}$ quantity to the 
conventional $M_{T2}$ variable. The latter is nothing but the special case of
$n=p$ {\em and} $c=0$:
\begin{equation}
M_{T2} \equiv M_{T2}^{(n,n,0)}\ ,
\label{mt2old}
\end{equation}
i.e. the conventional $M_{T2}$ is simply characterized by a single integer n, 
which indicates the length of the decay chain. We see that we are generalizing the
conventional $M_{T2}$ variable in two different aspects: first, we are allowing 
the parents $X_p$ to be different from the particles $X_n$ originally produced in the event 
(the grandparents),
and second, we are allowing the children $X_c$ to be different from the dark matter
particles $X_0$ appearing at the end of the cascade chain and responsible for the 
missing energy. The benefits of this generalization will become apparent 
in the next section, where we shall discuss the available measurements
from the different subsystem $M_{T2}^{(n,p,c)}$ variables.

In conclusion of this section, let us derive the result (\ref{Nm_mt2}) used in 
Section~\ref{sec:mt2}. We count how many different subsystem $M_{T2}^{(n,p,c)}$ quantities
(\ref{mt2inc}) exist for a given maximum value $n$ of the grandparent index. 
First, pick a parent index $p$, which can range from $1$ to $n$. 
Then, for this fixed value of $p$, the child index $c$ can take a total of 
$p$ values: $0\le c\le p-1$, while the grandparent index
can take\footnote{Note that different values of the grandparent index $n$ correspond to 
different types of events. For example, in order to study the $M_{T2}^{(n_1,p,c)}$
variables for some given $n_1$, we must look at events of $X_{n_1}$ pair-production, 
while in order to form the $M_{T2}^{(n_2,p,c)}$ distributions for another
value $n_2<n_1$, we must look at events of $X_{n_2}$ pair-production. 
Because of the mass hierarchy (\ref{masshierarchy}), the observation of 
events of the former type ($X_{n_1}$ pair-production) guarantees that the 
collider will eventually be able to also produce events of the latter type 
($X_{n_2}$ pair-production). Therefore, the relevant integer for our 
count is the maximum value of $n$ achievable at a given collider experiment.} 
a total of $n-p+1$ values. Therefore, the total number of allowed combinations $(n,p,c)$ is
\begin{equation}
\sum_{p=1}^{n} p\, (n-p+1) = 
\frac{1}{6}\, n\, (n+1)\, (n+2)\ ,
\label{mt2count}
\end{equation}
in agreement with (\ref{Nm_mt2}).

\section{A short decay chain $X_2\to X_1\to X_0$}
\label{sec:example}

As we already discussed in Sec.~\ref{sec:methods}, a relatively long
($n\ge 3$) new physics decay chain can be handled by a variety of mass 
measurement methods, and in principle a complete determination of the 
mass spectrum in that case {\em is possible} at a hadron collider. 
We also showed that a relatively short ($n=1$ or $n=2$)
decay chain would present a major challenge, and a complete mass 
determination might be possible only through $M_{T2}$ methods. 
From now on we shall therefore concentrate only on this most
problematic case of $n\le 2$. 

First let us summarize what types of subsystem $M_{T2}^{(n,p,c)}$
measurements are available in the case of $n\le 2$. According to
eq.~(\ref{mt2count}), there exist a total of 4 different 
$M_{T2}^{(n,p,c)}$ quantities, which are illustrated in Fig.~\ref{fig:n1n2}.
\FIGURE[ht]{
{
\unitlength=1.5 pt
\SetScale{1.5}
\SetWidth{1.0}      
\normalsize    
{} \qquad\allowbreak
\begin{picture}(350,110)(0,-10)
\CBoxc(84,50)(42,100){Blue}{White}
\CBoxc(235,55)(103,120){Blue}{White}
\CBoxc(213,50)(46,100){Blue}{White}
\CBoxc(260,50)(42,100){Blue}{White}
\SetColor{Gray}
\Line(10,65)(50,65)
\Line(10,35)(50,35)
\Line(130,65)(170,65)
\Line(130,35)(170,35)
\Line( 30,65)( 50,95)
\Line( 30,35)( 50, 5)
\Line(150,65)(170,95)
\Line(150,35)(170, 5)
\Line( 80,65)(100,95)
\Line( 80,35)(100, 5)
\Line(210,65)(230,95)
\Line(210,35)(230, 5)
\Line(250,65)(270,95)
\Line(250,35)(270, 5)
\SetColor{Red}
\Text( 60, -15)[c]{(a)}
\Text(237, -15)[c]{(b)}
\Text( 84, 50)[c]{\Blue{$M_{T2}^{(1,1,0)}$}}
\Text(213, 50)[c]{\Blue{$M_{T2}^{(2,2,1)}$}}
\Text(260, 50)[c]{\Blue{$M_{T2}^{(2,1,0)}$}}
\Text(243,108)[c]{\Blue{$M_{T2}^{(2,2,0)}$}}
\Text( 70, 70)[c]{\Red{$X_1$}}
\Text( 70, 40)[c]{\Red{$X_1$}}
\Text( 95, 70)[c]{\Red{$X_0$}}
\Text( 95, 40)[c]{\Red{$X_0$}}
\Text(200, 70)[c]{\Red{$X_2$}}
\Text(200, 40)[c]{\Red{$X_2$}}
\Text(230, 70)[c]{\Red{$X_1$}}
\Text(230, 40)[c]{\Red{$X_1$}}
\Text(265, 70)[c]{\Red{$X_0$}}
\Text(265, 40)[c]{\Red{$X_0$}}
\SetWidth{1.2}      
\Line(50,65)(80,65)
\Line(50,35)(80,35)
\Line(80,65)(110,65)
\Line(80,35)(110,35)
\Line(170,65)(290,65)
\Line(170,35)(290,35)
\Text( 45,95)[r]{\Black{ISR}}
\Text( 45, 5)[r]{\Black{ISR}}
\Text( 97,95)[r]{\Black{$x_1$}}
\Text( 97, 5)[r]{\Black{$x_1$}}
\Text(165,95)[r]{\Black{ISR}}
\Text(165, 5)[r]{\Black{ISR}}
\Text(227,95)[r]{\Black{$x_2$}}
\Text(227, 5)[r]{\Black{$x_2$}}
\Text(267,95)[r]{\Black{$x_1$}}
\Text(267, 5)[r]{\Black{$x_1$}}
\Text(20,70)[c]{\Black{$p(\bar{p})$}}
\Text(20,40)[c]{\Black{$p(\bar{p})$}}
\Text(140,70)[c]{\Black{$p(\bar{p})$}}
\Text(140,40)[c]{\Black{$p(\bar{p})$}}
\COval(50,50)(30,10)(0){Blue}{Green}
\COval(170,50)(30,10)(0){Blue}{Green}
\end{picture}
}
\caption{The subsystem $M_{T2}^{(n,p,c)}$ 
variables which are available for (a) $n=1$ and (b) $n=2$ events.}
\label{fig:n1n2} 
}
Each $M_{T2}^{(n,p,c)}$ distribution would exhibit an 
upper endpoint $M_{T2,max}^{(n,p,c)}$, whose measurement
would provide one constraint on the physical masses.
In order to be able to invert and solve for the masses
of the new particles in terms of the measured endpoints, 
we need to know the analytical expressions
relating the endpoints $M_{T2,max}^{(n,p,c)}$ to the
physical masses $M_i$. In this section we summarize those
relations for each $M_{T2}^{(n,p,c)}$ quantity 
with $n\le 2$. Some of these results 
(e.g.~portions of Secs.~\ref{sec:mt2_110} and Secs.~\ref{sec:mt2_220})
have already appeared in the literature,
and we include them here for completeness. 
The discussion in Secs.~\ref{sec:mt2_221} and 
Secs.~\ref{sec:mt2_210}, on the other hand, is new.
In all cases, we shall allow for the presence of an arbitrary
transverse momentum $p_T$ due to ISR. This represents a 
generalization of all existing results in the literature, 
which have been derived in the two special cases
$p_T=0$ \cite{Cho:2007dh} or $p_T=\infty$ \cite{Barr:2007hy}.

We shall find it convenient to write the
formulas for the endpoints $M_{T2,max}^{(n,p,c)}$
not in terms of the actual masses, but in terms of the 
mass parameters
\begin{equation}
\mu_{(n,p,c)} \equiv \frac{M_n}{2}\, \left( 1-\frac{M_c^2}{M_p^2}  \right)\ .
\label{mu_npc}
\end{equation}
The advantage of using this shorthand notation will become 
apparent very shortly. Notice that not all of the $\mu$ parameters
defined in (\ref{mu_npc}) are independent. For a given maximum value 
of $n$, the total number of $\mu$ parameters from (\ref{mu_npc})
is the same as the total number of subsystem $M_{T2}$ variables and
is given by (\ref{Nm_mt2}). All of those $\mu$ parameters 
are functions of just $n+1$ masses $M_i$, $0\le i\le n$,
as indicated by eq.~(\ref{Np_mt2}). Therefore, the $\mu$ 
parameters must obey certain relations, whose number is given by 
(\ref{Ndif_mt2}). For example, for 
$n\le 2$, we have a total of four $\mu$ parameters: 
$\mu_{(1,1,0)}$, $\mu_{(2,1,0)}$, $\mu_{(2,2,0)}$ and $\mu_{(2,2,1)}$,
and only three masses: $M_0$, $M_1$ and $M_2$, so that there is one constraint:
\begin{equation}
\mu_{(2,1,0)}\, \left( \mu_{(2,2,0)}-\mu_{(2,2,1)} \right) = \mu^2_{(1,1,0)}\ .
\end{equation}


\subsection{The subsystem variable $M_{T2}^{(1,1,0)}$}
\label{sec:mt2_110}

We start with the simplest case of $n=1$ shown in Fig.~\ref{fig:n1n2}(a).
Here $M_{T2}^{(1,1,0)}$ is the only possibility, and it coincides with the
conventional $M_{T2}$ variable, as indicated by (\ref{mt2old}).
Therefore, the previous results in the literature which have been derived for the
conventional $M_{T2}$ variable (\ref{mt2old}), would still apply. 
In particular, in the limit of $p_T=0$,
the upper endpoint $M_{T2,max}^{(1,1,0)}$
depends on the test mass $\tilde M_0$ as follows \cite{Cho:2007dh}
\begin{equation}
M_{T2,max}^{(1,1,0)} (\tilde M_0,p_T=0) = \mu_{(1,1,0)} + \sqrt{\mu_{(1,1,0)}^2 + \tilde M_0^2}\ , 
\label{end110}
\end{equation}
where the parameter $\mu_{(1,1,0)}$ is defined in terms of the
physical masses $M_1$ and $M_0$ according to eq.~(\ref{mu_npc}):
\begin{equation}
\mu_{(1,1,0)} \equiv \frac{M_1}{2}\, \left( 1-\frac{M_0^2}{M_1^2}  \right)
= \frac{M_1^2-M_0^2}{2M_1}\ .
\label{mu110}
\end{equation}
As usual, the endpoint (\ref{end110}) can be interpreted as the 
mass $M_1$ of the parent particle $X_1$, so that eq.~(\ref{end110})
provides a relation between the masses of $X_0$ and $X_1$.
In the early literature on $M_{T2}$, this relation had to be derived numerically, 
by building the $M_{T2}$ distributions for different values 
of the test mass $\tilde M_0$, and reading off their endpoints.
Nowadays, with the work of Ref.~\cite{Cho:2007dh}, the
relation is known analytically, and, as seen from (\ref{end110}),
is parameterized by a single parameter $\mu_{(1,1,0)}$.
Therefore, in order to extract the value of this parameter,
we only need to perform a single measurement, i.e.
we only need to study the $M_{T2}$ distribution for 
{\em one} particular choice of the test mass $\tilde M_0$.
We shall find it convenient to choose $\tilde M_0=0$,
in which case eqs.~(\ref{end110}) and (\ref{mu110}) give
\begin{equation}
M_{T2,max}^{(1,1,0)} (\tilde M_0=0,p_T=0) = 2\, \mu_{(1,1,0)}
= \frac{M_1^2-M_0^2}{M_1}\ ,
\label{mT2110}
\end{equation}
providing the required measurement of the parameter $\mu_{(1,1,0)}$.
Eq.~(\ref{mT2110}) demonstrates the usefulness of the $M_{T2}$ concept --
just a single measurement of the endpoint of the $M_{T2}$ distribution
for a single fixed value of the test mass $\tilde M_0$ is sufficient to
provide us with one constraint among the unknown masses 
($M_1$ and $M_0$ in this case).

Unfortunately, one single measurement (\ref{mT2110}) 
is not enough to pin down two different masses.
In order to measure {\em both} $M_0$ and $M_1$, 
without any theoretical assumptions or prejudice,
we obviously need additional experimental input.
From the general expression (\ref{end110})
it is clear that measuring other $M_{T2,max}^{(1,1,0)}$
endpoints, for different values of the test mass $\tilde M_0$,
will not help, since we will simply be measuring the same
combination of masses $\mu_{(1,1,0)}$ over and over again,
obtaining no new information. Another possibility might be to
consider events with the next longest decay chain ($n=2$), which,
as advertised in the Introduction and shown below in
Section~\ref{sec:our}, will be able to provide enough information 
for a complete mass determination of all particles $X_0$, $X_1$ and $X_2$.
However, the existence and the observation of the $n=2$ decay 
chain is certainly not guaranteed -- to begin with, 
the particles $X_2$ may not exist, 
or they may have too low cross-sections. It is therefore 
of particular importance to ask the question whether
the $n=1$ process in Fig.~\ref{fig:n1n2}(a) alone can
allow a determination of both $M_0$ and $M_1$.
As shown in Ref.~\cite{Barr:2007hy},
the answer to this question, at least in principle,  
is ``Yes'', and what is more, one can achieve this 
using the very same $M_{T2}$ variable $M_{T2}^{(1,1,0)}$.

The key is to realize that in reality at any collider, 
and especially at hadron colliders
like the Tevatron and the LHC, there will be sizable contributions
from initial state radiation (ISR) with nonzero $p_T$, where one or more jets
are radiated off the initial state, before the hard scattering 
interaction. (In Figs.~\ref{fig:metevent}, \ref{fig:metevent2} 
and \ref{fig:n1n2} the green ellipse represents the hard scattering,
while ``ISR'' stands for a generic ISR jet.).
This effect leads to a drastic change in the 
behavior of the $M_{T2,max}^{(1,1,0)}(\tilde M_0,p_T)$ function,
which starts to exhibit a kink at the true location 
of the child mass $\tilde M_0=M_0$:
\begin{equation}
\left(\frac{\partial M_{T2,max}^{(1,1,0)}(\tilde M_0,p_T)}{\partial \tilde M_0}\right)_{\tilde M_0=M_0-\epsilon}
\ne
\left(\frac{\partial M_{T2,max}^{(1,1,0)}(\tilde M_0,p_T)}{\partial \tilde M_0}\right)_{\tilde M_0=M_0+\epsilon} ,
\label{kink110}
\end{equation}
and furthermore, the value of $M_{T2,max}^{(1,1,0)}$ at that point
reveals the true mass of the parent as well:
\begin{equation}
M_{T2,max}^{(1,1,0)}(\tilde M_0=M_0,p_T) = M_1\ .
\label{true110}
\end{equation}
This kink feature (\ref{kink110},\ref{true110}) was observed and illustrated 
in Ref.~\cite{Barr:2007hy} (see their Sec.~4.4). 
We find that it can also be understood analytically, by
generalizing the result (\ref{end110}) to account 
for the additional ISR transverse momentum $\vec{p}_T$. 
Recall that eq.~(\ref{end110}) was derived in Ref.~\cite{Cho:2007dh}
under the assumption that
the missing transverse momentum due to the two escaping particles $X_0$ 
is exactly balanced by the transverse momenta of the two visible particles 
$x_1$ used to form $M_{T2}^{(1,1,0)}$:
\begin{equation}
\vec{P}^{(1)}_{0T} +  
\vec{P}^{(2)}_{0T} +
\vec{p}^{\,(1)}_{1T}  +
\vec{p}^{\,(2)}_{1T}  = 0\ .
\label{balmom110}
\end{equation}
We may sometimes refer to this situation as a ``balanced'' momentum 
configuration\footnote{This should not be confused with the term ``balanced'' 
used for the analytic $M_{T2}$ solutions discussed in \cite{Lester:2007fq,Cho:2007dh}.}.
In the presence of ISR with some non-zero transverse momentum $\vec{p}_T$, 
eq.~(\ref{balmom110}) in general ceases to be valid, and is modified to
\begin{equation}
\vec{P}^{(1)}_{0T} +  
\vec{P}^{(2)}_{0T} +
\vec{p}^{\,(1)}_{1T}  +
\vec{p}^{\,(2)}_{1T}  = -\vec{p}_T\ ,
\end{equation}
in accordance with (\ref{balmom}).
Including the ISR effects, we find that the expression (\ref{end110})
for the $M_{T2,max}^{(1,1,0)}$ endpoint splits into two branches
\begin{equation}
M_{T2,max}^{(1,1,0)}(\tilde M_0,p_T) =\left\{   
\begin{array}{ll}
F^{(1,1,0)}_{L}(\tilde M_0,p_T)\, , & ~~~{\rm if}\ \tilde M_0 \le M_0\, , \\ [2mm]
F^{(1,1,0)}_{R}(\tilde M_0,p_T)\, , & ~~~{\rm if}\ \tilde M_0 \ge M_0\, , 
\end{array}
\right.
\label{end110PT} 
\end{equation}
where 
\begin{eqnarray}
F_{L}^{(1,1,0)}(\tilde M_0,p_T) &=&
\left\{ \left[
\mu_{(1,1,0)}(p_T) + \sqrt{ \left(\mu_{(1,1,0)}(p_T)+\frac{p_T}{2}\right)^2 + \tilde M_0^2} 
\, \right]^2
- \frac{p_T^2}{4}   \right\}^{\frac{1}{2}},~~ \label{FL110} \\
F_{R}^{(1,1,0)}(\tilde M_0,p_T) &=&
\left\{ \left[
\mu_{(1,1,0)}(-p_T) + \sqrt{ \left(\mu_{(1,1,0)}(-p_T)-\frac{p_T}{2}\right)^2 + \tilde M_0^2} 
\, \right]^2
- \frac{p_T^2}{4}   \right\}^{\frac{1}{2}},~~~~
\label{FR110}
\end{eqnarray}
and the $p_T$-dependent parameter $\mu_{(1,1,0)}(p_T)$ is defined as 
\begin{equation}
\mu_{(1,1,0)}(p_T) = \mu_{(1,1,0)}\, 
\left( \sqrt{ 1+\left(\frac{p_T}{2M_1}\right)^2} - \frac{p_T}{2M_1} \right) \, .
\label{mu110PT}
\end{equation}       
Both branches correspond to extreme momentum configurations in which all 
three transverse vectors $\vec{p}_{1T}^{\,(1)}$, $\vec{p}_{1T}^{\,(2)}$ and 
$\vec{p}_T$ are collinear. The difference is that the left branch $F_{L}^{(1,1,0)}$
corresponds to the configuration 
$\left(\vec{p}_{1T}^{\,(1)} \uparrow\uparrow \vec{p}_{1T}^{\,(2)}\right) \uparrow\uparrow \vec{p}_T$,
while the right branch $F_{R}^{(1,1,0)}$ corresponds to 
$\left(\vec{p}_{1T}^{\,(1)} \uparrow\uparrow \vec{p}_{1T}^{\,(2)}\right) \uparrow\downarrow \vec{p}_T$. 
Therefore, the two branches are simply related as
\begin{equation}
F_{R}^{(1,1,0)}(\tilde M_0,p_T) = F_{L}^{(1,1,0)}(\tilde M_0,-p_T).
\end{equation}
It is easy to verify that in the absence of ISR, (i.e.~for $p_T=0$)
our general result (\ref{end110PT}) reduces to the previous 
formula (\ref{end110}). 

\FIGURE[t]{
\epsfig{file=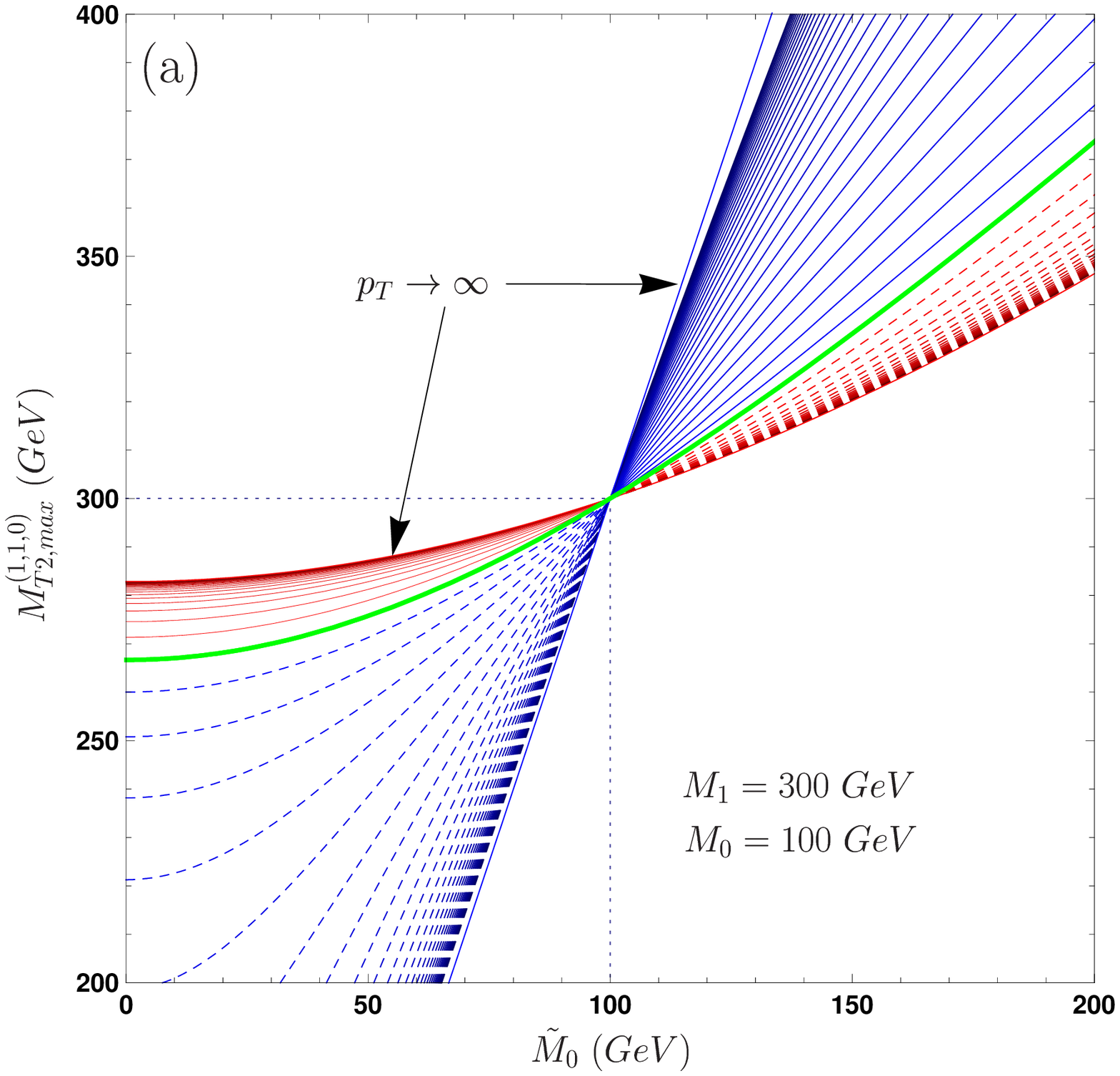,height=6.5cm}
\epsfig{file=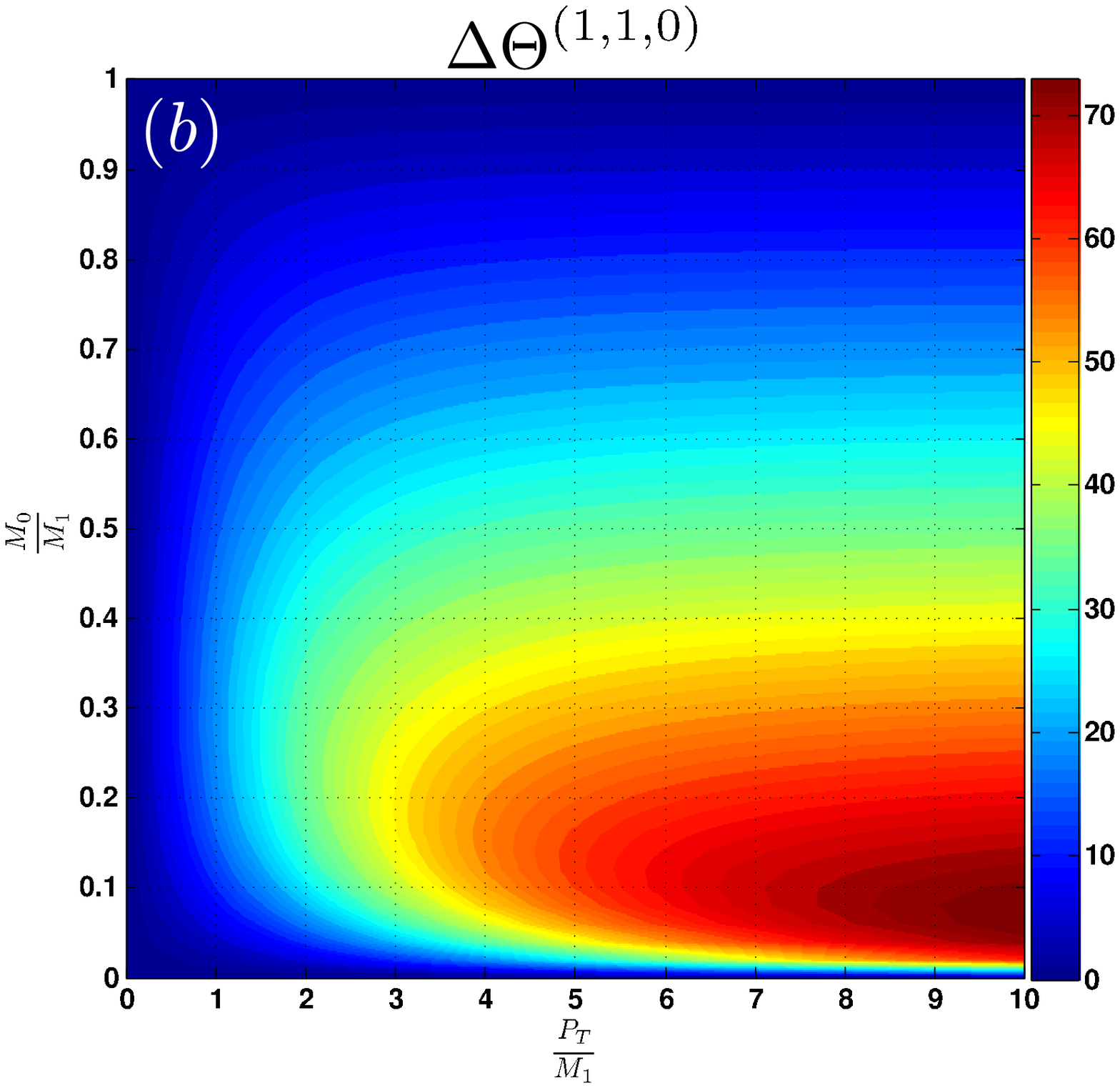,height=5.5cm}
\caption{\sl (a) Dependence of the $M_{T2,max}^{(1,1,0)}$ 
upper kinematic endpoint (solid lines) on the value of the test mass 
$\tilde M_0$, for $M_1=300$ GeV,  and $M_0=100$ GeV,
and for different values of the transverse momentum $p_T$
of the ISR jet, starting from $p_T=0$ (green line),
and increasing up to $p_T=3$ TeV in increments of $\Delta p_T=100$ GeV,
from bottom to top. The uppermost line corresponds to the limiting 
case $p_T\to \infty$.
The horizontal (vertical) dotted line denotes the
true value of the parent (child) mass. 
Solid (dashed) lines indicate true (false) endpoints.
The red lines correspond to the function
$F_L^{(1,1,0)}$ defined in eq.~(\ref{FL110}), while the 
blue lines correspond to the function
$F_R^{(1,1,0)}$ defined in eq.~(\ref{FR110}).
(b) The value of the kink $\Delta\Theta^{(1,1,0)}$
defined in (\ref{dtheta_110}), as a function of
the dimensionless ratios $\frac{p_T}{M_1}$ and $\frac{M_0}{M_1}$.}
\label{fig:kink110}}

Our result (\ref{end110PT}) for the $M_{T2,max}^{(1,1,0)}$ 
upper kinematic endpoint as a function of the test mass $\tilde M_0$
is illustrated in Fig.~\ref{fig:kink110}(a).
We have chosen the same mass spectrum ($M_0=100$ GeV and $M_1=300$ GeV)
as the one used in Ref.~\cite{Barr:2007hy}, so that
our Fig.~\ref{fig:kink110}(a) can be directly compared to
Fig.~9 of Ref.~\cite{Barr:2007hy}. We consider a single ISR jet and
show results for several different values of its transverse momentum $p_T$,
starting from $p_T=0$ (the green solid line) and increasing 
the value of $p_T$ in increments of $\Delta p_T=100$ GeV.
The uppermost solid line corresponds to the limiting 
case $P_{T}\to \infty$. The true value of the parent (child) 
mass is marked by the horizontal (vertical) dotted line.
The red (blue) lines correspond to the function $F_L^{(1,1,0)}$
($F_R^{(1,1,0)}$). The solid portions of those lines 
correspond to the true $M_{T2,max}^{(1,1,0)}$ endpoint,
while the dashed segments are simply the extension 
of $F_L^{(1,1,0)}$ and $F_R^{(1,1,0)}$ into the ``wrong''
region for $\tilde M_0$, giving a false endpoint.

Fig.~\ref{fig:kink110}(a) reveals that the two branches
(\ref{FL110}) and (\ref{FR110}) always cross at the point $(M_0,M_1)$,
in agreement with eq.~(\ref{true110}). Interestingly,
the sharpness of the resulting kink at $\tilde M_0=M_0$
depends on the hardness of the ISR jet, as can be seen 
directly from (\ref{end110PT}). For small $p_T$,
the kink is barely visible, and in the limit $p_T\to 0$ 
we obtain the old result (\ref{end110}) for the ``balanced'' 
momentum configuration, shown with the green solid line,
which does not exhibit any kink. 
In the other extreme, at very large $p_T$, we see 
a pronounced kink, which has a well-defined limit as $p_T\to \infty$.
Our results in this regard are in agreement with the findings 
of Ref.~\cite{Barr:2007hy}. 

The $M_{T2,max}^{(1,1,0)}$ kink exhibited in eq.~(\ref{end110PT})
and in Fig.~\ref{fig:kink110}(a) is our first, but not last,
encounter with a kink feature in an $M_{T2}^{(n,p,c)}$ variable.
Below we shall see that the $M_{T2}$ kinks are rather common 
phenomena, and we shall encounter at least two other kink types
by the end of Sec.~\ref{sec:example}. Therefore, we find it convenient to
quantify the sharpness of any such kink as follows.
Consider a generic subsystem $M_{T2}^{(n,p,c)}$ variable 
whose endpoint $M_{T2,max}^{(n,p,c)}(\tilde M_c,p_T)$ exhibits a kink:
\begin{equation}
M_{T2,max}^{(n,p,c)}(\tilde M_c,p_T) =\left\{   
\begin{array}{ll}
F^{(n,p,c)}_{L}(\tilde M_c,p_T)\, , & ~~~{\rm if}\ \tilde M_c \le M_c\, , \\ [2mm]
F^{(n,p,c)}_{R}(\tilde M_c,p_T)\, , & ~~~{\rm if}\ \tilde M_c \ge M_c\, . 
\end{array}
\right.
\label{endnpc} 
\end{equation}
The kink appears because $M_{T2,max}^{(n,p,c)}(\tilde M_c,p_T)$
is not given by a single function, but has two separate branches.
The first (``low'') branch applies for $\tilde M_c\le M_c$, and is given by 
some function $F_L^{(n,p,c)}(\tilde M_c,p_T)$, while 
the second (``high'') branch is valid for $\tilde M_c\ge M_c$, 
and is given by a different function, $F_R^{(n,p,c)}(\tilde M_c,p_T)$.
The function $M_{T2,max}^{(n,p,c)}(\tilde M_c,p_T)$ itself is continuous and 
the two branches coincide at $\tilde M_c=M_c$:
\begin{equation}
F_L^{(n,p,c)}(M_c,p_T) = F_R^{(n,p,c)} (M_c,p_T)\ ,
\end{equation}
but their {\em derivatives} do not match:
\begin{equation}
\left( \frac{\partial F_L^{(n,p,c)}}{\partial \tilde M_c}\right)_{\tilde M_c=M_c} \ne 
\left( \frac{\partial F_R^{(n,p,c)}}{\partial \tilde M_c}\right)_{\tilde M_c=M_c}\ ,
\end{equation}
leading to the appearance of the kink. Let us define
the left and right slope of the $M_{T2,max}^{(n,p,c)}(\tilde M_c,p_T)$ function
at $\tilde M_c=M_c$ in terms of two angles $\Theta_L^{(n,p,c)}$ and 
$\Theta_R^{(n,p,c)}$, correspondingly:
\begin{eqnarray}
\tan\Theta_L^{(n,p,c)} &\equiv& 
\left( \frac{\partial F_L^{(n,p,c)}(\tilde M_c)}{\partial \tilde M_c}\right)_{\tilde M_c= M_c}
\ ,  \label{ThetaL} \\ [2mm]
\tan\Theta_R^{(n,p,c)} &\equiv& 
\left( \frac{\partial F_R^{(n,p,c)}(\tilde M_c)}{\partial \tilde M_c}\right)_{\tilde M_c= M_c}
\ .  \label{ThetaR}
\end{eqnarray}
Now we shall define the amount of kink as the angular difference $\Delta \Theta^{(n,p,c)}$
between the two branches:
\begin{equation}
\Delta \Theta^{(n,p,c)} \equiv \Theta_R^{(n,p,c)} - \Theta_L^{(n,p,c)} 
= \arctan \left( \frac{\tan\Theta_R^{(n,p,c)}-\tan\Theta_L^{(n,p,c)}}
{1+\tan\Theta_R^{(n,p,c)}\tan\Theta_L^{(n,p,c)}} \right)\ .
\label{DTheta}
\end{equation}
A large value of $\Delta\Theta^{(n,p,c)}$ implies
that the relative angle between the low and high branches 
at the point of their junction $\tilde M_c=M_c$ is also large,
and in that sense the kink would be more pronounced and 
relatively easier to see. 

This definition can be immediately applied to the $M_{T2,max}^{(1,1,0)}$ kink 
that we just discussed. Substituting the formulas (\ref{FL110}) and (\ref{FR110})
for the two branches $F^{(1,1,0)}_{L}$ and $F^{(1,1,0)}_{R}$
into the definitions (\ref{ThetaL},\ref{ThetaR})
and subsequently into (\ref{DTheta}), we obtain an expression for 
the size $\Delta\Theta^{(1,1,0)}$ of the $M_{T2,max}^{(1,1,0)}$ kink:
\begin{equation}
\Delta\Theta^{(1,1,0)} = 
\arctan\left( \frac{M_0\,(M_1^2-M_0^2)\,p_T\,\sqrt{4M_1^2+p_T^2}}{M_1\,(M_1^2-M_0^2)^2+2M_0^2M_1\,(4M_1^2+p_T^2)} \right)\ .
\label{dtheta_110}
\end{equation}
The result (\ref{dtheta_110}) is illustrated numerically in Fig.~\ref{fig:kink110}(b).
As can be seen from (\ref{dtheta_110}), $\Delta\Theta^{(1,1,0)}$
depends on the two masses $M_0$ and $M_1$, as well as the
size of the ISR $p_T$. However, since $\Delta\Theta^{(1,1,0)}$ is a dimensionless 
quantity, its dependence on those three parameters can be simply illustrated in 
terms of the dimensionless ratios $\frac{p_T}{M_1}$ and $\frac{M_0}{M_1}$.
This is why in Fig.~\ref{fig:kink110}(b) we plot
$\Delta\Theta^{(1,1,0)}$ (in degrees) 
as a function of $\frac{p_T}{M_1}$ and $\frac{M_0}{M_1}$. 

Fig.~\ref{fig:kink110}(b) confirms that the kink develops at large $p_T$,
and is completely absent at $p_T=0$, a result which may have already 
been anticipated on the basis of Fig.~\ref{fig:kink110}(a).  
For any given mass ratio $\frac{M_0}{M_1}$,
the kink is largest for the hardest possible $p_T$.
In the limit $p_T\to \infty$ we obtain
\begin{equation}
\lim_{p_T\to \infty} \Delta\Theta^{(1,1,0)} = \arctan\left( \frac{M_1^2-M_0^2}{2M_0M_1}\right)\ ,
\label{ptinf}
\end{equation}
in agreement with the result obtained in \cite{Barr:2007hy}.
From Fig.~\ref{fig:kink110}(b) one can see that at sufficiently large $p_T$,
the $\Delta\Theta^{(1,1,0)}$ contours become almost horizontal, i.e. 
the size of the kink $\Delta\Theta^{(1,1,0)}$ becomes very weakly
dependent on $p_T$. A careful examination of the figure
reveals that the asymptotic behavior at $p_T\to \infty$
is in agreement with the analytical result (\ref{ptinf}).
Notice that the maximum possible value of any kink of the 
type (\ref{endnpc}) is $\Delta\Theta_{max}^{(n,p,c)}=90^{\circ}$. According
to Fig.~\ref{fig:kink110}(b) and eq.~(\ref{ptinf}), in the case of
$\Delta\Theta^{(1,1,0)}$ the absolute maximum
can be obtained only in the $p_T\to \infty$ and $M_0\to 0$ limit.
The former condition will never be realized in a realistic experiment, 
while the latter condition makes the observation of the kink
rather problematic, since the ``low'' branch $F_L$ of the $M_{T2,max}^{(1,1,0)}(\tilde M_0,p_T)$
function is too short to be observed experimentally.
Therefore, under realistic circumstances, we would expect the 
size of the kink $\Delta \Theta^{(1,1,0)}$ to be only on the order of 
a few tens of degrees, which are the more typical values 
seen in Fig.~\ref{fig:kink110}(b).

According to Fig.~\ref{fig:kink110}(b), for a given fixed $p_T$, 
the sharpness of the $\Delta \Theta^{(1,1,0)}$ kink depends on the mass hierarchy of the particles 
$X_1$ and $X_0$. When they are relatively degenerate, i.e.~their mass ratio
$\frac{M_0}{M_1}$ is large, the kink is relatively small. Conversely,
when $X_0$ is much lighter than $X_1$, the kink is more pronounced.
The optimum mass ratio $\frac{M_0}{M_1}$ which maximizes the kink for a given $p_T$,
is rather weakly dependent on the $p_T$, and for $p_T\to \infty$ 
eventually goes to zero, in agreement with eq.~(\ref{ptinf}). 
However, for more reasonable values of 
$p_T$ as the ones shown on the left half of the plot, 
the optimal ratio $\frac{M_0}{M_1}$
varies between 0.3 (at $p_T\sim 0$) to 0.1 (at $p_T\sim 5M_1$).
In this sense, the value of $\frac{M_0}{M_1}=\frac{1}{3}$ which was chosen 
for the illustration in Fig.~\ref{fig:kink110}(a) 
(as well as Fig.~9 in Ref.~\cite{Barr:2007hy}) is rather typical. 

In conclusion of this subsection, it is worth summarizing
the main points from it. The good news is that the $\Delta \Theta^{(1,1,0)}$ 
kink in principle offers a second, independent piece of information 
about the masses of the particles $X_0$ and $X_1$.
When taken together with the $M_{T2,max}^{(1,1,0)}$
endpoint measurement (\ref{mT2110}), it will allow us 
to determine {\em both} masses $M_0$ and $M_1$, in a completely 
model-independent way. Our analytical results regarding
the $\Delta \Theta^{(1,1,0)}$ kink complement the study
of Ref.~\cite{Barr:2007hy}, where this kink was first discovered.
However, on the down side, we should mention that
much of our discussion regarding the $\Delta \Theta^{(1,1,0)}$ 
kink may be of limited practical interest, for several reasons. 
First, as seen in Fig.~\ref{fig:kink110}, the kink becomes visible 
only for sufficiently large values of the $p_T$. Since the ISR $p_T$ 
spectrum is falling rather steeply, one would need to collect 
relatively large amounts of data, in order to guarantee the presence of
events with sufficiently hard ISR jets. Even then, the collected events
may not contain the momentum configuration 
required to give the maximum value of $M_{T2}^{(1,1,0)}$.
An alternative approach to make use of the kink structure would be to
measure the endpoint function $M_{T2,max}^{(1,1,0)}(\tilde M_0,p_T)$
for several different $p_T$ ranges, 
and then fit it to the analytical formula (\ref{end110PT}).
Whether and how well this can work in practice, remains to be seen, 
but the results of \cite{Barr:2007hy} from a toy exercise 
in the absence of any backgrounds and detector resolution effects 
do not appear very encouraging. Nevertheless, while the kink structure
$\Delta\Theta^{(1,1,0)}$ may be difficult to observe, the 
measurement (\ref{mT2110})
of the endpoint $M_{T2,max}^{(1,1,0)}(\tilde M_0=0,p_T=0)$ should 
be relatively straightforward. In Secs.~\ref{sec:ourmt2} 
and \ref{sec:ourim} we shall see that the additional $M_{T2}$
information from events with $n=2$ decay chains will eventually 
allow us to determine all the unknown masses.

\subsection{The subsystem variable $M_{T2}^{(2,2,1)}$}
\label{sec:mt2_221}

The subsystem variable $M_{T2}^{(2,2,1)}$ is illustrated in Fig.~\ref{fig:n1n2}(b),
where we use the subchain within the smaller rectangle on the left.
$M_{T2}^{(2,2,1)}$ is a genuine subchain variable in the sense that 
we only use the SM decay products $x_2$, and ignore any remaining
objects arising from the two $x_1$'s.  
In the absence of ISR ($p_T=0$) one can adapt the results from \cite{Cho:2007dh} 
and show that the formula for the $M_{T2}^{(2,2,1)}$ endpoint is 
\begin{equation}
M_{T2,max}^{(2,2,1)} (\tilde M_1,p_T=0) = 
\mu_{(2,2,1)} + \sqrt{\mu_{(2,2,1)}^2 + \tilde M_1^2}\ , 
\label{end221}
\end{equation}
where the parameter $\mu_{(2,2,1)}$ was defined in eq.~(\ref{mu_npc}):
\begin{equation}
\mu_{(2,2,1)} \equiv \frac{M_2}{2}\, \left( 1-\frac{M_1^2}{M_2^2}  \right)
= \frac{M_2^2-M_1^2}{2M_2}\ .
\label{mu221}
\end{equation}
Almost all of our discussion from the previous Section \ref{sec:mt2_110}
can be directly applied here as well. For example, 
in order to measure the parameter $\mu_{(2,2,1)}$,
we only need to extract the endpoint of {\em a single}
distribution, for a single fixed value of the test mass $\tilde M_1$.
As before, we choose to use $\tilde M_1=0$. The resulting endpoint
measurement
\begin{equation}
M_{T2,max}^{(2,2,1)} (\tilde M_1=0,p_T=0) = 2\, \mu_{(2,2,1)}
= \frac{M_2^2-M_1^2}{M_2}
\label{mT2221}
\end{equation}
provides the required measurement of the parameter $\mu_{(2,2,1)}$
appearing in eq.~(\ref{end221}), as well as one constraint 
on the masses $M_1$ and $M_2$ involved in the problem. 
More importantly, the new constraint (\ref{mT2221}) 
is independent of the relation (\ref{mT2110}) found 
previously in Sec.~\ref{sec:mt2_110}.

The new variable $M_{T2}^{(2,2,1)}$ will also exhibit a kink in 
the plot of its endpoint $M_{T2,max}^{(2,2,1)}$
as a function of the test mass $\tilde M_1$. 
This is the same type of kink as the one discussed 
in the previous subsection, therefore all of our previous 
results would apply here as well. In particular, 
the analytical expression for the kink is given by
\begin{equation}
M_{T2,max}^{(2,2,1)}(\tilde M_1,p_T) =\left\{   
\begin{array}{ll}
F^{(2,2,1)}_{L}(\tilde M_1,p_T)\, , & ~~~{\rm if}\ \tilde M_1 \le M_1\, , \\ [2mm]
F^{(2,2,1)}_{R}(\tilde M_1,p_T)\, , & ~~~{\rm if}\ \tilde M_1 \ge M_1\, , 
\end{array}
\right.
\label{end221PT} 
\end{equation}
where 
\begin{eqnarray}
F_{L}^{(2,2,1)}(\tilde M_1,p_T) &=&
\left\{ \left[
\mu_{(2,2,1)}(p_T) + \sqrt{ \left(\mu_{(2,2,1)}(p_T)+\frac{p_T}{2}\right)^2 + \tilde M_1^2} 
\, \right]^2
- \frac{p_T^2}{4}   \right\}^{\frac{1}{2}},~~ \label{FL221} \\
F_{R}^{(2,2,1)}(\tilde M_1,p_T) &=&
\left\{ \left[
\mu_{(2,2,1)}(-p_T) + \sqrt{ \left(\mu_{(2,2,1)}(-p_T)-\frac{p_T}{2}\right)^2 + \tilde M_1^2} 
\, \right]^2
- \frac{p_T^2}{4}   \right\}^{\frac{1}{2}},~~~~
\label{FR221}
\end{eqnarray}
and the $p_T$-dependent parameter $\mu_{(2,2,1)}(p_T)$ is defined in analogy to
(\ref{mu110PT})
\begin{equation}
\mu_{(2,2,1)}(p_T) = \mu_{(2,2,1)}\, 
\left( \sqrt{ 1+\left(\frac{p_T}{2M_2}\right)^2} - \frac{p_T}{2M_2} \right) \, .
\label{mu221PT}
\end{equation}       
The size of the new kink $\Delta\Theta^{(2,2,1)}$ 
can be easily read off from eq.~(\ref{dtheta_110}),
where one should make the obvious replacements 
$M_0\to M_1$ and $M_1\to M_2$.

We can now generalize the two examples discussed so far
($M_{T2}^{(1,1,0)}$ and $M_{T2}^{(2,2,1)}$) to the case of an
arbitrary grandparent index $n$, with $p=n$ and $c=n-1$.
We get
\begin{equation}
M_{T2,max}^{(n,n,n-1)}(\tilde M_{n-1},p_T) =\left\{   
\begin{array}{ll}
F^{(n,n,n-1)}_{L}(\tilde M_{n-1},p_T)\, , & ~~~{\rm if}\ \tilde M_{n-1} \le M_{n-1}\, , \\ [2mm]
F^{(n,n,n-1)}_{R}(\tilde M_{n-1},p_T)\, , & ~~~{\rm if}\ \tilde M_{n-1} \ge M_{n-1}\, , 
\end{array}
\right.
\label{endnnn-1PT} 
\end{equation}
where 
\begin{equation}
F_{L}^{(n,n,n-1)}(\tilde M_{n-1},p_T) =
\left\{ \left[
\mu_{(n,n,n-1)}(p_T) + \sqrt{ \left(\mu_{(n,n,n-1)}(p_T)+\frac{p_T}{2}\right)^2 + \tilde M_{n-1}^2} 
\, \right]^2
- \frac{p_T^2}{4}   \right\}^{\frac{1}{2}},
\label{FLnnn-1} 
\end{equation}
\begin{equation}
F_{R}^{(n,n,n-1)}(\tilde M_{n-1},p_T) =
\left\{ \left[
\mu_{(n,n,n-1)}(-p_T) + \sqrt{ \left(\mu_{(n,n,n-1)}(-p_T)-\frac{p_T}{2}\right)^2 + \tilde M_{n-1}^2} 
\, \right]^2
- \frac{p_T^2}{4}   \right\}^{\frac{1}{2}},
\label{FRnnn-1}
\end{equation}
and the $p_T$-dependent parameter $\mu_{(n,n,n-1)}(p_T)$ is simply the
generalization of eqs.~(\ref{mu110PT}) and (\ref{mu221PT}):
\begin{equation}
\mu_{(n,n,n-1)}(p_T) = \mu_{(n,n,n-1)}\, 
\left( \sqrt{ 1+\left(\frac{p_T}{2M_n}\right)^2} - \frac{p_T}{2M_n} \right) \, .
\label{munnn-1PT}
\end{equation}
For $n=1$ or $n=2$, the general formula (\ref{endnnn-1PT}) reproduces
our previous results (\ref{end110PT}) and (\ref{end221PT}), correspondingly.

\subsection{The subsystem variable $M_{T2}^{(2,2,0)}$}
\label{sec:mt2_220}

The variable $M_{T2}^{(2,2,0)}$ is illustrated in Fig.~\ref{fig:n1n2}(b),
where we use the whole chain within the larger rectangle. 
As long as we ignore the effects of any ISR, we have a balanced\footnote{In the sense 
of eq.~(\ref{balmom110}). See the discussion following eq.~(\ref{balmom110}).} 
momentum configuration and the analytical 
results from Ref.~\cite{Cho:2007dh} would apply. In particular,
the endpoint $M_{T2,max}^{(2,2,0)}(\tilde M_0,p_T=0)$ is given by \cite{Cho:2007dh}
\begin{equation}
M_{T2,max}^{(2,2,0)}(\tilde M_0,p_T=0) =\left\{   
\begin{array}{ll}
F^{(2,2,0)}_{L}(\tilde M_0,p_T=0)\, , & ~~~{\rm if}\ \tilde M_0 \le M_0\, , \\ [2mm]
F^{(2,2,0)}_{R}(\tilde M_0,p_T=0)\, , & ~~~{\rm if}\ \tilde M_0 \ge M_0\, , 
\end{array}
\right.
\label{end220} 
\end{equation}
where
\begin{eqnarray}
F_{L}^{(2,2,0)}(\tilde M_0,p_T=0) &=&
\mu_{(2,2,0)} + \sqrt{ \mu_{(2,2,0)}^2 + \tilde M_0^2}\, , \label{FL220} \\
F_{R}^{(2,2,0)}(\tilde M_0,p_T=0) &=&
\mu_{(2,2,1)} + \mu_{(2,1,0)}
+ \sqrt{ \left(\mu_{(2,2,1)} - \mu_{(2,1,0)}\right)^2  + \tilde M_0^2}\, ,
\label{FR220}
\end{eqnarray}
and the various parameters $\mu_{(n,p,c)}$ are defined in
(\ref{mu_npc}). Notice that these expressions are valid only for
$p_T=0$. We have also derived the corresponding generalized expression
for $M_{T2,max}^{(2,2,0)}(\tilde M_0,p_T)$ for arbitrary values 
of $p_T$, which we list in Appendix~\ref{app:mt2max}.

The most striking feature of the endpoint function (\ref{end220})
is that it will also exhibit a kink $\Delta\Theta^{(2,2,0)}$ 
at the true value of the test mass $\tilde M_0=M_0$. However, 
as emphasized in \cite{Barr:2007hy},
the physical origin of this kink is different from the kinks
$\Delta\Theta^{(1,1,0)}$ and $\Delta\Theta^{(2,2,1)}$ which
we encountered previously in Secs.~\ref{sec:mt2_110} and 
\ref{sec:mt2_221}. This is easy to understand --
in Secs.~\ref{sec:mt2_110} and \ref{sec:mt2_221} we saw 
that the kinks $\Delta\Theta^{(1,1,0)}$ 
and $\Delta\Theta^{(2,2,1)}$ arise due to ISR effects, while
eq.~(\ref{end220}) holds in the absence of any ISR.
The explanation for the $\Delta\Theta^{(2,2,0)}$ kink
has actually already been provided in \cite{Cho:2007dh}. In essence,
one can treat the SM decay products $x_1$ and $x_2$
in each chain as a composite particle of variable mass, 
and the two branches $F_{L}^{(2,2,0)}$ and $F_{R}^{(2,2,0)}$ 
correspond to the two extreme values for the mass
of this composite particle.

In spite of its different origin, the kink in the function
(\ref{end220}) shares many of the same properties.
Let us use a specific example as an illustration.
Consider a popular example from supersymmetry, such as
gluino pair-production, followed by sequential two-body decays to
squarks and the lightest neutralinos. This is precisely 
a cascade of the type $n=2$, in which $X_2$ is the gluino $\tilde g$,
$X_1$ is a squark $\tilde q$, 
and $X_0$ is the lightest neutralino $\tilde\chi^0_1$.
Let us choose the superpartner masses according to the
SPS1a mass spectrum, which was also used in Ref.~\cite{Cho:2007dh}:
\begin{equation}
M_2 = 613\ {\rm GeV}, \qquad
M_1 = 525\ {\rm GeV}, \qquad
M_0 = 99\ {\rm GeV}.
\label{SPS1a}
\end{equation}
The resulting function $M_{T2,max}^{(2,2,0)}(\tilde M_0,p_T=0)$
is plotted in Fig.~\ref{fig:mt2kink}(a) with the upper set of lines
(compare to Fig.~12(b) in Ref.~\cite{Cho:2007dh}). 

\FIGURE[t]{
\epsfig{file=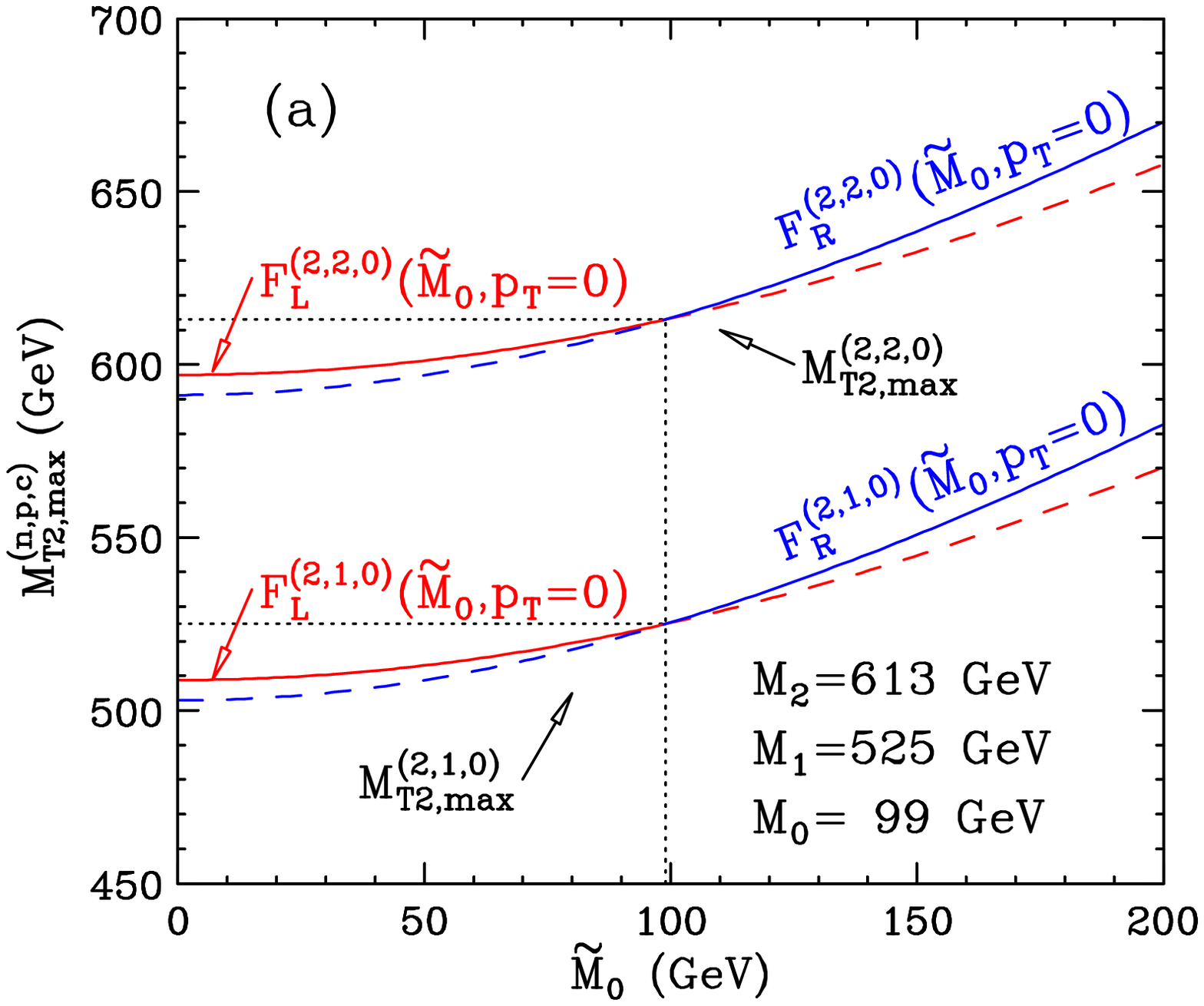,width=7.5cm}
\epsfig{file=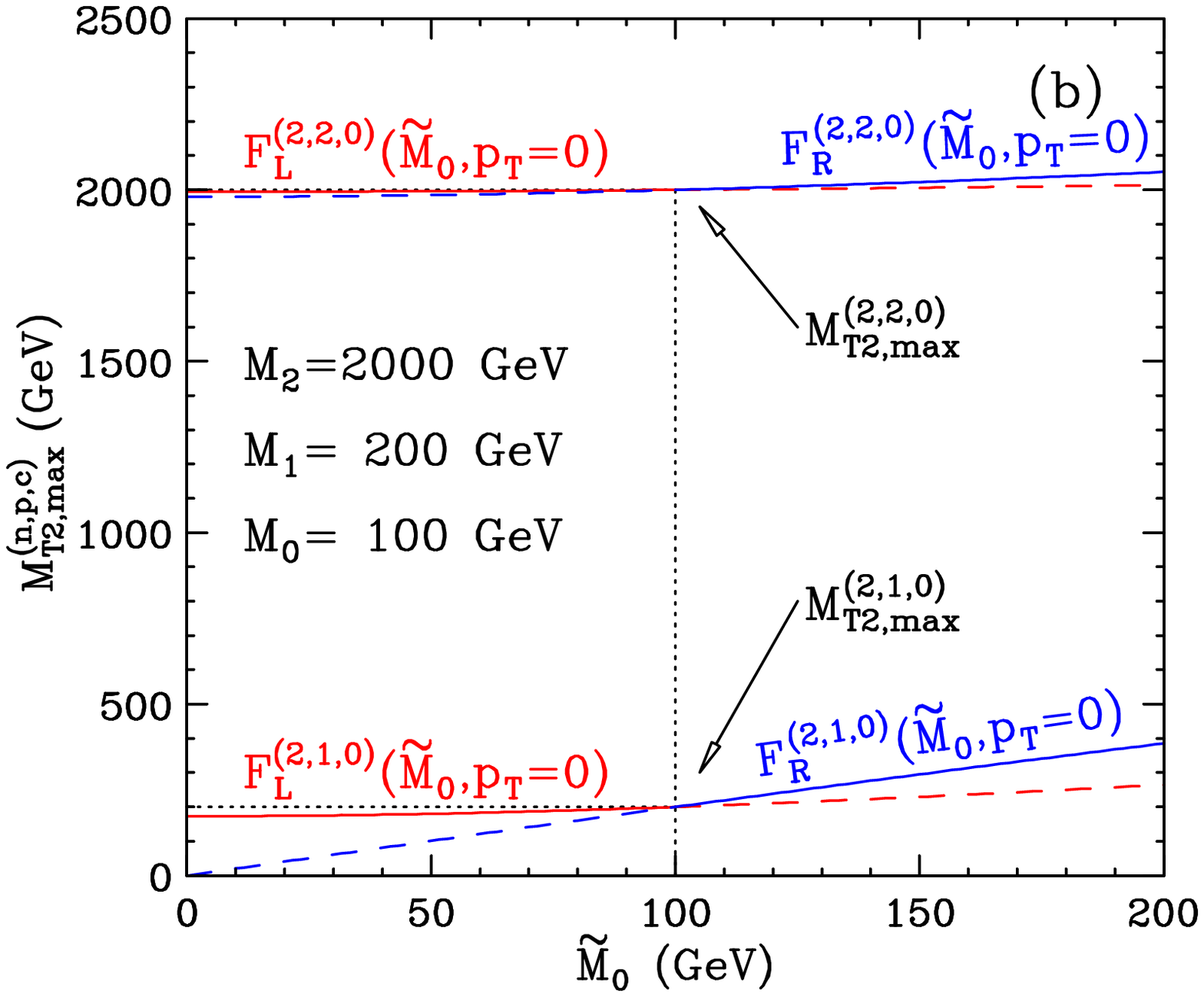,width=7.5cm}
\caption{\sl Dependence of the $M_{T2,max}^{(2,2,0)}$ 
and $M_{T2,max}^{(2,1,0)}$ upper kinematic endpoints
on the value of the test mass $\tilde M_0$, for
(a) the SPS1a parameter point in MSUGRA: 
$M_2=613$ GeV, $M_1=525$ GeV,  and $M_0=99$ GeV;
or (b) a split spectrum 
$M_2=2000$ GeV, $M_1=200$ GeV,  and $M_0=100$ GeV.
The horizontal (vertical) dotted lines denote the
true value of the parent (child) mass for each case. 
Solid (dashed) lines indicate true (false) endpoints, while 
red (blue) lines correspond to $F_L^{(n,p,c)}$ ($F_R^{(n,p,c)}$) branches.
}
\label{fig:mt2kink}}

There are several noteworthy features of $M_{T2,max}^{(2,2,0)}(\tilde M_0,p_T=0)$
which are evident from Fig.~\ref{fig:mt2kink}(a).
First, when the test mass $\tilde M_0$ is equal to the true
child mass $M_0$, the $M_{T2}$ endpoint yields the true
parent mass, in this case $M_2$:
\begin{equation}
M_{T2,max}^{(2,2,0)}(\tilde M_0=M_0,p_T=0) = M_2\ .
\label{kink220}
\end{equation}
This property of $M_{T2}$
is true by design, and is confirmed by the dotted lines in 
Fig.~\ref{fig:mt2kink}(a).
Second, as seen from eq.~(\ref{end220}), $M_{T2,max}^{(2,2,0)}(\tilde M_0,p_T=0)$
is not given by a single function, but has two separate branches.
The first (``low'') branch $F_L^{(2,2,0)}$ 
applies for $\tilde M_0\le M_0$, and is shown 
in Fig.~\ref{fig:mt2kink}(a) with red lines.
The second (``high'') branch $F_R^{(2,2,0)}$ 
is valid for $\tilde M_0\ge M_0$ and is shown in
blue in Fig.~\ref{fig:mt2kink}(a). While the two branches
coincide at $\tilde M_{0}=M_0$:
\begin{equation}
F_L^{(2,2,0)}(M_0,p_T=0) = F_R^{(2,2,0)} (M_0,p_T=0)\ ,
\end{equation}
their {\em derivatives} do not match:
\begin{equation}
\left( \frac{\partial F_L^{(2,2,0)}}{\partial \tilde M_0}\right)_{\tilde M_0=M_0} \ne 
\left( \frac{\partial F_R^{(2,2,0)}}{\partial \tilde M_0}\right)_{\tilde M_0=M_0}\ ,
\end{equation}
leading to a kink $\Delta\Theta^{(2,2,0)}$ 
in the function $M_{T2,max}^{(2,2,0)}(\tilde M_0,p_T=0)$
\cite{Cho:2007qv,Gripaios:2007is,Barr:2007hy,Cho:2007dh}.
As before, let us try to investigate quantitatively 
the size of this kink. Applying the general definition (\ref{DTheta}),
we obtain
\begin{equation}
\Delta\Theta^{(2,2,0)} = 
\arctan \left( \frac{2(1-y)(1-z)\sqrt{yz}}{(y+z)(1+yz)+4yz}\right) \ ,
\label{DTheta220}
\end{equation}
where we have defined the squared mass ratios
\begin{equation}
y\equiv \frac{M_1^2}{M_2^2}\, , \qquad 
z\equiv \frac{M_0^2}{M_1^2}\, .
\label{yz}
\end{equation}
The result (\ref{DTheta220}) is plotted in Fig.~\ref{fig:kink}(a)
as a function of the mass ratios $\sqrt{y}$ and $\sqrt{z}$.
\FIGURE[t]{
\epsfig{file=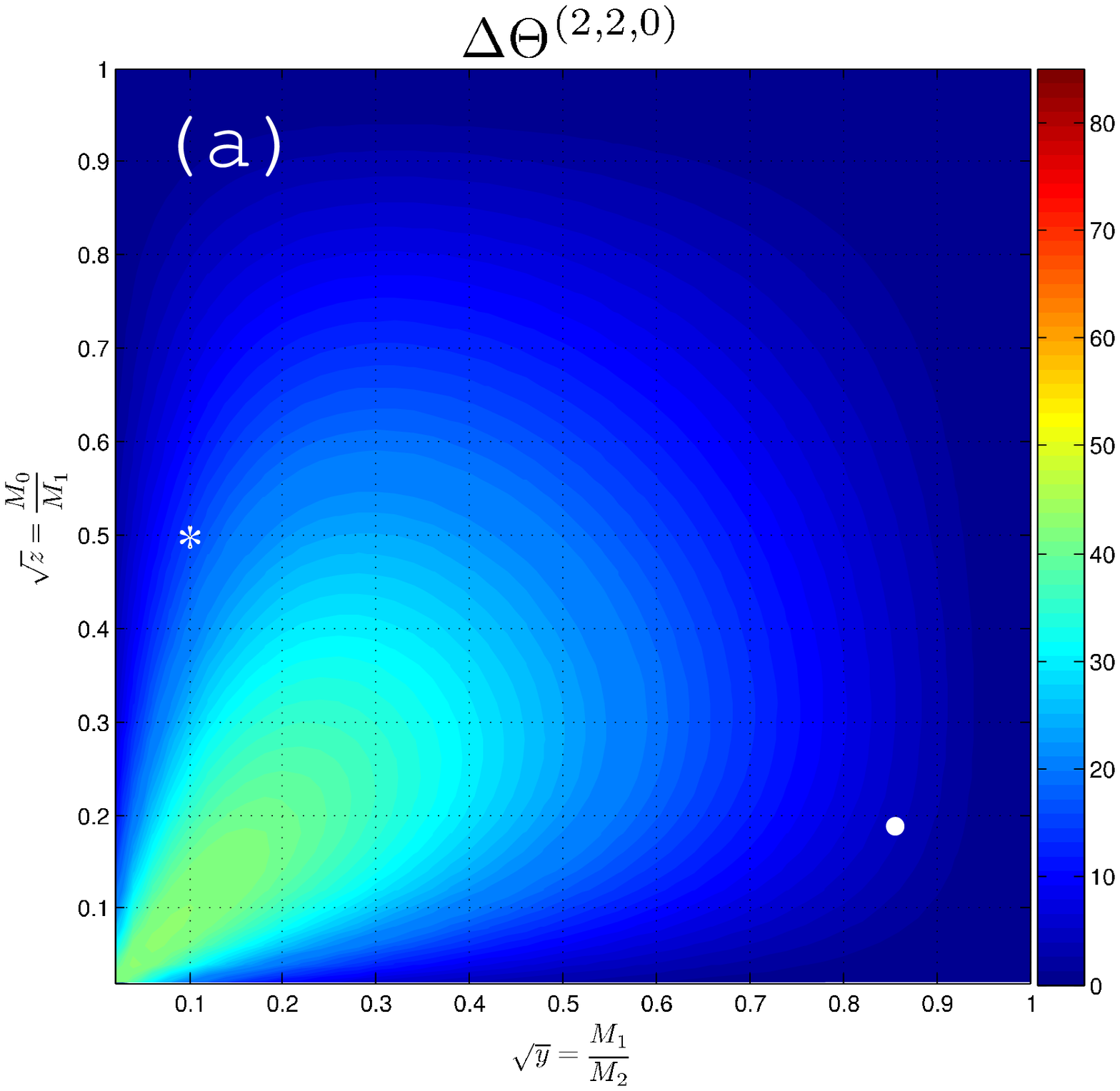,height=6cm}~~
\epsfig{file=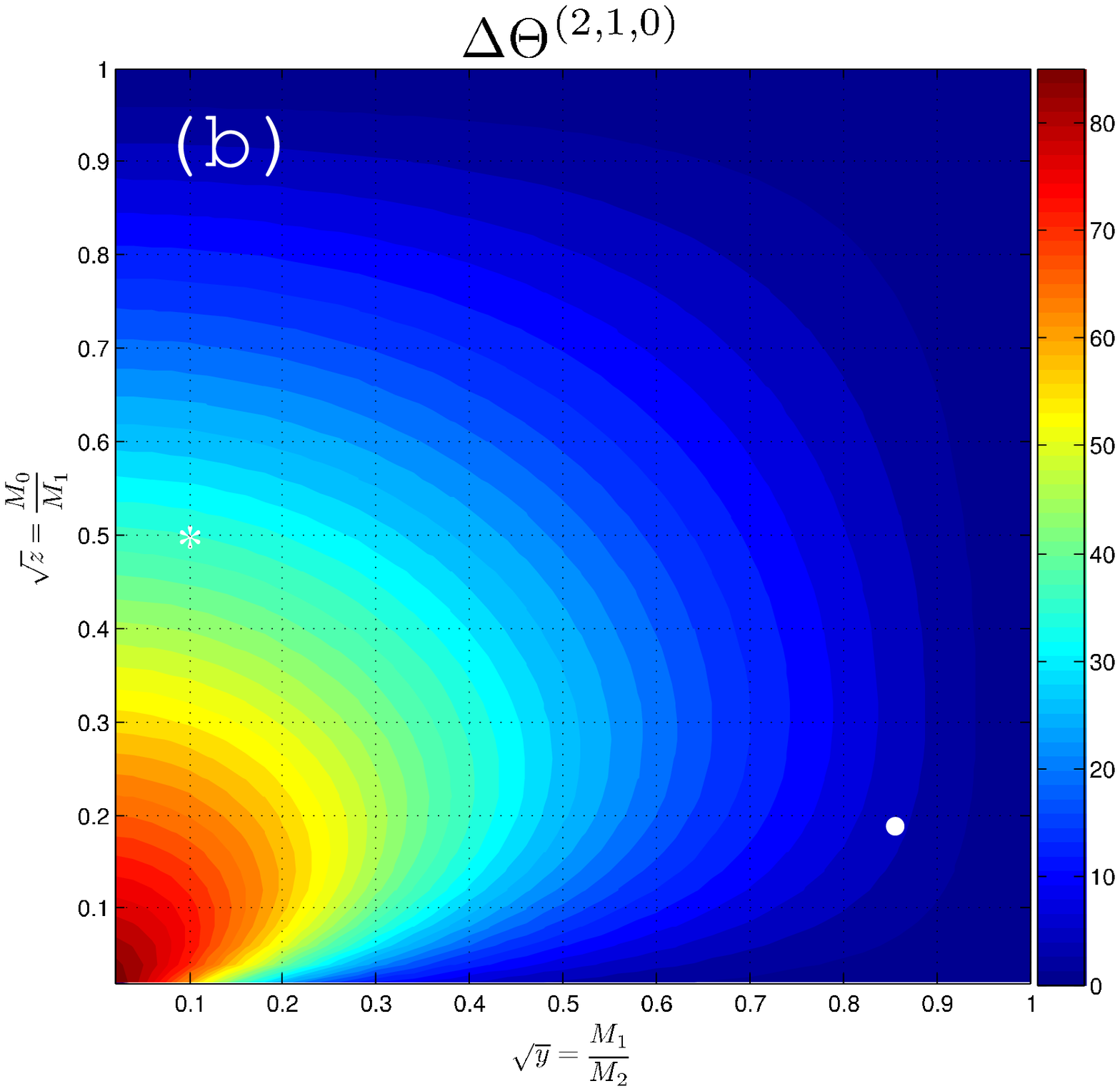,height=6cm}
\caption{\sl The amount of kink: (a) $\Delta\Theta^{(2,2,0)}$ and
(b) $\Delta\Theta^{(2,1,0)}$ in degrees, as a function of the mass ratios
$\sqrt{y}$ and $\sqrt{z}$. The white dot and the white asterisk 
denote the locations in this $(\sqrt{y},\sqrt{z})$ parameter space
of the two sample spectra (\ref{SPS1a}) and (\ref{split})
used for Figs.~\ref{fig:mt2kink}(a) and \ref{fig:mt2kink}(b),
correspondingly.}
\label{fig:kink}}
Fig.~\ref{fig:kink}(a) demonstrates that as both $y$ and $z$ become small,
the kink $\Delta\Theta^{(2,2,0)}$ gets more pronounced. 
Fig.~\ref{fig:kink}(a) also shows that the kink $\Delta\Theta^{(2,2,0)}$
is a symmetric function of $y$ and $z$, as can also be seen directly from 
eq.~(\ref{DTheta220}). Therefore, the kink $\Delta\Theta^{(2,2,0)}$ will be 
best observable in those cases where $y$ and $z$ are both small, and in addition,
the mass spectrum happens to obey the relation\footnote{Notice that for this
special value of $M_1=\sqrt{M_0M_2}$, the upper endpoint of the invariant mass
distribution $M_{x_1x_2}$ is the same as in the case when the intermediate
particle $X_1$ is off-shell, i.e.~when $M_1>M_2$. Then we find that the 
$M_{T2,max}^{(2,2,0)}$ formulas and corresponding kink structures are identical
in the on-shell and off-shell cases. For details, see Appendix~\ref{app:mt2max}.}
$y=z$, i.e.~$M_1=\sqrt{M_0M_2}$. Unfortunately, the SPS1a
study point is rather far from this category -- the spectrum (\ref{SPS1a})
corresponds to the values $\sqrt{y}=0.856$ and $\sqrt{z}=0.189$, which are indicated 
in Fig.~\ref{fig:kink}(a) by a white dot. This conclusion is also 
supported by Fig.~\ref{fig:mt2kink}(a), which shows a rather mild kink
in the SPS1a case.

We shall be rather ambivalent in our attitude toward the $\Delta\Theta^{(2,2,0)}$
kink as well. While the interpretation of the kink is straightforward,
its observation in the actual experiment is again an open issue.
On the one hand, the experimental precision would depend
on the particular signature, 
i.e.~the type of the SM particles $x_1$ and $x_2$. 
If those are leptons, their 4-momenta $p_1^{(k)}$ and $p_2^{(k)}$
will be measured relatively well and the kink might be observable.
However, when $x_1$ and $x_2$ are jets, the experimental 
resolution may not be sufficient. Secondly, as seen in 
Fig.~\ref{fig:mt2kink}(a), the kink itself may not be very pronounced,
and its observability will in fact depend on the particular mass spectrum.

The main lesson from the above discussion is that while the existence
of the kink is without a doubt, its actual observation is by no means
guaranteed. Therefore, our main mass measurement method, described 
later in Sec.~\ref{sec:ourmt2}, will not use any information related 
to the kink. In fact in Sec.~\ref{sec:ourmt2} we shall show that one can 
completely reconstruct the mass spectrum of the new particles, using 
just measurements of $M_{T2}$ endpoints, each done at a single fixed
value of the corresponding test mass. It is worth noting that,
in general, an endpoint in a spectrum is a sharper feature than 
a kink of the type (\ref{DTheta}). Therefore, we would expect that the
experimental precision on the extracted endpoints will be much better 
than the corresponding precision on the kink location. The kink will 
also not play any role 
in our hybrid method, described in Sec.~\ref{sec:ourim}.
Only for the method described in Sec.~\ref{sec:ourkink},
we shall try to make use of the kink information.
 
Let us now return to our original discussion of the $M_{T2}^{(2,2,0)}$ 
endpoint (\ref{end220}). Following our previous approach from 
Secs.~\ref{sec:mt2_110} and \ref{sec:mt2_221}, we would choose a
fixed value of the test mass $\tilde M_0$ and measure the 
corresponding $M_{T2}$ endpoint. However, the presence of two branches
(\ref{FL220}) and (\ref{FR220}) leads to a slight complication:
for a randomly chosen value of $\tilde M_0$, we will not know whether 
we should use (\ref{FL220}) or (\ref{FR220}) when interpreting the
endpoint measurement. This requires us to make very special choices 
for the fixed value of $\tilde M_0$, which would remove this ambiguity.
It is easy to see that by choosing $\tilde M_0=0$, we can ensure that
the endpoint is always described by the ``low'' branch (\ref{FL220}),
and the $M_{T2,max}^{(2,2,0)}$ measurement can then be uniquely interpreted as
\begin{equation}
M_{T2,max}^{(2,2,0)} (\tilde M_0=0,p_T=0) = 2\, \mu_{(2,2,0)}
= \frac{M_2^2-M_0^2}{M_2}\ .
\label{mT2220_low}
\end{equation}
However, we could also design a special choice of $\tilde M_0$, 
which would select the ``high'' branch (\ref{FL220})
and again uniquely remove the branch ambiguity.
For this purpose, we must choose a value for the test mass $\tilde M_0$ 
which is sufficiently large, in order to safely guarantee that 
it is well beyond the true mass $M_0$.
Since the true mass $M_0$ can never exceed the beam energy
$E_b$, one obvious safe and rather conservative choice for $\tilde M_0$ 
could be $\tilde M_0=E_b$, in which case from (\ref{end220}) we get
\begin{eqnarray}
M_{T2,max}^{(2,2,0)} (\tilde M_0=E_b,p_T=0) &=&
\mu_{(2,2,1)} + \mu_{(2,1,0)}
+ \sqrt{ \left(\mu_{(2,2,1)} - \mu_{(2,1,0)}\right)^2  + E_b^2} 
\label{mT2220_high1}
\\
&=& M_2 - \frac{M_2}{2}\left( \frac{M_1^2}{M_2^2}+ \frac{M_0^2}{M_1^2}\right)
+\sqrt{ \frac{M_2^2}{4}\left( \frac{M_1^2}{M_2^2}- \frac{M_0^2}{M_1^2}\right)^2  + E_b^2 } \, .
\nonumber
\end{eqnarray}
Notice that the high branch function $F_R^{(2,2,0)}$
in eq.~(\ref{FR220}) is rather unique in one very important aspect:
it depends not just on one, but on {\em two} mass parameters, 
namely the combinations
$\mu_{(2,2,1)} + \mu_{(2,1,0)}$ and $\mu_{(2,2,1)} - \mu_{(2,1,0)}$.
In contrast, the ``low'' branch $F_L^{(2,2,0)}$,
as well as the previously discussed endpoint functions 
$M_{T2,max}^{(n,n,n-1)}(\tilde M_0,p_T=0)$, each contained a single 
$\mu$ parameter. As a result, in those cases we did not benefit 
from any extra measurements for different values of the test 
mass $\tilde M_0$ -- had we done that, we would have been measuring 
the same $\mu$ parameter over and over again. However, the situation with 
$F_R^{(2,2,0)}$ is different, and here we {\em will}
benefit from an additional measurement for a different value of 
$\tilde M_0$. For example, let us choose $\tilde M_0=E'_b$, with
$E'_b>E_b$, which will still keep us on the high branch.
We obtain another constraint 
\begin{eqnarray}
M_{T2,max}^{(2,2,0)} (\tilde M_0=E'_b,p_T=0) &=&
\mu_{(2,2,1)} + \mu_{(2,1,0)}
+ \sqrt{ \left(\mu_{(2,2,1)} - \mu_{(2,1,0)}\right)^2  + {E'_b}^2} 
\label{mT2220_high2}
\\
&=& M_2 - \frac{M_2}{2}\left( \frac{M_1^2}{M_2^2}+ \frac{M_0^2}{M_1^2}\right)
+\sqrt{ \frac{M_2^2}{4}\left( \frac{M_1^2}{M_2^2}- \frac{M_0^2}{M_1^2}\right)^2  + {E'_b}^2 } \, .
\nonumber 
\end{eqnarray}
It is easy to check that the constraints (\ref{mT2220_low}-\ref{mT2220_high2})
are all independent, thus providing {\em three} independent
equations\footnote{In practice, instead of relying on individual
endpoint measurements for three different values of $\tilde M_0$,
one may prefer to use the experimental information 
for the whole function $M_{T2,max}^{(2,2,0)}(\tilde M_0,p_T=0)$ and simply fit 
to it the analytical expression (\ref{end220}) for the
three floating parameters $M_0$, $M_1$ and $M_2$,
as was done in Ref.~\cite{Cho:2007dh}. As we shall see shortly,
this method does not lead to any new information, 
and may only improve the statistical error on the mass determination.
Therefore, to keep our discussion as simple as possible, we prefer to 
talk about the three individual measurements (\ref{mT2220_low}-\ref{mT2220_high2})
as opposed to fitting the whole distribution (\ref{end220}).} 
for the three unknown masses $M_0$, $M_1$ and $M_2$. 
These three equations (\ref{mT2220_low}-\ref{mT2220_high2})
can be solved rather easily\footnote{The general solution 
for $M_2$, $M_1$ and $M_0$ in terms of the measured
endpoints (\ref{mT2220_low}-\ref{mT2220_high2}) is rather messy 
and not very illuminating, therefore we do not list it here.}, 
and one obtains the proper solution for the masses $M_0$, $M_1$ 
and $M_2$, up to a two-fold ambiguity:
\begin{equation}
M_2 \to M_2\, , \qquad M_1 \to \frac{M_0}{M_1} M_2\, , \qquad  M_0 \to M_0\, ,
\label{amb220}
\end{equation}
which is nothing but the interchange $y\leftrightarrow z$ at a fixed $M_2$.
The ambiguity arises because the expression (\ref{end220}) for
the endpoint $M_{T2,max}^{(2,2,0)}$ (and consequently,
the set of constraints (\ref{mT2220_low}-\ref{mT2220_high2}))
is invariant under the transformation (\ref{amb220}).
Because of this ambiguity, in addition to the original 
SPS1a input values (\ref{SPS1a}) for the mass spectrum, 
we obtain a second solution
\begin{equation}
M_2 = 613\ {\rm GeV}, \qquad
M_1 = 115.6\ {\rm GeV}, \qquad
M_0 = 99\ {\rm GeV}.
\label{SPS1afake}
\end{equation}
This second solution was missed in the analysis of Ref.~\cite{Cho:2007dh}.
It is easy to check that the alternative mass spectrum 
(\ref{SPS1afake}) gives an identical $M_{T2,max}^{(2,2,0)}(\tilde M_0,p_T=0)$
distribution as the one shown in Fig.~\ref{fig:mt2kink}(a),
so that it is impossible to rule it out on the basis of 
$M_{T2,max}^{(2,2,0)}$ measurements alone.

The previous discussion reveals an important and somewhat overlooked 
benefit from the existence of the kink -- one can make not one, not two,
but {\em three} independent endpoint measurements from a single
$M_{T2}^{(n,p,c)}$ distribution! In fact, we shall argue that 
the three measurements (\ref{mT2220_low}-\ref{mT2220_high2})
are much more robust than the kink measurement (\ref{DTheta}).
For example, when the child mass is relatively small,
the lower branch $F_{L}^{(2,2,0)}$ is relatively short
and the kink will be difficult to see, even under 
ideal experimental conditions. An extreme example of this sort
is presented in Section~\ref{sec:our}, where we discuss 
top quark events, in which the child (neutrino) mass $M_0$ 
is practically zero and the kink cannot be seen at all.
However, even under those circumstances, the 
endpoint measurements (\ref{mT2220_low}-\ref{mT2220_high2})
are still available. More importantly, the
constraints (\ref{mT2220_low}-\ref{mT2220_high2})
are independent of the previously found
relations (\ref{mT2110}) and (\ref{mT2221}), so that the 
latter can be used to resolve the two-fold ambiguity 
(\ref{amb220}).

Before we move on to a discussion of the last remaining subsystem
$M_{T2}$ quantity in the next Sec.~\ref{sec:mt2_210}, let us 
recap our main result derived in this subsection.
We showed that the $M_{T2}^{(2,2,0)}$ variable 
yields {\em three} independent endpoint measurements
(\ref{mT2220_low}-\ref{mT2220_high2}), and possibly a 
kink measurement (\ref{DTheta}). The $M_{T2}^{(2,2,0)}$
endpoint measurements {\em by themselves} are sufficient 
to determine all three masses $M_0$, $M_1$ and $M_2$, 
up to the two-fold ambiguity (\ref{amb220}). This represents 
a pure $M_{T2}$-based mass measurement method, which does 
not use any any kink or invariant mass information. 

\subsection{The subsystem variable $M_{T2}^{(2,1,0)}$}
\label{sec:mt2_210}

The variable $M_{T2}^{(2,1,0)}$ is illustrated in Fig.~\ref{fig:n1n2}(b),
where we use the subchain within the smaller rectangle on the right. 
This is another genuine subsystem quantity, since we only use the SM decay products 
$x_1$ and ignore the upstream objects $x_2$. However, the upstream
objects $x_2$ are important in the sense that they have some
non-zero transverse momentum, and as a result, the
sum of the transverse momenta $\vec{P}^{(k)}_{0T}$ 
of the children $X_0$ is {\em not} balanced 
by the sum of the transverse momenta of the SM objects $x_1$ 
used in the $M_{T2}$ calculation:
\begin{equation}
\vec{P}^{(1)}_{0T}+\vec{P}^{(2)}_{0T} + \vec{p}^{\,(1)}_{1T} + \vec{p}^{\,(2)}_{1T}
= - \vec{p}^{\,(1)}_{2T} - \vec{p}^{\,(2)}_{2T} - \vec{p}_T\ne 0\ .
\label{unbmom}
\end{equation}
Notice that even in the absence of any ISR $p_T$, this
is still an unbalanced configuration, due to the transverse momenta
$\vec{p}^{\,(1)}_{2T}$ and $\vec{p}^{\,(2)}_{2T}$ of the
upstream objects $x_2$. Therefore, we cannot use the 
existing analytical results on $M_{T2}$, since previous studies
always assumed that the right-hand side of eq.~(\ref{unbmom}) 
is exactly zero, due to the lack of any particles upstream.
We therefore need to generalize the previous
treatments of $M_{T2}$ and obtain the corresponding endpoint 
formulas for our new subsystem $M_{T2}^{(2,1,0)}$ variable.
The general mathematical properties of subsystem 
$M_{T2}^{(n,p,c)}$ variables will be presented in a forthcoming publication
\cite{BKMP}. Here we shall only use the results relevant for our 
example ($n=2$). In particular, in the absence of any intrinsic ISR
(i.e., for $p_T=0$), we find that the endpoint 
of the $M_{T2}^{(2,1,0)}$ distribution is given by
\begin{equation}
M_{T2,max}^{(2,1,0)}(\tilde M_0,p_T=0) =\left\{   
\begin{array}{ll}
F^{(2,1,0)}_{L}(\tilde M_0,p_T=0)\, , & ~~~{\rm if}\ \tilde M_0 \le M_0\, , \\ [2mm]
F^{(2,1,0)}_{R}(\tilde M_0,p_T=0)\, , & ~~~{\rm if}\ \tilde M_0 \ge M_0\, , 
\end{array}
\right.
\label{end210} 
\end{equation}
where
\begin{eqnarray}
F_{L}^{(2,1,0)}(\tilde M_0,p_T=0) &=& 
\left\{ \left[
\mu_{(2,2,0)} - \mu_{(2,2,1)} + \sqrt{ \mu_{(2,2,0)}^2 + \tilde M_0^2} 
\, \right]^2
-\mu^2_{(2,2,1)}   \right\}^{\frac{1}{2}}\, , \label{FL210} \\
F_{R}^{(2,1,0)}(\tilde M_0,p_T=0) &=& 
\left\{ \left[
  \mu_{(2,1,0)}
+ \sqrt{ \left(\mu_{(2,2,1)} - \mu_{(2,1,0)}\right)^2  + \tilde M_0^2}
\, \right]^2
-\mu^2_{(2,2,1)} \right\}^{\frac{1}{2}}\, ,
\label{FR210}
\end{eqnarray}
and the various parameters $\mu_{(n,p,c)}$ are defined in
(\ref{mu_npc}). The corresponding expressions for general $p_T$
(i.e., arbitrary intrinsic ISR) are listed in Appendix \ref{app:mt2max}.

From eq.~(\ref{end210}) we see that, once again, the endpoint function 
$M_{T2,max}^{(2,1,0)}(\tilde M_0,p_T=0)$ would
exhibit a kink $\Delta\Theta^{(2,1,0)}$ at the true value of the test mass $\tilde M_0=M_0$:
\begin{equation}
\left( \frac{\partial F_L^{(2,1,0)}}{\partial \tilde M_0}\right)_{\tilde M_0=M_0} \ne 
\left( \frac{\partial F_R^{(2,1,0)}}{\partial \tilde M_0}\right)_{\tilde M_0=M_0}\ .
\end{equation}
The existence of this kink should come as no surprise ---
Ref.~\cite{Barr:2007hy} 
showed (in the $p_T\to\infty$ limit) that {\em any} type of upstream momentum will
generate a kink in an otherwise smooth $M_{T2,max}$ function.
As before, the value of the $M_{T2}$ endpoint $M_{T2,max}^{(2,1,0)}$ at the
kink location reveals the true mass of the parent:
\begin{equation}
M_{T2,max}^{(2,1,0)}(\tilde M_0=M_0,p_T=0) = M_1\ .
\label{true210}
\end{equation}
At the same time, the physical origin of this kink is different from either 
of the two kink types ($\Delta\Theta^{(1,1,0)}$ and $\Delta\Theta^{(2,2,0)}$) 
discussed earlier. Clearly, the new kink is different from $\Delta\Theta^{(2,2,0)}$, which
was due to the varying invariant mass of the $\{x_1,x_2\}$ system.
Here we are using a single SM particle $x_1$ whose mass is constant.
Furthermore, the new kink $\Delta\Theta^{(2,1,0)}$ cannot be due
to any ISR like in the case of $\Delta\Theta^{(1,1,0)}$, 
since eq.~(\ref{end210}) does not account for any ISR effects.
The real reason for this new $\Delta\Theta^{(2,1,0)}$ kink is a third one,
namely, the kinematical restrictions placed by the decays of the 
upstream particles (in this case, the grandparents $X_2$). 

We now proceed to investigate the new kink $\Delta\Theta^{(2,1,0)}$
quantitatively. Using the same example of gluino pair-production 
for the SPS1a mass spectrum (\ref{SPS1a}), we plot the 
function (\ref{end210}) in Fig.~\ref{fig:mt2kink}(a).
Comparing the lower and the upper set of lines in the figure, 
we notice that the $M_{T2,max}^{(2,1,0)}$ and $M_{T2,max}^{(2,2,0)}$ variables
share several common characteristics.
They both exhibit a kink at the true location of the child mass 
$\tilde M_{0}=M_0$, while their values at that point reveal
the true parent mass in each case: $M_1$ for $M_{T2,max}^{(2,1,0)}$ 
and $M_2$ for $M_{T2,max}^{(2,2,0)}$. 
Using the definition (\ref{DTheta}), we find that the
size of the $\Delta \Theta^{(2,1,0)} $ kink is given by
\begin{equation}
\Delta\Theta^{(2,1,0)} = 
\arctan \left( \frac{(1-y^2)(1-z)\sqrt{z}}{2z(1+y^2)+y(1+z^2)+2yz}\right) \ ,
\label{DTheta210}
\end{equation}
where the parameters $y$ and $z$ were already defined in (\ref{yz}).
The kink  $\Delta \Theta^{(2,1,0)}$ is plotted in Fig.~\ref{fig:kink}(b)
as a function of $\sqrt{y}$ and $\sqrt{z}$. We notice that the kink
structure becomes more pronounced for relatively small $y$ and $z$.
Comparing Figs.~\ref{fig:kink}(a) and \ref{fig:kink}(b), we see that 
for any given set of values for $y$ and $z$, the $\Delta \Theta^{(2,1,0)}$ 
kink discussed here is more pronounced than the $\Delta \Theta^{(2,2,0)}$ 
kink from the previous subsection\footnote{This statement can also be verified
using the analytical formulas (\ref{DTheta220}) and (\ref{DTheta210}).}. 
The difference is particularly noticeable in the region of 
$\sqrt{y}\sim 0$ and $\sqrt{z}\sim 0.2$. The SPS1a mass spectrum (\ref{SPS1a})
in our previous example was rather far away from this region, as indicated by 
the white dots in Fig.~\ref{fig:kink}. Now let us choose a different mass 
spectrum, which is closer to the region where the difference between
the two kinks becomes more noticeable, for example
\begin{equation}
M_2 = 2000\ {\rm GeV}, \qquad
M_1 = 200\ {\rm GeV}, \qquad
M_0 = 100\ {\rm GeV},
\label{split}
\end{equation}
corresponding to the point marked with the white asterisk 
in Figs.~\ref{fig:kink}(a) and \ref{fig:kink}(b).
The resulting endpoint functions $M_{T2,max}^{(2,2,0)}$ and
$M_{T2,max}^{(2,1,0)}$ are plotted in Fig.~\ref{fig:mt2kink}(b).
Indeed we see that with this new spectrum the kink in the
$M_{T2,max}^{(2,1,0)}$ function is much more noticeable than 
the kink in the $M_{T2,max}^{(2,2,0)}$ function.
Therefore, our first conclusion regarding the $M_{T2}^{(2,1,0)}$ 
variable is that its kink is in general sharper 
and appears to be more promising than the previously discussed kink in the 
$M_{T2}^{(2,2,0)}$  variable from Sec.~\ref{sec:mt2_220}.

Following our previous strategy, we shall not dwell too long on the kink, 
but instead we shall discuss the available endpoint measurements
for various values of $\tilde M_0$. Again, the presence of two branches 
in eq.~(\ref{end210}) can be used to our advantage. As in Sec.~\ref{sec:mt2_220},
we first choose a test mass value $\tilde M_0=0$, which would ``select'' the 
low branch (\ref{FL210}) and result in an endpoint measurement
\begin{eqnarray}
M_{T2,max}^{(2,1,0)} (\tilde M_0=0, p_T=0) &=& 
2\, \sqrt{\mu_{(2,2,0)}\,(\mu_{(2,2,0)}-\mu_{(2,2,1)}) } \ .
\label{mT2210_low}
\end{eqnarray}
Just as before, we could also choose a rather large value for 
$\tilde M_0=E_b$, which would select the high branch (\ref{FR210})
and result in the measurement
\begin{equation}
M_{T2,max}^{(2,1,0)} (\tilde M_0=E_b, p_T=0) =
\left\{ \left[ \mu_{(2,1,0)}
+ \sqrt{ \left(\mu_{(2,2,1)} - \mu_{(2,1,0)}\right)^2  + E_b^2}
\, \right]^2
-\mu^2_{(2,2,1)} \right\}^{\frac{1}{2}}  . 
\label{mT2210_high1}
\end{equation}
A third choice, $\tilde M_0=E'_b$, with $E'_b>E_b$,
would yield yet another endpoint measurement
\begin{equation}
M_{T2,max}^{(2,1,0)} (\tilde M_0=E'_b, p_T=0) =
\left\{ \left[ \mu_{(2,1,0)}
+ \sqrt{ \left(\mu_{(2,2,1)} - \mu_{(2,1,0)}\right)^2  + {E'}_b^2}
\, \right]^2
-\mu^2_{(2,2,1)} \right\}^{\frac{1}{2}}. \qquad
\label{mT2210_high2}
\end{equation}
Again we obtained three equations (\ref{mT2210_low}-\ref{mT2210_high2})
for the three unknown $\mu$-parameters $\mu_{(2,2,0)}$, $\mu_{(2,2,1)}$ and $\mu_{(2,1,0)}$,
or equivalently, for the three unknown masses $M_0$, $M_1$ and $M_2$. 
These equations are all independent and can be easily solved,
giving a total of four solutions. However, three of the solutions 
are always unphysical, so that we end up with a single unique solution.
This represents an important advantage of the $M_{T2,max}^{(2,1,0)}$ 
variable in comparison with the $M_{T2,max}^{(2,2,0)}$ variable discussed 
in Sec.~\ref{sec:mt2_220}. There we found that $M_{T2,max}^{(2,2,0)}$ always gives 
rise two a two-fold ambiguity in the mass spectrum, while now we see that
$M_{T2,max}^{(2,1,0)}$ does not suffer from this problem and already by itself 
allows for a complete and unambiguous determination of the mass spectrum.

\section{$M_{T2}$-based mass measurement methods}
\label{sec:our}

In this section we use the analytical results derived in the previous 
section to propose three different strategies for determining the masses
in $n\le2$ decay chains. We shall illustrate each of our methods with
a specific example, for which we choose to consider the dilepton samples from
$W^+W^-$ and $t\bar{t}$ events. The former is an example of 
the $n=1$ decay chain exhibited in Fig.~\ref{fig:n1n2}(a), while the
latter is an example of the $n=2$ decay chain in Fig.~\ref{fig:n1n2}(b).
Most importantly, these samples already exist in the Tevatron data and 
will also be among the first to be studied at the LHC. 
Correspondingly, throughout this section we shall use 
the following mass spectrum
\begin{eqnarray}
M_2 &=& m_t \ \, = 173\ {\rm GeV}, \nonumber \\
M_1 &=& m_W = 80\ {\rm GeV},  \label{top} \\
M_0 &=& m_\nu \ \, = 0\ {\rm GeV}. \nonumber
\end{eqnarray}

\FIGURE[t]{
\epsfig{file=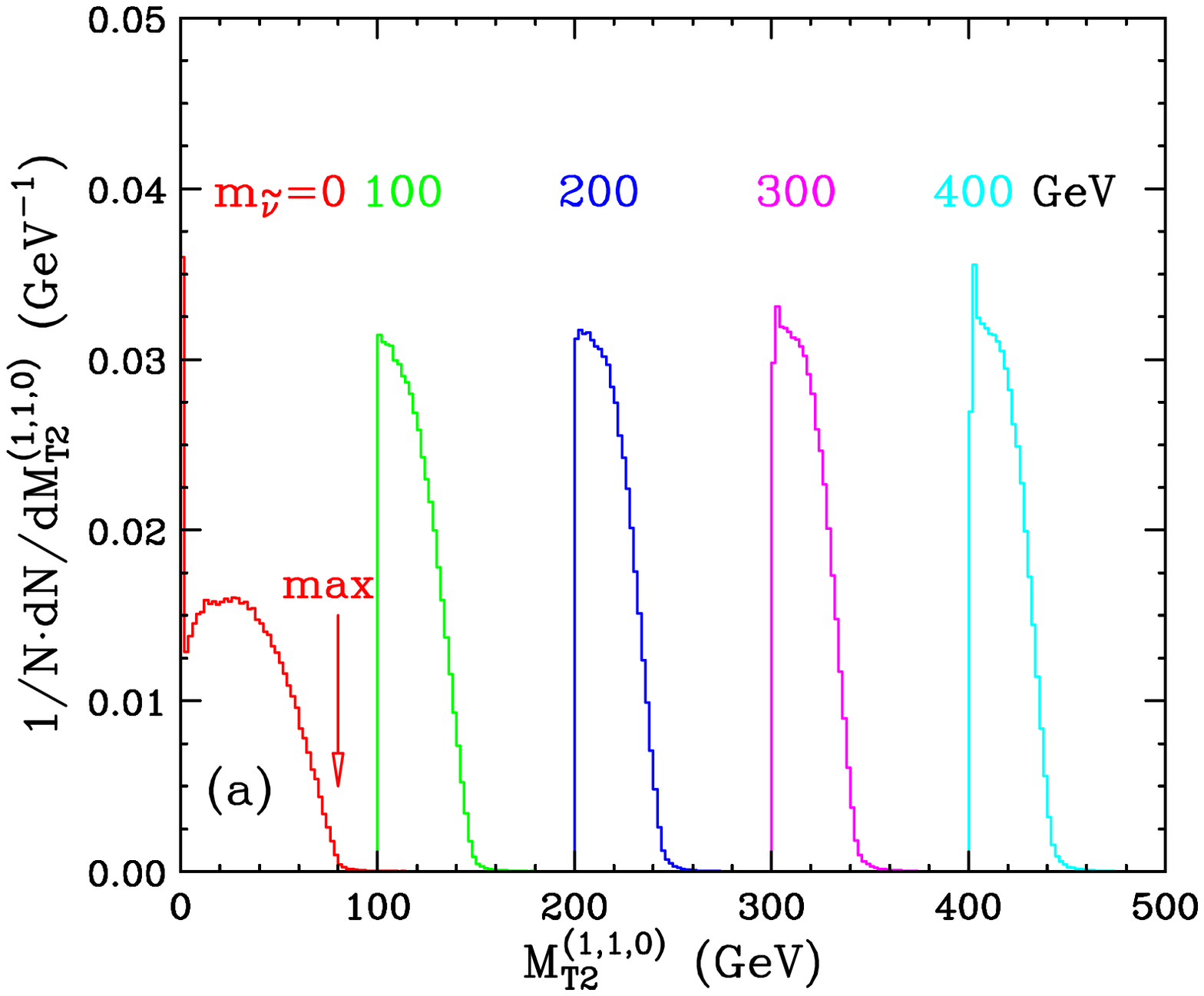,height=6.0cm}
\epsfig{file=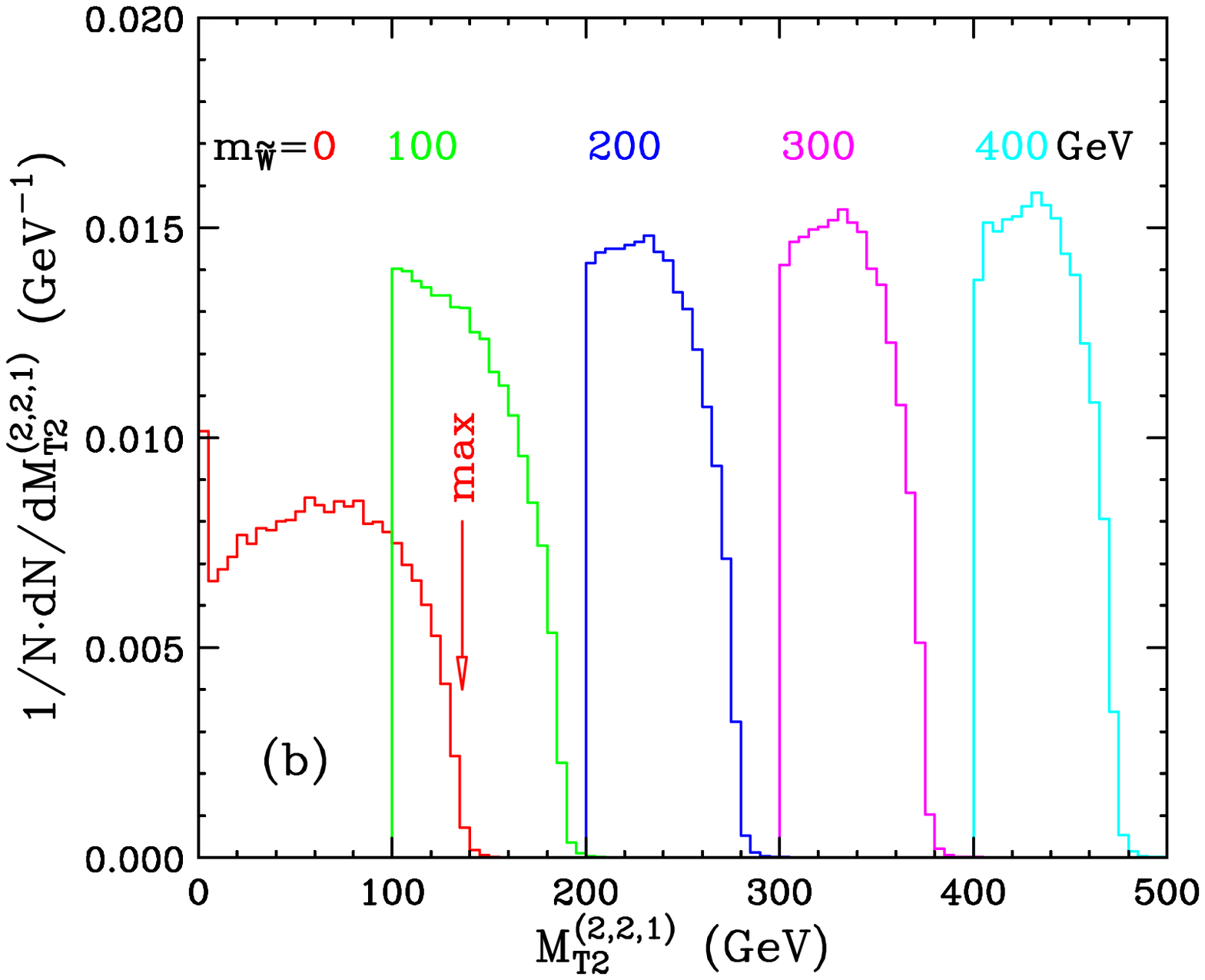,height=6.0cm}
\epsfig{file=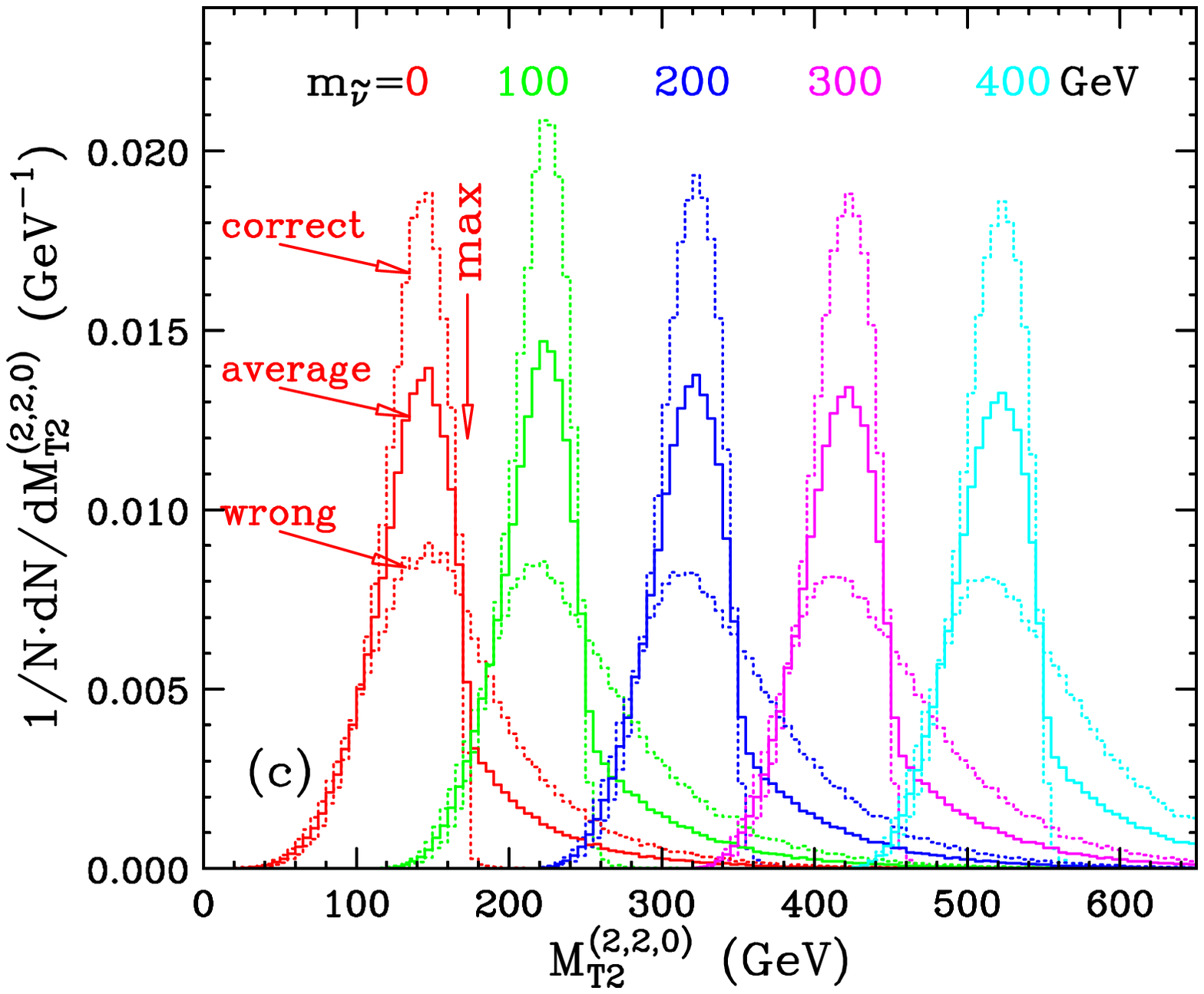,height=6.0cm}~
\epsfig{file=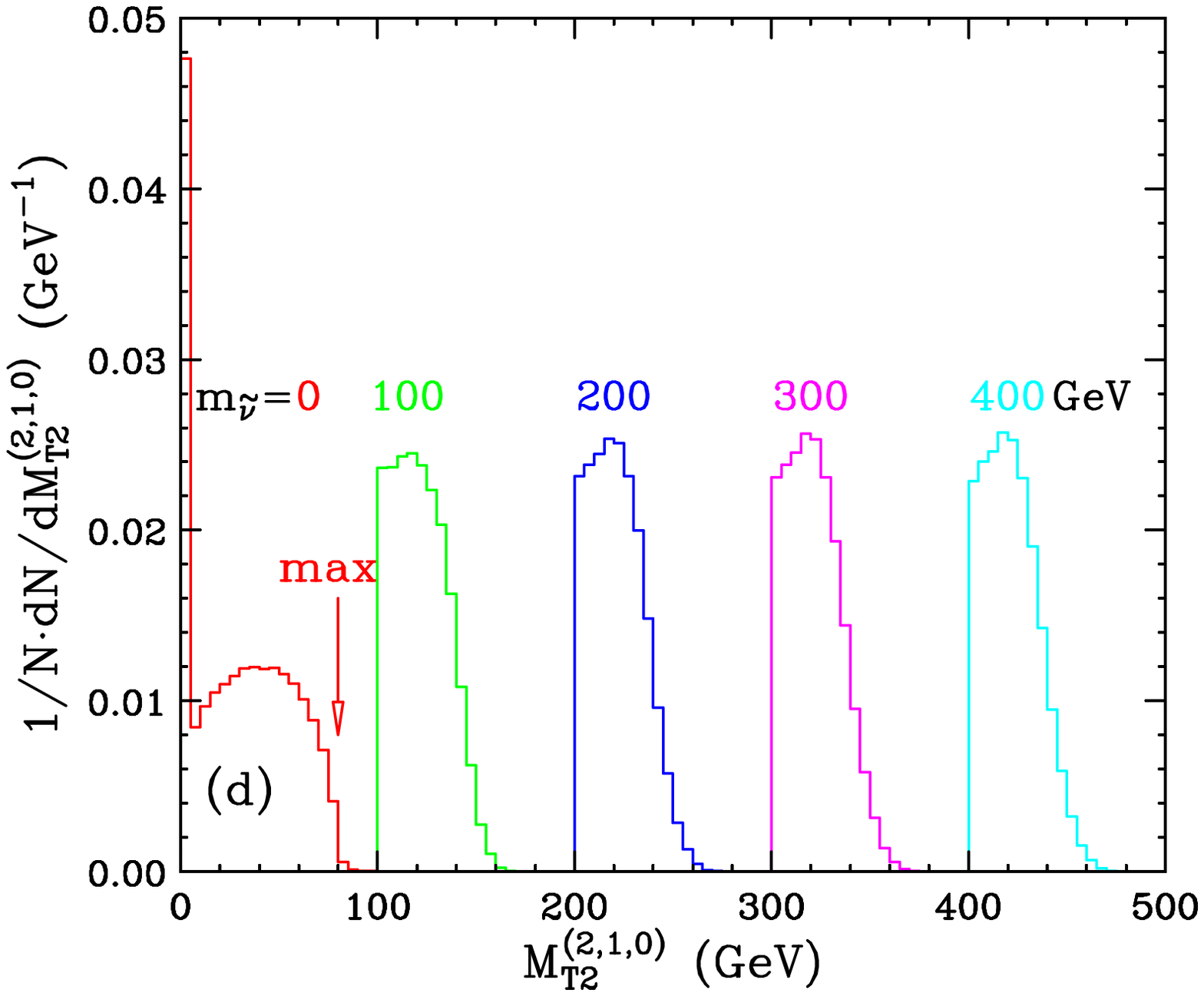,height=6.0cm}
\caption{\sl Unit-normalized distributions of $M_{T2}^{(n,p,c)}$ variables 
in dilepton events from (a) $W^+W^-$ pair production and 
(b-d) $t\bar{t}$ pair production. Each panel shows results for
five different values (0, 100, 200, 300 and 400 GeV)
of the corresponding test mass. The methods of Secs.~\ref{sec:ourmt2}
and \ref{sec:ourim} only make use of the $M_{T2}$
endpoint at zero test mass, $M_{T2,max}^{(n,p,c)}(\tilde M_c=0)$,
which is indicated by the vertical red arrow. In panel (c),
the two dotted line $M_{T2}^{(2,2,0)}$ distributions correspond to  
the correct and the wrong pairing of the two $b$-jets with the leptons,
while the solid line distribution is the average of these two.
}
\label{fig:mt2top}}

Before we begin, let us review the four different $M_{T2}^{(n,p,c)}$ 
variables which are in principle available in that case. 
Each one of them is plotted in Fig.~\ref{fig:mt2top} 
for five different values of the corresponding test mass
(0, 100, 200, 300 and 400 GeV). In Fig.~\ref{fig:mt2top}(a) we show
the $M_{T2}^{(1,1,0)}$ variable from $W^+W^-$ pair production events,
while in  Figs.~\ref{fig:mt2top}(b-d) we correspondingly show the 
$M_{T2}^{(2,2,1)}$, $M_{T2}^{(2,2,0)}$ and $M_{T2}^{(2,1,0)}$ variables  
from $t\bar{t}$ events. We used PYTHIA for event generation and 
did not impose any selection cuts, since they will not affect 
the location of the $M_{T2}$ endpoint\footnote{The cuts would have an impact 
on the overall acceptance and efficiency. This effect is not relevant 
here, since we are showing unit-normalized distributions. The cuts may 
also distort the shape of each distribution, but should preserve 
the location of the upper endpoint.}. The plots are made for the Tevatron
(a $p\bar{p}$ collider with a 2 TeV center-of-mass energy), where
the relevant data is already available.
The corresponding analysis for the LHC is very similar.
All of our plots in this Section have the full ISR effects.
As discussed in Section~\ref{sec:example}, the presence of 
ISR with nonzero $p_T$ will increase the nominal $M_{T2}^{(n,p,c)}$
endpoints:
\begin{equation}
M_{T2,max}^{(n,p,c)}(\tilde M_c,p_T) \ge M_{T2,max}^{(n,p,c)}(\tilde M_c,0)\, ,
\end{equation}
where the equality is obtained only when $\tilde M_c = M_c$.
ISR will therefore introduce some systematic error when 
one is trying to measure $M_{T2,max}^{(n,p,c)}(\tilde M_c,0)$.
The size of this error depends on the ISR $p_T$ spectrum,
which in turn depends on the type of collider (Tevatron or LHC). 
At the Tevatron, this will not be such a serious issue,
as evidenced from Fig.~\ref{fig:mt2top}, where the observed endpoints
in the presence of ISR match pretty well with their expected values 
for the $p_T=0$ case. On the other hand, at the LHC this may become 
a problem, which can be handled in one of two ways.
First, depending on the particular signature, one may be able 
to select a sample with $p_T\approx 0$ (at a certain cost in statistics), 
by imposing a suitably designed jet veto to remove jets from ISR.
Alternatively, one can use the full event sample (which would include 
ISR jets), and make use of our general formulas in Appendix~\ref{app:mt2max},
which contain the explicit $p_T$ dependence of $M_{T2,max}^{(n,p,c)}$.

In the previous Section~\ref{sec:example} 
we derived that in the case of  $n\le2$ cascades, there are 8 different 
$M_{T2}$ endpoint measurements: 
one for $M_{T2,max}^{(1,1,0)}$ (see eq.~(\ref{mT2110}) and Sec.~\ref{sec:mt2_110}),  
one for $M_{T2,max}^{(2,2,1)}$ (see eq.~(\ref{mT2221}) and Sec.~\ref{sec:mt2_221}),  
three for $M_{T2,max}^{(2,2,0)}$ (see eqs.~(\ref{mT2220_low}-\ref{mT2220_high2}) 
and Sec.~\ref{sec:mt2_220}), and 
three for $M_{T2,max}^{(2,1,0)}$ (see eqs.~(\ref{mT2210_low}-\ref{mT2210_high2})
and Sec.~\ref{sec:mt2_210}).
Given that we are trying to determine only three masses $M_0$, 
$M_1$ and $M_2$, it is clear that these 8 measurements should 
be sufficient to completely determine the spectrum. 
We also see that our previous count (\ref{mt2count}) has actually 
greatly underestimated the power of $M_{T2}$, and the number of
available measurements is in fact much larger than the number of 
$M_{T2,max}^{(n,p,c)}$ variables. Indeed, as 
shown in Sections~\ref{sec:mt2_220} and \ref{sec:mt2_210},
there are cases where we might be able to obtain more than 
one mass constraint from a given $M_{T2,max}^{(n,p,c)}$ variable.
Of course, the 8 measurements cannot all be independent among 
themselves, as they only depend on three parameters. Our three 
methods below will be distinguished based on which subset of 
these measurements we are using. 

\subsection{Pure $M_{T2}$ endpoint method}
\label{sec:ourmt2}

With this method, we use $M_{T2}$ endpoint measurements 
$E_{npc}$ at a single fixed value of the test mass, which 
for convenience we take to be $\tilde M_c=0$:
\begin{equation}
E_{npc} \equiv M_{T2,max}^{(n,p,c)}(\tilde M_c=0,p_T=0)\ .
\end{equation}
The corresponding formulas interpreting those 
measurements in terms of the physical masses
$M_0$, $M_1$ and $M_2$ were derived in Sec.~\ref{sec:example}: 
\begin{eqnarray}
E_{110} &\equiv& M_{T2,max}^{(1,1,0)}(0,0) = \frac{M_1^2-M_0^2}{M_1} = M_2 \sqrt{y}\, (1-z)\, , \label{meas110} \\
E_{221} &\equiv& M_{T2,max}^{(2,2,1)}(0,0) = \frac{M_2^2-M_1^2}{M_2} = M_2 \, (1-y) \, , \label{meas221} \\
E_{220} &\equiv& M_{T2,max}^{(2,2,0)}(0,0) = \frac{M_2^2-M_0^2}{M_2} = M_2 \, (1-yz) \, , \label{meas220} \\
E_{210} &\equiv& M_{T2,max}^{(2,1,0)}(0,0) = \frac{1}{M_2}\, \sqrt{(M_2^2-M_0^2)(M_1^2-M_0^2)}
= M_2 \sqrt{y(1-z)(1-yz)}\ . \label{meas210}
\end{eqnarray}
Using the mass spectrum (\ref{top}), the predicted 
locations of these four $M_{T2}$ endpoints are
\begin{eqnarray}
E_{110} &=& 
80\ {\rm GeV}\, , \label{eq:end110} \\
E_{221} &=& 
136\ {\rm GeV}\, , \label{eq:end221} \\
E_{220} &=& 
173\ {\rm GeV}\, , \label{eq:end220} \\
E_{210} &=& 
80 \ {\rm GeV}\ , \label{eq:end210}
\end{eqnarray}
which are marked with the vertical red arrows in Fig.~\ref{fig:mt2top}.
Given that we have four measurements (\ref{meas110}-\ref{meas210})
for only three parameters $M_0$, $M_1$ and $M_2$, one should be able
to uniquely determine all three of the unknown parameters.
Naively, it seems that using just three of the measurements 
(\ref{meas110}-\ref{meas210}) should be sufficient for this purpose, 
and furthermore, that {\em any} three of the measurements 
(\ref{meas110}-\ref{meas210}) will do the job. However, 
one should exercise caution, since not all four 
measurements (\ref{meas110}-\ref{meas210}) are independent.
It is easy to check that $E_{221}$,
$E_{220}$ and $E_{210}$ obey the following relation
\begin{equation}
E_{210}^2
= E_{220}\left(E_{220}-E_{221}  \right)\ .
\label{meas_constraint}
\end{equation}
This means that in order to be able to solve for the masses
from eqs.~(\ref{meas110}-\ref{meas210}), we must always 
make use of the $E_{110}$ measurement
in eq.~(\ref{meas110}), and then we have the freedom to choose
any two out of the remaining three measurements (\ref{meas221}-\ref{meas210}).
For example, using the set of three measurements $\{E_{110},E_{221},E_{220}\}$
(i.e. eqs.~(\ref{meas110}-\ref{meas220})), the masses are 
uniquely determined as
\begin{eqnarray}
M_0 &=& \frac{E_{110}\,\left\{ E_{221}(E_{220}-E_{221})\left[\,E_{220}\,(E_{220}-E_{221})-E^2_{110}\,\right]\right\}^{\frac{1}{2}}} {E^2_{110}-(E_{220}-E_{221})^2}\ , \\ [2mm] 
M_1 &=& \frac{E_{110}\,E_{221}\,(E_{220}-E_{221})} {E^2_{110}-(E_{220}-E_{221})^2}\ ,  \\ [2mm] 
M_2 &=& \frac{E^2_{110}\,E_{221}}{E^2_{110}-(E_{220}-E_{221})^2}\ .
\end{eqnarray}
Similarly, one can solve for $M_0$, $M_1$ and $M_2$ using the set
of measurements $\{E_{110},E_{220},E_{210}\}$, or alternatively,
the set of measurements $\{E_{110},E_{221},E_{210}\}$.
In each case, the remaining fourth unused measurement provides a 
useful consistency check on the mass determination.

\subsection{$M_{T2}$ endpoint shapes and kinks}
\label{sec:ourkink}

The method proposed in Sec.~\ref{sec:ourmt2} uses the measured 
endpoints from {\em several different} $M_{T2}^{(n,p,c)}$
variables. Now we discuss an alternative method which makes use
of {\em a single} $M_{T2}^{(n,p,c)}$ variable. 

Let us begin with the simplest case of $n=1$ as shown in Fig.~\ref{fig:n1n2}(a). 
In that case, we have only one $M_{T2}$ variable at our disposal,
namely $M_{T2}^{(1,1,0)}$. Its properties were discussed in 
Sec.~\ref{sec:mt2_110}, where we showed that its endpoint 
$M_{T2,max}^{(1,1,0)}$ can allow the determination of
{\em both} masses $M_0$ and $M_1$, at least as a matter of principle.
Indeed, the endpoint measurement (\ref{meas110}) at zero test mass provides 
one relation among $M_0$ and $M_1$. The key observation in Sec.~\ref{sec:mt2_110} 
(which was first done in \cite{Barr:2007hy})
was that with the inclusion of ISR effects, the endpoint function 
$M_{T2,max}^{(1,1,0)}(\tilde M_0,p_T)$ exhibits a kink at $\tilde M_0=M_0$, 
which can then be used to determine both masses $M_0$ and $M_1$.
The method can be readily applied to the existing dilepton event sample 
from $W^+W^-$ pair production, which will allow an independent measurement
of the $W$ mass $m_W$ and the neutrino mass $m_\nu$. While the precision of this
measurement will not be competitive with existing $W$ and neutrino mass
determinations, it is nevertheless useful to test the viability of this 
approach with real data.

Now let us discuss the more complicated case of $n=2$, which in our 
example corresponds to $t\bar{t}$ pair production with both tops decaying 
leptonically. As discussed in Secs.~\ref{sec:mt2_221}, \ref{sec:mt2_220}
and \ref{sec:mt2_210}, here we have a choice of three different
$M_{T2}$ variables: $M_{T2}^{(2,2,1)}$, $M_{T2}^{(2,2,0)}$, and 
$M_{T2}^{(2,1,0)}$. Because of the larger $t\bar{t}$ cross-section, 
we expect that the statistical precision on each one of those
three variables will be better than the $M_{T2}^{(1,1,0)}$
variable of the $n=1$ case. As shown in Secs.~\ref{sec:mt2_220} 
and \ref{sec:mt2_210}, each of the two variables $M_{T2}^{(2,2,0)}$ 
and $M_{T2}^{(2,1,0)}$ exhibits a kink in its endpoint $M_{T2,max}^{(n,p,c)}$
when considered as a function of the test mass $\tilde M_0$,
even when the transverse momentum of the intrinsic ISR in the event is zero, 
$p_T=0$.
Then, which of these two variables is better suited for a mass determination?
The case of $M_{T2,max}^{(2,2,0)}(\tilde M_0,p_T=0)$ was already
discussed in \cite{Cho:2007qv,Barr:2007hy,Cho:2007dh}.
Here we would like to propose the alternative measurement of 
$M_{T2,max}^{(2,1,0)}(\tilde M_0,p_T=0)$. What is more, we would like to
emphasize that our function $M_{T2,max}^{(2,1,0)}(\tilde M_0,p_T=0)$ 
offers several unique advantages over the previously considered case of
$M_{T2,max}^{(2,2,0)}(\tilde M_0,p_T=0)$:
\begin{enumerate}
\item The subsystem variable $M_{T2}^{(2,1,0)}$ does not suffer from the 
combinatorics problem which is present for $M_{T2}^{(2,2,0)}$.
Indeed, when constructing the $M_{T2}^{(2,2,0)}$
distribution, one has to decide how to pair up the $b$-jets 
with the two leptons. Because it is difficult to distinguish between 
a $b$ and a $\bar{b}$, there is a two-fold ambiguity which is
quite difficult to resolve by other means. In contrast, our 
subsystem variable $M_{T2}^{(2,1,0)}$ does not make direct use of the
$b$-jets, and is therefore free of such combinatorics issues.
\item As we already saw in Sec.~\ref{sec:mt2_220}, even under 
perfect experimental conditions, the fit to the $M_{T2,max}^{(2,2,0)}$
endpoint results in two separate solutions for the mass spectrum:
one solution (see (\ref{SPS1a})) is given by the true values of 
the input masses, while the second solution (see (\ref{SPS1afake}))
is obtained by the transformation (\ref{amb220}). Using 
$M_{T2,max}^{(2,2,0)}$ alone, there is no way to tell the 
difference between these two mass spectra. In contrast, our
variable $M_{T2}^{(2,1,0)}$ does not suffer from this ambiguity, 
and according to our results from Sec.~\ref{sec:mt2_210} 
the solution is always unique.
\item The third advantage of the subsystem variable $M_{T2}^{(2,1,0)}$
is related to the expected precision on the determination of the 
masses. As we pointed out in Sec.~\ref{sec:mt2_210} and illustrated
explicitly in Fig.~\ref{fig:kink}, the kink $\Delta\Theta^{(2,1,0)}$ 
in the $M_{T2,max}^{(2,1,0)}(\tilde M_0,p_T=0)$ function is much sharper 
than the corresponding kink $\Delta\Theta^{(2,2,0)}$
in the $M_{T2,max}^{(2,2,0)}(\tilde M_0,p_T=0)$ function. 
This can also be seen explicitly from the two examples shown 
in Fig.~\ref{fig:mt2kink}. As a result, we expect that 
the kink structure can be better identified in the case of 
$M_{T2,max}^{(2,1,0)}$, which would lead to smaller errors 
on the mass determination. 
\end{enumerate}
Of course, one could (and in fact should) use the experimental 
information from {\em both} $M_{T2,max}^{(2,2,0)}$ 
and $M_{T2,max}^{(2,1,0)}$, if available. Our main
goal here is simply to point out the obvious advantages of the
subsystem variable $M_{T2}^{(2,1,0)}$, which so far has not been 
used in the literature.

\subsection{Hybrid method: $M_{T2}$ endpoints plus an invariant mass endpoint}
\label{sec:ourim}

As discussed in Sec.~\ref{sec:imd}, any cascade with $n\ge 2$ 
will provide a certain number of measurements (\ref{Nm_im})
in addition to the $M_{T2}$ measurements discussed so far.
In particular, for the $n=2$ example considered here, there will be 
one measurement of the endpoint of the $M_{x_1x_2}$ invariant mass 
distribution. The formula for the endpoint $M_{x_1x_2,max}$
in terms of the unknown physical masses $M_0$, $M_1$ and $M_2$
is in general given by
\begin{equation}
E_{im} \equiv M_{x_1x_2,max}=\frac{1}{M_1}\sqrt{\left(M_2^2-M_1^2\right) \left(M_1^2-M_0^2\right) }
= M_2 \sqrt{(1-y)(1-z)}\ .
\label{measim}
\end{equation}
In the case of $t\bar{t}$ events considered here, this 
is simply the endpoint of the invariant mass distribution $m_{b\ell}$ of 
each lepton and its corresponding $b$-jet. This distribution 
(unit-normalized) is shown in Fig.~\ref{fig:mbl}.
Unfortunately, here one is facing the same combinatorial problem as 
with the $M_{T2}^{(2,2,0)}$ variable -- we cannot easily tell the
charge of the $b$-jet, therefore a priori it is not clear which
$b$-jet goes with which lepton. Fortunately, there are only two possibilities:
the result from the correct (wrong) pairing is shown in Fig.~\ref{fig:mbl}
with the green (blue) dotted line. We see that the green histogram
with the correct pairing has an endpoint at the expected location
\begin{equation}
E_{im} =\sqrt{\frac{\left(m_t^2-m_W^2\right) \left(m_W^2-m_\nu^2\right) }{m_W^2}}
= 153.4\ {\rm GeV} \, ,
\label{eq:endim}
\end{equation}
with a relatively small tail due to the finite width effects.
More importantly, the (blue) distribution from the wrong pairings 
is relatively smooth, and as a result the endpoint (\ref{eq:endim}) 
is preserved in the experimentally observable (red) distribution, 
which includes all possible $b\ell$ pairings.

\FIGURE[t]{
\epsfig{file=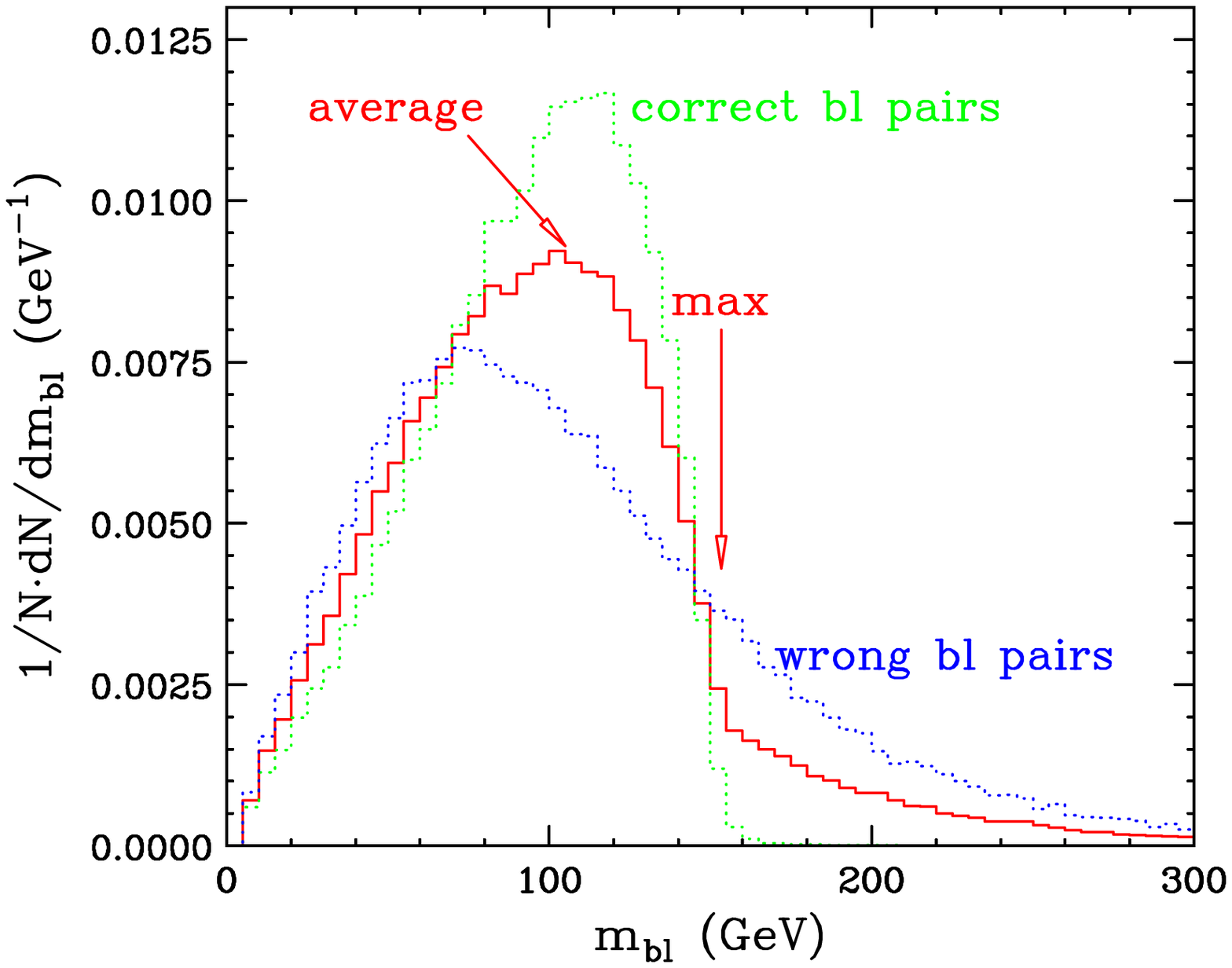,width=10cm}
\caption{\sl Unit-normalized $m_{b\ell}$ invariant mass-squared 
distributions in dilepton $t\bar{t}$ events. The green (blue) dotted line corresponds
to the correct (wrong) pairing of the leptons and the $b$-jets, while the 
red solid line is the average of those two distributions. The endpoint 
(\ref{measim}) of the $m_{b\ell}$ distribution is marked by the 
vertical red arrow.}
\label{fig:mbl}}

Now we can add the new measurement (\ref{measim}) to the previously discussed
set of measurements (\ref{meas110}-\ref{meas210}). We obtain a total of five
measurements for the three underlying parameters $M_0$, $M_1$ and $M_2$,
therefore there exist two relations among the measurements. The first relation
is already given by (\ref{meas_constraint}) and does not involve the
invariant mass endpoint (\ref{measim}). The second relation is given by
\begin{equation}
E^2_{im} = \frac{E_{221}E_{110}^2}{E_{220}-E_{221}}\ .
\label{meas_constraint2}
\end{equation}
We can now consider a hybrid method, which would make use of the invariant 
mass endpoint (\ref{measim}), plus any two of the $M_{T2}$ measurements 
(\ref{meas110}-\ref{meas210}). In principle, one again needs to be careful 
and make sure that the three used measurements are independent. 
Fortunately, as seen from eqs.~(\ref{meas_constraint},\ref{meas_constraint2}), 
the invariant mass endpoint $E_{im}$ is independent from {\em any}
pair of $M_{T2}$ measurements. There are 6 possible pairs among
the $M_{T2}$ measurements (\ref{meas110}-\ref{meas210}), and 
in principle each one can be used in combination with the invariant 
mass endpoint (\ref{measim}). What is the best choice?
We find that in all 6 of those cases one obtains a unique solution 
for the masses $M_0$, $M_1$ and $M_2$. Therefore, the optimal choice 
is dictated by the experimental precision on each of the measurements 
(\ref{meas110}-\ref{meas210}). We expect that the measurement
(\ref{end110}) of $M_{T2,max}^{(1,1,0)}$ will be less precise due to 
the smaller cross-section for $W^+W^-$ production. Similarly,
$M_{T2,max}^{(2,2,0)}$ suffers from the combinatorial problem already mentioned earlier.
Therefore for our illustration of the hybrid method we choose to use
the $M_{T2}^{(2,2,1)}$ endpoint (\ref{meas221}), 
the $M_{T2}^{(2,1,0)}$ endpoint (\ref{meas210}),
and the invariant mass endpoint (\ref{measim}).
The solution for the masses in terms of those three measurements is given by
\begin{eqnarray}
M_0 &=& \frac{\sqrt{2E_{221}}\, E_{im} \left( 2 E_{221} E_{210}^2 + E_{221} E_{im}^2 - E_{im}^2 \sqrt{E_{221}^2+4E_{210}^2} \right)^{\frac{1}{2}} }
{E_{221}^2 + 2 E_{im}^2 - E_{221} \sqrt{E_{221}^2+4E_{210}^2} } \ , \\ [2mm]
M_1 &=& \frac{\sqrt{2}\, E_{221}\, E_{im} \left( E_{221} \sqrt{E_{221}^2 + 4E_{210}^2} - E_{221}^2 \right)^{\frac{1}{2}} }
{E_{221}^2 + 2 E_{im}^2 - E_{221} \sqrt{E_{221}^2+4E_{210}^2} } \ , \\ [2mm]
M_2 &=& \frac{ 2 E_{221} E_{im}^2 }
{E_{221}^2 + 2 E_{im}^2 - E_{221} \sqrt{E_{221}^2+4E_{210}^2} }\ .
\end{eqnarray}
It is easy to check that substituting the measured values of the endpoints 
$E_{221}$, $E_{210}$ and $E_{im}$ from (\ref{eq:end221}), (\ref{eq:end210})
and (\ref{eq:endim}), into the three equations above yields the
values for the neutrino, $W$ and top quark mass, correspondingly.
Notice that this determination was done in a completely model-independent 
fashion and without any prior assumptions, unlike the analysis of 
Ref.~\cite{Cho:2008cu}, which made use the known values of the $W$ 
and/or neutrino masses. 

Before concluding this section, we should emphasize again that the $W^+W^-$ and $t\bar{t}$
data required to test each of the three methods described in this section, already exists 
at the Tevatron. Therefore we use this opportunity to encourage the CDF and D0
collaborations to perform a model-independent mass determination analysis along
the lines presented here.

\section{Summary and conclusions}
\label{sec:conclusions}

In this section we shall briefly summarize our main results.
\begin{itemize}
\item We compared the three main methods previously proposed for mass measurements 
in cascade decays with semi-invisibly decaying particles. We showed that the
endpoint method (Sec.~\ref{sec:imd}) and the polynomial method (Sec.~\ref{sec:exact})
are able to completely determine the mass spectrum only if the decay chain 
is sufficiently long, namely $n\ge 3$. The same conclusion applies if we 
consider a hybrid combination of the endpoint and polynomial methods.
As a corollary, when the decay chain happens to be relatively short, $n\le 2$, 
these two methods are not sufficient, and one {\em must} somehow resort to the 
third, $M_{T2}$, method, in order to completely pin down the mass spectrum. 
Then in Sec.~\ref{sec:mt2} we argued 
that the $M_{T2}$ method {\em by itself} is sufficient for a complete
mass spectrum determination, even in the problematic cases of $n=1$ or $n=2$.
In Sec.~\ref{sec:our}  we backed our claim with two explicit examples:
$W^+W^-$ pair production, which is an example of an $n=1$ chain, 
and $t\bar{t}$ pair production, which is an example of an $n=2$ chain.
We showed that the $M_{T2}$ method in principle provides more than 
enough measurements for the unambiguous determination of the complete 
mass spectrum.
\item When applying the $M_{T2}$ method, we generalized the concept
of $M_{T2}$ by introducing various subsystem (or subchain) $M_{T2}^{(n,p,c)}$ variables.
The latter are defined similarly to the conventional $M_{T2}$ variable, 
but are labelled by three integers $n$, $p$, and $c$, whose meaning is as follows.
The integer $n$ labels the ``grandparent'' particle originally 
produced in the hard scattering and initiating the decay chain. 
We then apply the usual $M_{T2}$ concept to the subchain 
starting at the ``parent'' particle labelled by $p$ and
terminating at the ``child'' particle labelled by $c$. 
In general, the ``child'' particle does not have to be the very last
(i.e. the missing) particle in the decay chain, just like the
``parent'' particle does not have to be the very first particle 
produced in the event. The introduction of the $M_{T2}^{(n,p,c)}$ 
subchain variables greatly proliferates the number of available $M_{T2}$-type 
measurements, and allows us to make full use of the power of the $M_{T2}$
concept. 
\item In Sec.~\ref{sec:example} and Appendix \ref{app:mt2max} 
we provided analytical expressions for the endpoints of all 
$n\le 2$ subsystem $M_{T2}^{(n,p,c)}$ variables,
as a function of the corresponding test mass $\tilde M_c$,
and for arbitrary values of the ISR $p_T$.
Such results for general $p_T$ are being presented here for the first time.
Only some special cases of our results have so far 
appeared in the literature, for example the functions 
$M_{T2}^{(1,1,0)}(\tilde M_0,p_T=0)$ and 
$M_{T2}^{(2,2,0)}(\tilde M_0,p_T=0)$ were already derived in \cite{Cho:2007dh},
from where one could also deduce the form of $M_{T2}^{(2,2,1)}(\tilde M_1,p_T=0)$.
Our result for $M_{T2}^{(2,1,0)}(\tilde M_0,p_T=0)$ discussed in
Sec.~\ref{sec:mt2_210} is also new.
\item In Sec.~\ref{sec:example} we showed that the endpoint functions
$M_{T2}^{(n,p,c)}(\tilde M_c,p_T)$ may exhibit up to three different types of kinks. 
All three kinks appear in the same location: $\tilde M_c=M_c$, at which point
the value of the subchain endpoint $M_{T2,max}^{(n,p,c)}$ coincides with the 
true parent mass:
\begin{equation}
M_{T2,max}^{(n,p,c)}(\tilde M_c = M_c, p_T)=M_p\ .
\end{equation}
(This equation generalizes eqs.~(\ref{true110}), (\ref{kink220})
and (\ref{true210}).) However, the physical origin of each type of kink is different.
\begin{enumerate}
\item The first type of kink was originally identified in \cite{Gripaios:2007is,Barr:2007hy}
and arises solely as an ISR effect, so that in principle it should always
be present at some level. The sharpness of this kink
depends on the transverse momentum $p_T$ of the ISR objects. This particular kink 
type was responsible for the kink feature $\Delta\Theta^{(1,1,0)}$ 
observed in our very first and simplest example in Sec.~\ref{sec:mt2_110}. 
There we generalized the
existing analytical formulas for the endpoint $M_{T2,max}^{(1,1,0)}$
by including the effects of the ISR. This in turn allowed us to analyze 
the amount of kink $\Delta\Theta^{(1,1,0)}$ as a function of the ISR $p_T$ and
the mass spectrum, see Fig.~\ref{fig:kink110}.
\item The second type of kink, encountered in Sec.~\ref{sec:mt2_220},
was originally discovered in \cite{Cho:2007qv,Cho:2007dh} and is due to 
the variable mass of the composite system of visible (SM) particles 
used in the $M_{T2}$ calculation. To be more precise, this type of kink 
requires the following relation between the parent 
and child indices:
\begin{equation}
p>c+1\ ,
\label{condkink2}
\end{equation}
and does not require any ISR, i.e. it is present even when $p_T=0$.
Among the four $M_{T2}^{(n,p,c)}$ variables discussed in Sec.~\ref{sec:example},
only $M_{T2}^{(2,2,0)}$ satisfies this condition. Not surprisingly,
we encountered this kink during our discussion 
of the $M_{T2,max}^{(2,2,0)}$ endpoint in Sec.~\ref{sec:mt2_220}.
There we quantified the amount of kink $\Delta\Theta^{(2,2,0)}$
as a function of the mass spectrum, as shown in Fig.~\ref{fig:mt2kink}(a).
\item The third type of kink, which we encountered in Sec.~\ref{sec:mt2_210}
during our discussion of the $M_{T2}^{(2,1,0)}$ variable,
is new, and to the best of our knowledge has not been discussed 
in the existing literature\footnote{Ref.~\cite{Cheng:2008hk} considered
the decay chain $\tilde q\to \tilde\chi^0_2\to\tilde \ell\to \tilde\chi^0_1$
and used the two leptons to form an $M_{T2}$ variable which 
in our notation would correspond to $M_{T2}^{(3,2,0)}$,
i.e. $n=3$, $p=2$ and $c=0$. Since this case satisfies both 
conditions (\ref{condkink2}) and (\ref{condkink3}), the kink
observed in \cite{Cheng:2008hk} is a combination of the second 
and third kink types according to our classification.}. 
This kink arises due to the decays of heavier particles above the 
parent level and exists even in the absence of any ISR: its presence
simply requires the following relation between the grandparent and
parent indices
\begin{equation}
n>p\ ,
\label{condkink3}
\end{equation}
while the child index $c$ is arbitrary. Because of the upstream objects,
this situation does {\em not} correspond to the balanced momentum 
configuration discussed in \cite{Lester:2007fq,Cho:2007dh}. 
Therefore, the analytical expressions for $M_{T2}$ derived in those
two papers are not applicable and will need to be generalized \cite{BKMP}.
In general, there are two sources of momentum imbalance in this case:
upstream objects from grandparent decays, as well as genuine ISR with $p_T\ne 0$. 
In Sec.~\ref{sec:mt2_210} we concentrated on the former effect and 
provided analytical expressions 
for the endpoint function $M_{T2,max}^{(2,1,0)}(\tilde M_0,p_T=0)$
and the associated kink $\Delta\Theta^{(2,1,0)}$, which allowed us to
quantify the sharpness of the kink as a function of the mass spectrum,
see Fig.~\ref{fig:mt2kink}(b). Comparing Fig.~\ref{fig:mt2kink}(b)
to Fig.~\ref{fig:mt2kink}(a), we saw that the new kink $\Delta\Theta^{(2,1,0)}$
is in general much more pronounced, and therefore offers better prospects for 
experimental detection. The corresponding formulas for the more 
general case, when both grandparent decay products
and genuine ISR with arbitrary $p_T$ are present, 
are listed in Appendix~\ref{app:mt2max}.
\end{enumerate}
Of course, there are cases when two or even all three of these kinks 
will be simultaneously present. For example, 
$M_{T2,max}^{(2,2,0)}(\tilde M_0,p_T\ne 0)$ will exhibit kinks 1 and 2,
$M_{T2,max}^{(2,1,0)}(\tilde M_0,p_T\ne 0)$ will exhibit kinks 1 and 3,
while
$M_{T2,max}^{(3,2,0)}(\tilde M_0,p_T\ne 0)$ will exhibit all three: 1, 2 and 3.
\item Our $M_{T2}$ analysis in Sec.~\ref{sec:example} revealed that 
in the case of an $n\le 2$ cascade, there exist 8 different measurements
of subsystem $M_{T2}$ endpoints
\begin{enumerate}
\item One measurement (\ref{mT2110}) of the endpoint $M_{T2,max}^{(1,1,0)}$
at zero test mass $\tilde M_0$.
\item One measurement (\ref{mT2221}) of the endpoint $M_{T2,max}^{(2,2,1)}$
at zero test mass $\tilde M_1$.
\item One measurement (\ref{mT2220_low}) of the endpoint $M_{T2,max}^{(2,2,0)}$
at zero test mass $\tilde M_0$.
\item Two measurements (\ref{mT2220_high1}) and (\ref{mT2220_high2})
of the endpoint $M_{T2,max}^{(2,2,0)}$ at two large values ($E_b$ and $E'_b$) of the 
test mass $\tilde M_0$.
\item One measurement (\ref{mT2210_low}) of the endpoint $M_{T2,max}^{(2,1,0)}$
at zero test mass $\tilde M_0$.
\item Two measurements (\ref{mT2210_high1}) and (\ref{mT2210_high2})
of the endpoint $M_{T2,max}^{(2,1,0)}$ at two large values ($E_b$ and $E'_b$) of the 
test mass $\tilde M_0$.
\end{enumerate}
In addition, we also have one measurement of the endpoint (\ref{measim})
of the invariant mass distribution $M_{x_1x_2}$, bringing the total number 
of available endpoint measurements to 9. Given that an $n=2$ chain contains 
only three unknown masses $M_0$, $M_1$ and $M_2$, it should be clear that 
the spectrum can be fully determined. In Sec.~\ref{sec:our} we proposed 
three different methods for mass determinations, depending on what subsets 
of all those measurements one decides to use. For example, the pure 
$M_{T2}$ endpoint method of Sec.~\ref{sec:ourmt2} makes use of different
$M_{T2,max}^{(n,p,c)}(\tilde M_c=0)$ endpoint measurements at zero test mass.
The method of Sec.~\ref{sec:ourkink} makes use of a single 
$M_{T2,max}^{(n,p,c)}(\tilde M_c)$ function, measured at different 
values of $\tilde M_c$. Finally, the hybrid method of Sec.~\ref{sec:ourim} 
combines some of the $M_{T2,max}^{(n,p,c)}(\tilde M_c=0)$ endpoint measurements 
at zero test mass with the invariant mass endpoint $M_{x_1x_2,max}$.
\end{itemize}

In conclusion, our work shows that, at least as a matter of principle, 
the $M_{T2}$ concept, when properly generalized to include the subsystem 
variables $M_{T2}^{(n,p,c)}$, can allow the measurement
of the masses of {\em all} particles in SUSY-like events 
with {\em arbitrary} decay chains at hadron colliders.

\bigskip

\acknowledgments
We are grateful to A.~Barr, B.~Gripaios and C.~Lester for useful discussions.
We dedicate this paper to our parents, grandparents and children.
This work is supported in part by a US Department of Energy 
grant DE-FG02-97ER41029. Fermilab is operated by Fermi Research Alliance, LLC under
Contract No. DE-AC02-07CH11359 with the U.S. Department of Energy.

\appendix
\section{Appendix: \ Analytical expressions for $M_{T2,max}^{(n,p,c)}(\tilde M_c, p_T)$}
\label{app:mt2max}
\allowdisplaybreaks
\renewcommand{\theequation}{A.\arabic{equation}}
\setcounter{equation}{0}

The purpose of this Appendix is to collect in one place 
all relevant formulas for the various subsystem $M_{T2}$ endpoints
$M_{T2,max}^{(n,p,c)}(\tilde M_c,p_T)$ in the presence of initial 
state radiation (ISR) with arbitrary transverse momentum $p_T$.
In all cases, we will find that $M_{T2,max}^{(n,p,c)}(\tilde M_c,p_T)$
is given by two branches:
\begin{equation}
M_{T2,max}^{(n,p,c)}(\tilde M_c,p_T) =\left\{   
\begin{array}{ll}
F^{(n,p,c)}_{L}(\tilde M_c,p_T)\, , & ~~~{\rm if}\ \tilde M_c \le M_c\, , \\ [2mm]
F^{(n,p,c)}_{R}(\tilde M_c,p_T)\, , & ~~~{\rm if}\ \tilde M_c \ge M_c\, .
\end{array}
\right.
\label{app:MT2npc} 
\end{equation}
In what follows we shall list the analytic expressions for 
each branch $F^{(n,p,c)}_{L}$ and $F^{(n,p,c)}_{R}$, for 
all possible $(n,p,c)$ cases with $n-c \le 2$. The grandparents
$X_n$, the parents $X_p$ and the children $X_c$ are always assumed 
to be on-shell. However, any intermediate particles $X_m$
with $n>m>p$ or $p>m>c$ may or may not be on-shell, 
and the two cases will have to be treated differently.
Such an example is provided by the endpoint function
$M_{T2,max}^{(n,n,n-2)}(\tilde M_{n-2},p_T)$ discussed below
in Section ~\ref{sec:MT2nnn-2}.
For convenience, our results will be written in terms of 
the mass parameters $\mu_{(n,p,c)}$ defined in (\ref{mu_npc})
\begin{equation}
\mu_{(n,p,c)} \equiv \frac{M_n}{2}\, \left( 1-\frac{M_c^2}{M_p^2}  \right)\ .
\label{app:mu_npc}
\end{equation}
These parameters represent certain combinations of the masses of the 
grandparents ($M_n$), parents ($M_p$) and children ($M_c$), and
do not contain any dependence on the ISR transverse momentum $p_T$.
As we discussed in Secs.~\ref{sec:example} and \ref{sec:our}, 
these are generally the quantities which are directly measured 
by experiment. Therefore, with the $M_{T2}$ method,
the goal of any experiment would be to perform a sufficient 
number of $\mu$-parameter measurements and then from those 
to determine the particle masses themselves.

In some special cases, namely $n=p$, 
we shall also define $p_T$-dependent $\mu$ parameters,
where the $p_T$ dependence is explicitly shown as an argument:
\begin{equation}
\mu_{(n,n,c)}(p_T) = \mu_{(n,n,c)}\, 
\left( \sqrt{ 1+\left(\frac{p_T}{2M_n}\right)^2} - \frac{p_T}{2M_n} \right) \, .
\label{app:munncPT}
\end{equation}
When $p_T = 0$, the $p_T$-dependent parameters (\ref{app:munncPT})
simply reduce to the $p_T$-independent ones (\ref{app:mu_npc}):
\begin{equation}
\mu_{(n,n,c)}(p_T=0) = \mu_{(n,n,c)}\ .
\end{equation}
We also remind the reader that test masses for the children are denoted
with a tilde: $\tilde M_c$, while the true mass of any particle does not 
carry a tilde sign.

\subsection{The subsystem variable $M_{T2,max}^{(n,n,n-1)}(\tilde M_{n-1}, p_T)$}

The corresponding expressions were already given in eqs.~(\ref{FLnnn-1}) 
and (\ref{FRnnn-1}) and we list them here for completeness:
\begin{eqnarray}
&&F_{L}^{(n,n,n-1)}(\tilde M_{n-1},p_T) = \nonumber \\ [2mm]
&=&\left\{ \left[
\mu_{(n,n,n-1)}(p_T) + \sqrt{ \left(\mu_{(n,n,n-1)}(p_T)+\frac{p_T}{2}\right)^2 + \tilde M_{n-1}^2} 
\, \right]^2
- \frac{p_T^2}{4}   \right\}^{\frac{1}{2}},
\label{app:FLnnn-1PT}  \\ [2mm]
&&F_{R}^{(n,n,n-1)}(\tilde M_{n-1},p_T) =  \nonumber \\ [2mm]
&=&\left\{ \left[
\mu_{(n,n,n-1)}(-p_T) + \sqrt{ \left(\mu_{(n,n,n-1)}(-p_T)-\frac{p_T}{2}\right)^2 + \tilde M_{n-1}^2} 
\, \right]^2
- \frac{p_T^2}{4}   \right\}^{\frac{1}{2}},
\label{app:FRnnn-1PT}
\end{eqnarray}
where the $p_T$-dependent parameter $\mu_{(n,n,n-1)}(p_T)$ was already defined in
(\ref{app:munncPT}):
\begin{equation}
\mu_{(n,n,n-1)}(p_T) = \mu_{(n,n,n-1)}\, 
\left( \sqrt{ 1+\left(\frac{p_T}{2M_n}\right)^2} - \frac{p_T}{2M_n} \right) \, .
\label{app:munnn-1PT}
\end{equation}
As already mentioned in Sec.~\ref{sec:mt2_110}, the
left branch $F_{L}^{(n,n,n-1)}$ corresponds to the momentum configuration 
$\left(\vec{p}_{nT}^{\,(1)} \uparrow\uparrow \vec{p}_{nT}^{\,(2)}\right) \uparrow\uparrow \vec{p}_T$,
while the right branch $F_{R}^{(n,n,n-1)}$ corresponds to 
$\left(\vec{p}_{nT}^{\,(1)} \uparrow\uparrow \vec{p}_{nT}^{\,(2)}\right) \uparrow\downarrow \vec{p}_T$.

\subsection{The subsystem variable $M_{T2,max}^{(n,n,n-2)}(\tilde M_{n-2}, p_T)$}
\label{sec:MT2nnn-2}

In this case there is an intermediate particle $X_{n-1}$
between the parent $X_n$ and the child $X_{n-2}$
(see Figs.~\ref{fig:metevent} and \ref{fig:metevent2}).
Our formulas below are written in such a way that 
they can be applied both in the case when
the intermediate particle 
$X_{n-1}$ is on shell ($M_n>M_{n-1}$)
and in the case when $X_{n-1}$
is off-shell  ($M_{n-1}\ge M_n$).

In both cases (off-shell or on-shell)
we find that the left branch of $M_{T2,max}^{(n,n,n-2)}(\tilde M_{n-2}, p_T)$ 
is given by
\begin{equation}
F_{L}^{(n,n,n-2)}(\tilde M_{n-2},p_T) 
=\left\{ \left[
\mu_{(n,n,n-2)}(p_T) + \sqrt{ \left(\mu_{(n,n,n-2)}(p_T)+\frac{p_T}{2}\right)^2 + \tilde M_{n-2}^2} 
\, \right]^2
- \frac{p_T^2}{4}   \right\}^{\frac{1}{2}},
\label{app:FLnnn-2PTon}
\end{equation}
where the $p_T$-dependent parameter $\mu_{(n,n,n-2)}(p_T)$ was already defined in
(\ref{app:munncPT}):
\begin{equation}
\mu_{(n,n,n-2)}(p_T) = \mu_{(n,n,n-2)}\, 
\left( \sqrt{ 1+\left(\frac{p_T}{2M_n}\right)^2} - \frac{p_T}{2M_n} \right) \, .
\label{app:munnn-2PT}
\end{equation}
The right branch $F_{R}^{(n,n,n-2)}$ is given by three different expressions, 
depending on the mass spectrum and the size of the ISR $p_T$:
\begin{eqnarray}
&&F_{R}^{(n,n,n-2)}(\tilde M_{n-2},p_T) =  \label{app:FRnnn-2PTon} \\ [2mm]
&=&\left\{   
\begin{array}{ll}
F^{(n,n,n-2)}_{L}(\tilde M_{n-2},-p_T)\, ,    & ~{\rm if}\ p_T > \frac{M_n^2-M_{n-2}^2}{M_{n-2}}\, , \\ [2mm]
F^{(n,n,n-2)}_{R,off}(\tilde M_{n-2},p_T)\, , & ~{\rm if}\ p_T \le\frac{M_n^2-M_{n-2}^2}{M_{n-2}}\ {\rm and}\ \Delta M_{n,n-2}(p_T) \le M_{x_{n-1}x_{n},max}\, , \\ [2mm]
F^{(n,n,n-2)}_{R,on} (\tilde M_{n-2},p_T)\, , & ~{\rm if}\ p_T \le\frac{M_n^2-M_{n-2}^2}{M_{n-2}}\ {\rm and}\ \Delta M_{n,n-2}(p_T) \ge M_{x_{n-1}x_{n},max}\, .
\end{array}
\right.
\nonumber
\end{eqnarray}
Here $\Delta M_{n,n-2}(p_T)$ is a $p_T$-dependent mass parameter defined as
\begin{equation}
\Delta M_{n,n-2}(p_T) \equiv 
\left\{  \left[\sqrt{ M_n^2+\frac{p_T^2}{4}} - M_{n-2} \right]^2
- \frac{p_T^2}{4}  \right\}^{\frac{1}{2}} ,
\label{app:Mhat}
\end{equation}
which in the limit $p_T\to 0$ reduces to
\begin{equation}
\Delta M_{n,n-2}(p_T=0) = M_n - M_{n-2},
\end{equation}
justifying its notation. Notice that $\Delta M_{n,n-2}(p_T)$
is always well-defined, since it is only used when the condition
$p_T \le (M_n^2-M_{n-2}^2)/M_{n-2}$ is satisfied and 
the expression under the square root in (\ref{app:Mhat}) is nonnegative.
The other mass parameter appearing in (\ref{app:FRnnn-2PTon}),
$M_{x_{n-1}x_{n},max}$, is the familiar endpoint of the invariant mass distribution of the 
$\{x_{n-1}, x_{n}\}$ SM particle pair:
\begin{equation}
M_{x_{n-1}x_{n},max} \equiv 
\left\{   
\begin{array}{ll}
\frac{1}{M_{n-1}}\sqrt{(M^2_{n}-M^2_{n-1})(M^2_{n-1}-M^2_{n-2})} \, ,    & ~{\rm if}\ M_{n-1} < M_n\, , \\ [2mm]
M_n - M_{n-2}\, ,                                                        & ~{\rm if}\ M_{n-1} \ge   M_n\, .
\end{array}
\right.
\label{app:Minvnn-1}
\end{equation} 
For example, in the special case of $n=2$ and the intermediate particle 
$X_1$ on-shell, eq.~(\ref{app:Minvnn-1}) reduces to eq.~(\ref{measim}).
The two expressions $F^{(n,n,n-2)}_{R,off}$ and $F^{(n,n,n-2)}_{R,on}$
appearing in (\ref{app:FRnnn-2PTon}) are given by
\begin{equation}
F_{R,off}^{(n,n,n-2)}(\tilde M_{n-2},p_T) 
 = \left\{ \left[
\tilde M_{n-2} + \sqrt{ {\Delta M}^2_{n,n-2}(p_T) +\frac{p_T^2}{4}} 
\, \right]^2
- \frac{p_T^2}{4}   \right\}^{\frac{1}{2}},
\label{app:FRoff}
\end{equation}
\begin{equation}
F_{R,on}^{(n,n,n-2)}(\tilde M_{n-2},p_T) 
 = \left\{ \left[
\sqrt{ M^2_{x_{n-1}x_{n},max}+p^2_{vis}(p_T)} + \sqrt{ \tilde M^2_{n-2} 
+ \left( p_{vis}(p_T) -\frac{p_T}{2}\right)^2} 
\, \right]^2
- \frac{p_T^2}{4}   \right\}^{\frac{1}{2}},
\label{app:FRon}
\end{equation}
where $\Delta M_{n,n-2}(p_T)$ and $M_{x_{n-1}x_{n},max}$ 
were already defined in
(\ref{app:Mhat}) and (\ref{app:Minvnn-1}), correspondingly.
The subscripts ``off'' and ``on'' in eqs.~(\ref{app:FRoff})
and (\ref{app:FRon}) can be understood as follows.
When the intermediate particle $X_{n-1}$ is off-shell and
$M_{n-1}\ge M_n$, from (\ref{app:Mhat}) and (\ref{app:Minvnn-1}) we get
\begin{equation}
\Delta M^2_{n,n-2}(p_T) = M_n^2 + M_{n-2}^2 - 2 M_n M_{n-2}\sqrt{1+\frac{p_T^2}{4M_n^2}}
\le (M_n-M_{n-2})^2 = M^2_{x_{n-1}x_{n},max}.
\end{equation}
Now returning to the logic of eq.~(\ref{app:FRnnn-2PTon}), we see that
in the off-shell case at low $p_T$ one would always use 
the expression $F_{R,off}^{(n,n,n-2)}(\tilde M_{n-2},p_T)$
defined in eq.~(\ref{app:FRoff}),
and never its alternative $F_{R,on}^{(n,n,n-2)}(\tilde M_{n-2},p_T)$
from eq.~(\ref{app:FRon}).
To put it another way, the expression $F_{R,on}^{(n,n,n-2)}(\tilde M_{n-2},p_T)$
in eq.~(\ref{app:FRon}) is only relevant when the intermediate particle $X_{n-1}$ is on-shell.

Finally, the quantity $p_{vis}(p_T)$ appearing in eq.~(\ref{app:FRon}) 
is a shorthand notation for the total transverse momentum 
of the visible particles $x_n$ and $x_{n-1}$ in each leg:
$$p_{vis}\equiv |\vec{p}_{nT}^{\,(k)} + \vec{p}_{(n-1)T}^{\,(k)}|\ .$$
In the case relevant for $F_{R,on}^{(n,n,n-2)}$, the value of $p_{vis}$
is given by
\begin{equation}
p_{vis}(p_T)\equiv 
(\mu_{(n,n,n-1)}+\mu_{(n,n-1,n-2)})\, \frac{p_T}{2M_n}
+|\mu_{(n,n,n-1)}-\mu_{(n,n-1,n-2)}|\, 
\sqrt{1+\frac{p_T^2}{4M_n^2}}\ .
\end{equation}
It is easy to check that 
in the limit of $p_T \to 0$ our eqs.~(\ref{app:FLnnn-2PTon}) 
and (\ref{app:FRnnn-2PTon}) reduce to the known results for
the case of no ISR (eqs.~(70) and (74) in Ref.~\cite{Cho:2007dh}).

The left branch $F_{L}^{(n,n,n-2)}$ in (\ref{app:FLnnn-2PTon})
corresponds to the momentum configuration 
$$\left(\vec{p}_{nT}^{\,(k)} + \vec{p}_{(n-1)T}^{\,(k)}\right) \uparrow\uparrow \vec{p}_T\ ,$$
while the right branch $F_{R}^{(n,n,n-2)}$ in (\ref{app:FRnnn-2PTon})
corresponds to 
$$\left(\vec{p}_{nT}^{\,(k)} + \vec{p}_{(n-1)T}^{\,(k)}\right) \uparrow\downarrow \vec{p}_T\ .$$
In the latter case, $F_{R,off}^{(n,n,n-2)}$ is obtained when $X_{n-2}$ is at rest:
$P^{(k)}_{(n-2)T}=0$, while $F_{R,on}^{(n,n,n-2)}$ corresponds to the case when 
$P^{(k)}_{(n-2)T}=\frac{1}{2}p_T-p_{vis}(p_T)$.

\subsection{The subsystem variable $M_{T2,max}^{(n,n-1,n-2)}(\tilde M_{n-2}, p_T)$}

Here we generalize our $p_T=0$ result (\ref{end210})
from Sec.~\ref{sec:mt2_210} to the case of arbitrary ISR $p_T$:
\begin{eqnarray}
&&F_{L}^{(n,n-1,n-2)}(\tilde M_{n-2},p_T) = \nonumber \\ [2mm]
&=&\left\{ \left[
\mu_{(n-1,n-1,n-2)}(\hat p_T) + \sqrt{ \left(\mu_{(n-1,n-1,n-2)}(\hat p_T)+\frac{\hat p_T}{2}\right)^2 + \tilde M_{n-2}^2} 
\, \right]^2
- \frac{\hat p_T^2}{4}   \right\}^{\frac{1}{2}},
\label{app:FLnn-1n-2PT}  \\ [2mm]
&&F_{R}^{(n,n-1,n-2)}(\tilde M_{n-2},p_T) =  \nonumber \\ [2mm]
&=&\left\{ \left[
\mu_{(n-1,n-1,n-2)}(-\hat p_T) + \sqrt{ \left(\mu_{(n-1,n-1,n-2)}(-\hat p_T)-\frac{\hat p_T}{2}\right)^2 + \tilde M_{n-2}^2} 
\, \right]^2
- \frac{\hat p_T^2}{4}   \right\}^{\frac{1}{2}}, ~~~~
\label{app:FRnn-1n-2PT}
\end{eqnarray}
where we have introduced the shorthand notation
\begin{equation}
\hat p_T \equiv p_T + 2\mu_{(n,n,n-1)}(p_T)\ .
\label{app:pthatPT}
\end{equation}
Notice that the second term on the right-hand side contains the $p_T$-dependent 
$\mu$ parameter defined in (\ref{app:munnn-1PT}).

The left branch $F_{L}^{(n,n-1,n-2)}$ in (\ref{app:FLnn-1n-2PT})
corresponds to the momentum configuration 
$$\vec{p}_{(n-1)T}^{\,(k)} \uparrow\uparrow \left( \vec{p}_{nT}^{\,(k)} \uparrow\uparrow \vec{p}_T\right)\ ,$$
while the right branch $F_{R}^{(n,n-1,n-2)}$ in (\ref{app:FRnn-1n-2PT})
corresponds to 
$$\vec{p}_{(n-1)T}^{\,(k)} \uparrow\downarrow \left(\vec{p}_{nT}^{\,(k)} \uparrow\uparrow \vec{p}_T\right)\ .$$

It is worth checking that our general $p_T$-dependent results 
(\ref{app:FLnn-1n-2PT}) and (\ref{app:FRnn-1n-2PT})
reduce to our previous formulas (\ref{FL210}) and (\ref{FR210})
in the $p_T\to 0$ limit and in the special case of $n=2$.
First taking the limit $p_T\to 0$ from (\ref{app:pthatPT}) 
and (\ref{app:munnn-1PT}) we get
\begin{equation}
\lim_{p_T\to 0} \hat p_T = 2\mu_{(n,n,n-1)}\ ,
\label{app:pthat}
\end{equation}
\begin{equation}
\lim_{p_T\to 0} \mu_{(n-1,n-1,n-2)}(\hat{p_T}) =
\mu_{(n-1,n-1,n-2)}(2\mu_{(n,n,n-1)}) = \mu_{(n,n,n-2)} - \mu_{(n,n,n-1)} \ , ~~~
\label{app:mupT0L}
\end{equation}
\begin{equation}
\lim_{p_T\to 0} \mu_{(n-1,n-1,n-2)}(-\hat{p_T}) =
\mu_{(n-1,n-1,n-2)}(-2\mu_{(n,n,n-1)}) = \mu_{(n,n-1,n-2)} \ .
\label{app:mupT0R}
\end{equation}
Substituting (\ref{app:pthat}-\ref{app:mupT0R})
into (\ref{app:FLnn-1n-2PT}) and (\ref{app:FRnn-1n-2PT}), we get 
\begin{eqnarray}
&& F_{L}^{(n,n-1,n-2)}(\tilde M_{n-2},p_T=0) = \nonumber \\ [2mm]
&=& \left\{ \left[
\mu_{(n,n,n-2)} - \mu_{(n,n,n-1)} + \sqrt{ \mu_{(n,n,n-2)}^2 + \tilde M_{n-2}^2} 
\, \right]^2
-\mu^2_{(n,n,n-1)}   \right\}^{\frac{1}{2}}\, , \label{FLnn-1n-2} \\ [4mm]
&& F_{R}^{(n,n-1,n-2)}(\tilde M_{n-2},p_T=0) = \nonumber \\ [2mm]
&=& \left\{ \left[
  \mu_{(n,n-1,n-2)}
+ \sqrt{ \left(\mu_{(n,n,n-1)} - \mu_{(n,n-1,n-2)}\right)^2  + \tilde M_{n-2}^2}
\, \right]^2
-\mu^2_{(n,n,n-1)} \right\}^{\frac{1}{2}}\, ,
\label{FRnn-1n-2}
\end{eqnarray}
which are nothing but the generalizations
of (\ref{FL210}) and (\ref{FR210}) for arbitrary $n$.

\listoftables           
\listoffigures          


\end{document}